\documentclass{article}

\usepackage{arxiv}

\usepackage{amsmath,amssymb,amsfonts}
\usepackage[ruled,linesnumbered,vlined]{algorithm2e}
\usepackage{multirow}
\usepackage[utf8]{inputenc} 
\usepackage[T1]{fontenc}    
\usepackage{hyperref}       
\usepackage{url}            
\usepackage{booktabs}       
\usepackage{amsfonts}       
\usepackage{nicefrac}       
\usepackage{microtype}      
\usepackage{graphicx}
\usepackage[numbers,square]{natbib}
\usepackage{doi}
\usepackage{booktabs}
\usepackage{wrapfig}
\usepackage{longtable}
\usepackage{array}
\usepackage{tabularx}
\usepackage{adjustbox}

\title{iGRAIL: Blockchain-Based Generative Federated Learning for Trustworthy Healthcare Intelligence}

\author{%
\normalfont
\begin{tabular}{c}
\href{https://orcid.org/0000-0003-4422-3196}{%
\includegraphics[scale=0.07]{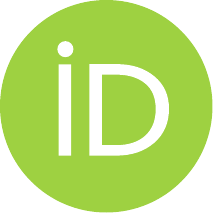}\hspace{1mm}\textbf{Debashis Das}}\\
\normalfont Department of Computer Science\\
Benedict College\\
Columbia, SC, USA\\
\texttt{debashis.das@ieee.org}
\end{tabular}
\\[1.2cm]
\begin{tabular}{@{}c@{\hspace{0.6cm}}c@{}}
\begin{tabular}{c}
\href{https://orcid.org/0000-0002-0104-967X}{%
\includegraphics[scale=0.07]{orcid.pdf}\hspace{1mm}\textbf{Sourav Banerjee}}\\
Department of Computer Science and Engineering\\
Alipurduar Government Engineering\\
and Management College\\
Alipurduar, WB, India\\
\texttt{mr.sourav.banerjee@ieee.org}
\end{tabular}
&
\begin{tabular}{c}
\href{https://orcid.org/0000-0003-2959-6541}{%
\includegraphics[scale=0.07]{orcid.pdf}\hspace{1mm}\textbf{Saraju P. Mohanty}}\\
Department of Computer Science and Engineering\\
University of North Texas\\
Denton, TX, USA\\
\texttt{smohanty@ieee.org}
\end{tabular}
\end{tabular}
}

\date{}

\renewcommand{\headeright}{}
\renewcommand{\undertitle}{}
\renewcommand{\shorttitle}{iGRAIL: Blockchain-Based Generative Federated Learning for Trustworthy Healthcare Intelligence}

\begin{document}
\maketitle

\begin{abstract}
Healthcare organizations collect huge amounts of patient data from clinical records, medical tests, monitoring systems, and healthcare services. Such data can underpin intelligent healthcare analytics but remains distributed across institutions and cannot be shared directly due to privacy, security, and trust concerns. Federated learning (FL) enables collaborative learning without sharing raw patient records, but the exchanged model information remains vulnerable to privacy leakage, tampering, replay, and unauthorized access. In this paper, we propose iGRAIL, a blockchain-enabled generative FL framework for trustworthy healthcare intelligence. iGRAIL combines local generative learning for data augmentation with differential privacy to protect sensitive information during local training. Model communication is secured using cryptographic mechanisms, and blockchain-based verification and smart-contract access control ensure that only authorized, authentic, fresh, and integrity-preserving updates are accepted for federated aggregation. The framework is evaluated on four healthcare datasets and compared with the conventional FL and blockchain-based FL (BC-FL) under predictive, privacy, non-IID, scalability, malicious-update, security-verification, and runtime settings. For the two large diabetes datasets, iGRAIL improves accuracy by about 5.4-5.6\%, the F1-score by 14.6-16.9\%, and MCC by 16.0-18.1\% over the best FL or BC-FL baseline. Increasing the strength of differential privacy reduces the privacy-loss parameter \(\epsilon\) by about 72.5\% for the cancer dataset and 87.5\% for the diabetes dataset, with only a limited loss in prediction performance. These results indicate that iGRAIL supports secure and scalable federated computing while preserving privacy, model integrity, and controlled participation, at the cost of additional computational overhead.
\end{abstract}


\section{Introduction}
\label{intro}

Healthcare is a significant application area of ML and AI, especially for disease prediction, patient monitoring, and clinical decision-making \cite{11289508}. Hospitals collect a large amount of patients’ data; learning from data across multiple institutions can help to develop more reliable and effective models \cite{9669196}.
Patient records are sensitive and cannot be shared freely. Their collection, storage, and use have to follow strict privacy regulations \cite{11551848}.
Security is another main concern as healthcare systems are often targeted by cyber-attacks and data breaches \cite{10839304}. Such incidents can lead to patient information exposure and disruption of healthcare services \cite{10971535}. These concerns restrict the utility of standard centralized learning where data from various hospitals are gathered in one location. FL addresses this issue by enabling hospitals to train a shared model on their local data without sharing patient records directly \cite{10839304}.

Local patient data storage limits direct data exposure, but privacy, security, and trust risks remain in federated learning \cite{11631765}. Unauthorized participants could submit tampered or replayed updates, or information could still be inferred from the exchanged model updates. Also, a malicious authenticated client can issue a semantically harmful update that cannot be detected by cryptographic verification alone. These limitations show that traditional FL alone cannot ensure privacy, model integrity, update provenance, freshness, and controlled participation in sensitive healthcare applications \cite{chatterjee2023approach}. There are a number of techniques that tackle parts of this problem. Generative adversarial networks (GANs) can learn the local data distribution and generate synthetic samples without directly transferring patient records \cite{10634770}. Differential privacy (DP) can control information leakage from individual training records through perturbation \cite{10645744}. Model information can be cryptographically protected during communication \cite{9258958} and blockchain \cite{9050459} can enable verifiable update records and decentralized coordination. Additionally, intelligent contracts can enforce predetermined authorization and participation rules \cite{10130092,10049604}. Existing approaches typically treat privacy, safe communication, blockchain verification, participation control, or federated aggregation as separate mechanisms, or apply them only at specific stages of the federated learning pipeline. Conversely, we propose a unified end-to-end framework that unites privacy-preserving local learning, secure model exchange, blockchain-assisted verification, controlled client participation, and federated aggregation into a single coordinated learning workflow.

Existing approaches typically address only certain privacy and security aspects of federated healthcare learning \cite{9353683}. Some methods focus on the protection of patient data; others on safe communication, blockchain-based transaction recording, or access control. Consequently, generative learning, differential privacy, protected model exchange, update verification, controlled participation, and federated aggregation are often implemented as separate functionalities rather than as coordinated steps within a single learning workflow. In particular, blockchain is often employed as a passive record-keeping mechanism \cite{11657487}, with little use in validating model updates prior to their aggregation. Thus, conventional frameworks may protect local data but not guarantee that the exchanged updates are authorized, authentic, fresh, and integrity-preserving for global model construction. This enables a unified federated learning architecture that combines the privacy-preserving local generative learning with secure model exchange, verification-gated aggregation, and smart-contract-based participation control. Closing this gap can help distributed healthcare settings move toward a more trusted collaborative learning process by preserving data locality and reducing risks during model exchange and aggregation.

We propose iGRAIL, a blockchain-based generative federated learning framework for trustworthy healthcare intelligence to overcome these limitations. Each participating institution conducts local generative learning with differential privacy and retains the raw patient data locally. Model updates are cryptographically secured and validated before aggregation for authorization, integrity, origin authenticity, and freshness. The blockchain is the coordination and verification layer. The intelligent contracts are the rules of participation and access control. Federated aggregation only permits verified updates, which enables privacy-preserving learning, secure model exchange, controlled participation, and auditable collaboration within a single workflow.

The main contributions of this work are as follows:
\begin{itemize}

\item We propose a secure federated aggregation protocol that guarantees authorization, integrity, origin authentication, and freshness of client updates prior to their acceptance.

\item We propose a blockchain-based coordination and verification layer to integrate protected model exchange, pre-aggregation verification, smart-contract-based participation control, and auditability into the federated learning workflow.

\item We develop a privacy-preserving generative federated learning mechanism that combines local synthetic data generation with differential privacy, enabling distributed healthcare intelligence to be learned without directly sharing raw patient records.

\item We provide a comprehensive evaluation under heterogeneous data distributions, varying federation sizes, differential-privacy settings, malicious-update conditions, security-verification cases, and runtime constraints.

\item Experimental results show that iGRAIL improves accuracy by approximately 5.4-5.6\%, F1-score by 14.6-16.9\%, and MCC by 16.0-18.1\% on the two large diabetes datasets compared with the strongest FL or BC-FL baseline. Stronger differential privacy also reduces the privacy-loss parameter $\epsilon$ by approximately 72.5\% for the Cancer dataset and 87.5\% for the diabetes dataset, while retaining useful predictive performance.

\end{itemize}

The rest of this paper is organized as follows. In Section~\ref{relatedwork}, we review the state-of-the-art studies on federated learning, privacy-preserving generative learning, and blockchain-enabled security in distributed healthcare systems. It also discusses the drawbacks of current approaches and points to the research gaps that motivate the proposed framework. In Section~\ref{pm}, we introduce the overall architecture of iGRAIL and elaborate on the key components, including privacy-preserving generative federated learning, differential privacy, secure model-update exchange, blockchain-based verification, and trustworthy aggregation. It further explains how these mechanisms work together to enable secure, reliable collaborative learning across distributed healthcare environments. The experimental setup, healthcare datasets, baseline methods, and evaluation metrics are described in Section~\ref{sec:experiments}. It then presents and discusses the results in terms of predictive performance, data heterogeneity, client scalability, differential privacy, robustness to malicious updates, security verification, and computational overhead. Finally, Section~\ref{conclu} concludes with a summary of the main results, the limitations of the proposed framework, and directions for future research.

\section{Related Work} 
\label{relatedwork} 
The privacy and security mechanisms supported by existing frameworks and the proposed iGRAIL approach are compared in Table~\ref{table:comparison}. Authentication and integrity checks are not enough to detect a malicious update to a model from an authorized client. Likewise, active man-in-the-middle attacks and collusion attacks might require additional trust assumptions, key management precautions, or specialized defenses \cite{10615611}. Therefore, iGRAIL is not considered to provide complete protection from those threats but only partial mitigation. The comparison shows how iGRAIL combines privacy and security mechanisms that complement each other. The main contribution is the integration of privacy-preserving generative learning, secure model communication, blockchain-based verification, and controlled admission of model updates into a unified federated learning workflow.

\subsection{Blockchain-Enabled Federated Learning}
Blockchain has been integrated with federated learning (FL) to enhance privacy, trust, and coordination among distributed participants. Das et al.~\cite{DAS2026129916} proposed FLIT, a blockchain-secured federated learning framework to build trust among the participants in the Internet of Medical Things. The framework studies secure collaborative learning, smart contract validation, and privacy protection in distributed healthcare environments.
In this paper, we continue this line of work by studying the incorporation of local generative learning, differential privacy, protected model exchange, and verification-gated aggregation in a joint healthcare learning workflow. Wan et al. \cite{c111} proposed a blockchain-enabled FL framework for privacy-preserving beyond-fifth-generation (B5G) edge computing. They suggest a combination of decentralized model training and blockchain-based coordination to allow secure collaboration among edge devices. But, the framework does not address both privacy-preserving synthetic data generation and comprehensive model-update verification. Liu et al. \cite{c112} proposed a blockchain-aided FL framework for secure IoMT applications. Their work is focused on supporting collaborative healthcare learning and improving trust and security in distributed medical devices. In spite of that, the combination of generative learning, differential privacy, and fine-grained participation control is still limited.

\subsection{Blockchain-Based Authentication and Access Control}
In distributed systems, blockchain-based authentication and access control mechanisms have been proposed to control participation and secure sensitive information \cite{1194883}. SAID is an elliptic curve cryptography (ECC) based authentication and incentive distribution scheme for blockchain-based data sharing proposed by Rizwan et al. \cite{c211}. It enhances the authentication of participants and enables controlled interaction between distributed entities. It does not, however, address privacy-preserving federated model training and verification of model updates before aggregation directly. Intelligent contract-based access control for cloud-based healthcare systems has been investigated by Saini et al. \cite{c212}. They use blockchain-based access policies to control access to healthcare resources and maintain accountability in their framework. It offers controlled access to sensitive information. However, it is not mainly used to protect the training and aggregation of collaborative models. MedChain: A smart contract and decentralized technology-based dynamic healthcare data management system. Priyadarshini et al. \cite{c613} offered a decentralized approach for managing health information, reducing the reliance on centralized administration. But the main objectives are not the secure control of participation in FL or the integrity check of the exchanged model parameters.

\subsection{Blockchain and crypto protection}
Also, cryptographic protection is combined with blockchain \cite{alkhodair2022flexichain} for better confidentiality, authentication, and integrity in distributed applications \cite{11100329}. Mahajan and Junnarkar \cite{c411} proposed a security approach that combines lightweight ECC and private blockchain for medical multimedia data. They offer cryptographic protection of sensitive medical information and employ blockchain to enable trusted data management. It does not, however, directly handle the exchange and verification of model parameters in federated training. Dwivedi et al. \cite{c412} propose ECC and blockchain for secure data storage and authentication in Internet of Drones systems. Their work shows that cryptographic authentication and decentralized coordination can be achieved in distributed environments. However, its security mechanisms are more suitable for drone-based applications than for privacy-preserving healthcare FL. Bakro et al. \cite{c413} proposed a cloud security scheme using blockchain, ECC, and the Advanced Encryption Standard (AES). The framework uses cryptographic protection and blockchain-based mechanisms to improve data security in cloud environments. However, it does not explore privacy-preserving generative learning or verification-gated federated aggregation in particular. Suganthi and Kavitha \cite{c311} studied the combination of blockchain, encryption, and deep learning for healthcare security. Their methodology combines smart analysis with cryptographic techniques that shield healthcare information. However, the focus is not on how to integrate those mechanisms into a complete privacy-preserving FL workflow. Raju et al. \cite{c312} also employed blockchain, encryption, and deep learning for improving security in healthcare applications. Their work shows the promise of merging cryptographic protection with learning-based approaches. However, it does not consider joint issues of secure exchange of federated models, decentralized participant authorization, and pre-aggregation verification.

\begin{table}[!tb]
\caption{Comparison of Privacy and Security Mechanisms in Related Frameworks}
\label{table:comparison}
\footnotesize
\centering
\setlength{\tabcolsep}{11pt}
\renewcommand{\arraystretch}{1.15}

\begin{adjustbox}{max width=\textwidth}%
\begin{tabular}{llcccccccc}
\toprule
\textbf{Ref.} &
\textbf{Technologies used} &
\textbf{IA} &
\textbf{MIA} &
\textbf{TA} &
\textbf{RA} &
\textbf{UA} &
\textbf{CA} &
\textbf{MITM} &
\textbf{DPA} \\
\midrule

Proposed
& Blockchain, ECC, GAN, DP
& \checkmark
& \checkmark
& \checkmark
& \checkmark
& \checkmark
& $\triangle$
& $\triangle$
& $\triangle$ \\

\cite{c111, c112}
& Blockchain, GAN, DP
& \checkmark
& \checkmark
& \checkmark
& \checkmark
& \checkmark
& $\times$
& $\times$
& \checkmark \\

\cite{c211, c212}
& Blockchain, ECC, DP
& $\times$
& \checkmark
& \checkmark
& \checkmark
& \checkmark
& \checkmark
& \checkmark
& \checkmark \\

\cite{c311, c312}
& Blockchain, ECC, GAN
& \checkmark
& $\times$
& \checkmark
& \checkmark
& \checkmark
& \checkmark
& \checkmark
& $\times$ \\

\cite{c411, c412, c413}
& Blockchain, ECC
& $\times$
& $\times$
& \checkmark
& \checkmark
& \checkmark
& \checkmark
& \checkmark
& $\times$ \\

\cite{c511, c512}
& Blockchain, DP
& $\times$
& \checkmark
& \checkmark
& \checkmark
& $\times$
& $\times$
& $\times$
& \checkmark \\

\cite{c611, c612, c613}
& Blockchain, GAN
& \checkmark
& $\times$
& \checkmark
& \checkmark
& \checkmark
& $\times$
& $\times$
& $\times$ \\

\cite{c711, c712, c713}
& GAN, DP
& \checkmark
& \checkmark
& $\times$
& $\times$
& $\times$
& $\times$
& $\times$
& \checkmark \\

\cite{c811, c812}
& Blockchain
& $\times$
& $\times$
& \checkmark
& \checkmark
& \checkmark
& $\times$
& $\times$
& $\times$ \\

\cite{c911, c912}
& GAN
& \checkmark
& $\times$
& $\times$
& $\times$
& $\times$
& $\times$
& $\times$
& $\times$ \\

\cite{c101, c102, c103, c104}
& DP
& $\times$
& \checkmark
& $\times$
& $\times$
& $\times$
& $\times$
& $\times$
& \checkmark \\

\bottomrule
\end{tabular}%
\end{adjustbox}

\vspace{2mm}
\begin{minipage}{\linewidth}
\footnotesize
\textit{Abbreviations:}
IA = Inference attacks;
MIA = Membership inference attacks;
TA = Tampering attacks;
RA = Replay attacks;
UA = Unauthorized access;
CA = Collusion attacks;
MITM = Man-in-the-Middle attacks;
DPA = Data poisoning attacks.

\par\smallskip
\textit{Note:}
$\checkmark$ indicates that the framework includes a direct
mechanism for the listed threat,
$\triangle$ indicates partial mitigation
under the stated system assumptions, and
$\times$ indicates that no specific defense
mechanism is provided.
Collusion, active man-in-the-middle attacks,
and coordinated model poisoning require
additional identity, authentication, or
robust aggregation mechanisms.
\end{minipage}
\end{table}

\subsection{Blockchain-Based Healthcare Data Security and Sharing}
Also, blockchain has been widely studied as a solution for secure healthcare data management and information sharing \cite{10971535}. A blockchain-based system for sharing electronic medical records while preserving privacy was proposed by Wu et al. \cite{c811}. This approach allows for safe data sharing and reduces the need for centralized healthcare data management. But it is mostly about the sharing of medical records, and not about the security of distributed model training \cite{chatterjee2023approach}. Zhang et al. \cite{c812} applied a consortium blockchain to exchange medical data in a secure manner among the participating healthcare organizations. They depend on a permissioned blockchain environment to allow trusted information sharing and coordination. It provides improved security for sharing health data, but it does not natively support generative FL or model update checks. Park et al. \cite{c512} combined local DP with blockchain for privacy-preserving management of genomic data. The approach employs local privacy preservation to minimize exposure of sensitive genomic data and enable blockchain-based data coordination. However, the framework is not intended for generative federated training or for secure aggregation of distributed model parameters. Chidambaranathan and Geetha \cite{c612} used deep learning and blockchain to detect attacks in healthcare. Healthcare-related threats are mitigated by combining learning-based security analysis with blockchain-supported mechanisms. But the main goal is attack detection, not security of the whole federated learning process.

\subsection{Generative learning for health care data}
Generative learning is of great interest for generating synthetic healthcare records and reducing reliance on direct patient information. Generative adversarial networks (GANs) \cite{9767588} propose a learning framework where a generator and a discriminator are trained to approximate the underlying data distribution. This mechanism has enabled several ways to generate synthetic health data, especially when access to real patient records is limited. DP-CTGAN was proposed by Fang et al. \cite{c711} for generating medical data with differential privacy. The method synthesizes records by combining conditional generation of tabular data with differential privacy to bound the disclosure of sensitive information. However, the method focuses mainly on private data generation rather than the secure coordination of distributed federated model updates. Torfi and Fox \cite{c712} proposed CorGAN for generating synthetic healthcare records while preserving correlations among data attributes. Their work emphasizes the necessity of maintaining relationships in medical data for the usefulness of synthetic records in downstream analyses. However, the direct combination of such generative mechanisms with blockchain-based verification and federated aggregation has not been discussed.

Imtiaz et al. \cite{c713} investigated the generation of synthetic healthcare data preserving privacy. Their work looks at how generative learning can be used to allow the analysis of healthcare data without exposing sensitive records. However, it does not support secure exchange and verification of model updates among multiple healthcare institutions. Tufail et al. \cite{c912} addressed class imbalance for Alzheimer’s disease classification using GAN-based learning. This is achieved through the synthesis of data to better represent underrepresented classes and to help predictive learning. It shows the benefits of generative methods for healthcare classification but does not address the security of decentralized learning and blockchain-based coordination. Neela and Kavitha \cite{c611} obtained medical images using Blockchain and GAN-based learning. Their research is focused on synergies of generative learning and distributed security mechanisms for protection of medical data. No integration with privacy-preserving federated aggregation, participant authorization, or verification of distributed updates to the model.

\subsection{Differential Privacy for Health Care Learning}
Differential privacy is a formal framework for limiting the effect of individual records on published data or learned models. It has been widely researched in healthcare applications where sensitive patient data require confidentiality in data analysis and collaborative learning. Choudhury et al. \cite{c101} explored privacy-preserving FL for sensitive healthcare data. Their work examines collaborative learning protocols that enable participants in different geographic locations to train models without directly sharing their local patient records. However, privacy-preserving training cannot ensure the authenticity, authorization, and safety of the exchanged model updates from malicious modification. Islam et al. \cite{c103} also explored privacy-preserving FL in healthcare applications. The approach seeks to support the development of distributed models while mitigating the risks associated with centralized access to sensitive medical data. But the combination of generative learning, blockchain-based coordination, and pre-aggregation security verification is limited. Vadavalli and Subhashini \cite{c102} used DP in the problem of healthcare truth discovery. Their approach leverages privacy protections to enable useful information to be derived from sensitive healthcare data while limiting the disclosure of individual information. However, it does not directly solve the problem of participant authentication and secure exchange of model parameters in FL.

Kavitha et al. \cite{c104} used DP for LSTM-based IoT healthcare processing. Their work combines sequential learning and privacy mechanisms to allow the analysis of healthcare data collected from connected devices. However, the main goals are not to achieve blockchain-based model verification and secure federated coordination while improving privacy protection for intelligent healthcare processing. Studies show that DP can help reduce privacy leakage during learning and data analysis. DP by itself does not provide authentication, message integrity, replay protection, or access control. Therefore, additional security mechanisms are needed to secure the entire collaborative learning workflow.

\subsection{Federated Learning for Distributed Medical Data}
FL has emerged as an important approach for healthcare applications in which medical data are distributed across institutions and cannot be centralized. It allows participating hospitals and medical device manufacturers to develop predictive models together while keeping their original datasets locally. However, the trustworthiness of collaborative learning can be influenced by heterogeneity in data distributions, computing resources, and security. Yan et al. \cite{c911} proposed a variation-aware federated learning approach for decentralized multi-source medical image data. They tackle cross-client data heterogeneity and study the effects of heterogeneous medical data on collaborative model training. The framework can learn in heterogeneous settings, but it does not support blockchain verification, smart contract participation, and privacy-preserving generative learning.

Prior work has studied important issues in healthcare data privacy, generative learning, cryptographic protection, and decentralized trust. However, these mechanisms are often studied in isolation or only combined for some stages of distributed learning \cite{11655293}. One important open research challenge is combining privacy-preserving synthetic data generation with differential privacy, cryptographically secure model exchange, blockchain-based update verification, and controlled federated aggregation \cite{11655301}. To address these limitations, the proposed iGRAIL framework integrates privacy-preserving generative learning, differential privacy, ECC-protected model exchange, blockchain-based verification, smart-contract-controlled participation, and verification-gated aggregation into a unified healthcare FL workflow. Local data augmentation is possible with the generative learning mechanism, and direct sharing of patient records is avoided. Further protection against information leakage is provided by DP. Cryptographic and blockchain mechanisms guarantee the security of model exchange, verify the authenticity and integrity of updates, and control participants' access before aggregation. iGRAIL aims to leverage the synergy of these complementary mechanisms to enhance the predictive performance, privacy protection, and trustworthiness in distributed healthcare learning scenarios.

\section{Proposed Methodology}
\label{pm}
\begin{figure*}[!tb]
\centering
\includegraphics[width=\linewidth,keepaspectratio]{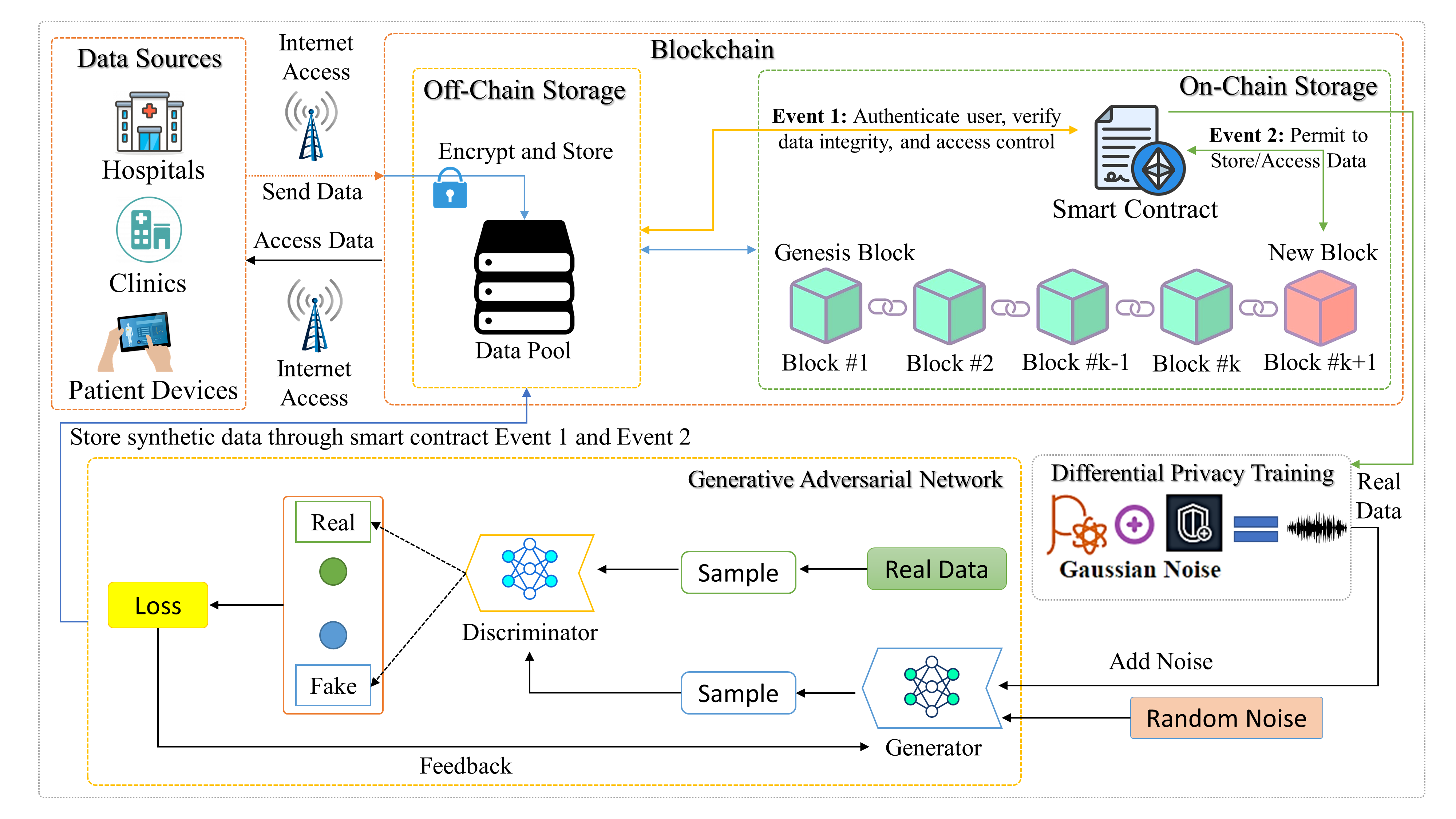}
\caption{Proposed iGRAIL framework for secure and privacy-preserving healthcare analytics.}
\label{fig:propsoed}
\end{figure*}

We present iGRAIL, a generative learning and orchestration framework for secure healthcare analytics. As shown in Figure \ref{fig:propsoed}, each healthcare institution learns from its local patient data without the transfer of original records. The whole process links local learning, privacy protection, federated collaboration, and trusted coordination.

\subsection{System Model and Learning Flow}
Participating healthcare institutions: $\mathcal{H}=\{H_1,H_2,\ldots,H_K\}$. Each institution $H_k$ has its private-labeled healthcare dataset $\mathcal{D}_k=\{(x_i,y_i)\}_{i=1}^{n_k}$, where $x_i\in\mathbb{R}^{d}$ denotes the clinical feature vector and $y_i\in\{0,1\}$ denotes the corresponding prediction label. Thus the complete logical dataset is $\mathcal{D}=\bigcup_{k=1}^{K}\mathcal{D}_k, \qquad \mathcal{D}_k\cap\mathcal{D}_l=\emptyset,\quad k\neq l,$ but $\mathcal{D}$ is never physically assembled at any central location. The institution $H_k$ can only see $\mathcal{D}_k$. This setting is a real-life situation in which patient records are still controlled by the institution that collects them. Learning has two interrelated objectives. The first goal is to learn the distribution of the distributed healthcare data with a federated generator $\mathcal{G}$ with parameters $\theta_{\mathcal{G}}$. The second objective is predictive learning
using a classifier $\mathcal{F}$ with parameters
$\theta_{\mathcal{F}}$. The generator produces synthetic labeled
samples that augment the limited local observations before the
classifier is trained. Thus, for institution $H_k$, the effective
training set at round $t$ is:
\begin{equation}
\widetilde{\mathcal{D}}_k^{(t)}
=
\mathcal{D}_k
\cup
\mathcal{S}_k^{(t)},
\qquad
\mathcal{S}_k^{(t)}
=
\left\{
(\tilde{x}_j,\tilde{y}_j)
\right\}_{j=1}^{m_k}
\label{eq:augmented_data}
\end{equation}

where $\mathcal{S}_k^{(t)}$ denotes the synthetic samples produced
from the learned generative distribution. The classifier estimates $\hat{y}_i
= \mathcal{F} \left(x_i;\theta_{\mathcal{F}}\right)$, and its local objective is:
\begin{equation}
\mathcal{L}_{\mathrm{cls}}^{k}
=
-\frac{1}{|\widetilde{\mathcal{D}}_k|}
\sum_{(x_i,y_i)\in\widetilde{\mathcal{D}}_k}
\left[
y_i\log\hat{y}_i
+
(1-y_i)\log(1-\hat{y}_i)
\right]
\label{eq:classification_loss}
\end{equation}

At the federated round $t$, $H_k$ receives the current global model
parameters, performs local privacy-preserving learning, and releases
only protected model information. Before aggregation, each submitted
update must pass authorization, freshness, origin authentication, and
integrity verification. Let
$\mathcal{V}^{(t)}\subseteq\{1,\ldots,K\}$ denote the verified client
set. Only updates from $k\in\mathcal{V}^{(t)}$ participants in
aggregation.

\subsection{Local Generative Learning}
\label{comp1}
Each institution $H_k$ learns the distribution of its local healthcare records without releasing the original observations. Since the downstream objective is supervised prediction, the generative process must preserve both the clinical feature distribution and its relationship with the class label. The target local distribution can therefore be written as $p_k(x,y) = p_k(x\mid y)p_k(y)$. Given latent noise $z\sim p_z(z)$ and class condition $y$, the generator produces $\tilde{x}
=
\mathcal{G}
\left(
z,y;\theta_{\mathcal{G}}^k
\right),
\qquad
(\tilde{x},y)\sim\hat{p}_k(x,y)$, where $\hat{p}_k$ approximates the private distribution $p_k$.
Synthetic observations are used only for local augmentation and do
not replace the original patient observations. The implementation uses WGAN-GP because the Wasserstein objective provides more stable optimization than the original GAN objective. For the local institution $H_k$, the critic objective is:
\begin{equation}
\begin{aligned}
\mathcal{L}_{\mathcal{D}}^{k}
=&\;
\mathbb{E}_{\tilde{x}\sim p_g^k}
[\mathcal{D}(\tilde{x},y)]
-
\mathbb{E}_{x\sim p_k}
[\mathcal{D}(x,y)]
\\
&+
\lambda_{\mathrm{GP}}
\mathbb{E}_{\hat{x}}
\left[
\left(
\|\nabla_{\hat{x}}\mathcal{D}(\hat{x},y)\|_2-1
\right)^2
\right]
\end{aligned}
\label{eq:wgangp}
\end{equation}

where $\hat{x}=\alpha x+(1-\alpha)\tilde{x}$,
$\alpha\sim U(0,1)$, and $\lambda_{\mathrm{GP}}$ controls the
gradient penalty. The corresponding generator objective is $\mathcal{L}_{\mathcal{G}}^{k}
=
-
\mathbb{E}_{z\sim p_z,\;y\sim p_k(y)}
\left[
\mathcal{D}
\left(
\mathcal{G}(z,y),y
\right)
\right]$. The generated conditional distribution $\hat{p}_k(x\mid y)$ is attracted by the local distribution $p_k(x\mid y)$ by minimizing $\mathcal{L}_{\mathcal{G}}^{k}$. This is important under heterogeneous healthcare data because clients can have different class proportions and local feature distributions.

\subsection{Differentially Private Local Training}
\label{comp2}
\begin{figure}[!tb]
\centering
\includegraphics[width=0.90\columnwidth]{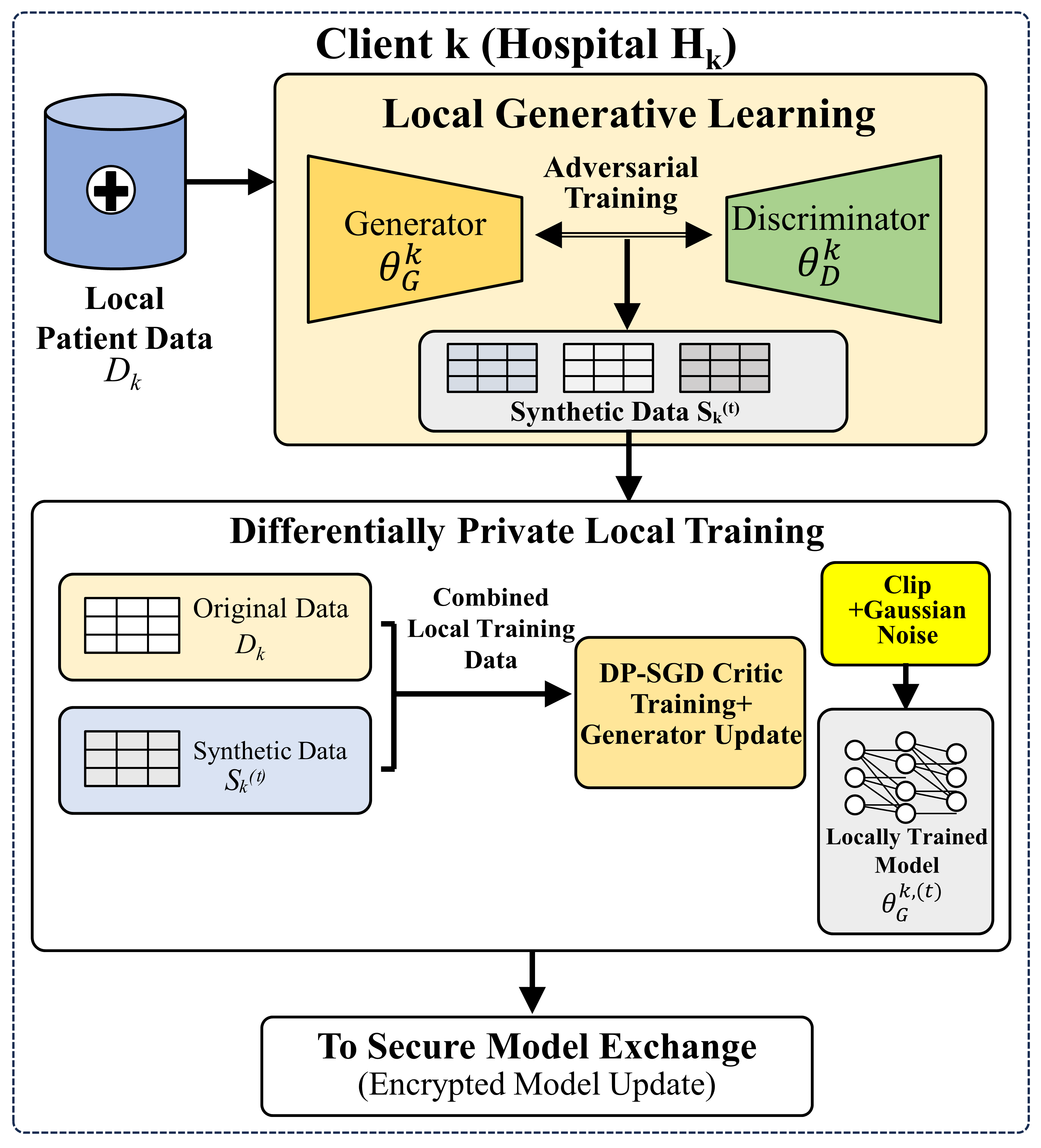}
\caption{Local generative learning and differentially private training at healthcare institution \(H_k\). The dataset \(D_k\) supports synthetic sample generation \(S_k^{(t)}\) and differential privacy protects the local update \(\theta_G^{k,(t)}\) before secure model exchange.}
\label{fig}
\vspace{-0.3cm}
\end{figure}

In Figure~\ref{fig}, we summarize local generative learning and differentially private training at institution $H_k$. The conditional generator $G(z,y;\theta^G_k)$ is trained on the private dataset $D_k$ to generate synthetic samples $S_k(t)$ for local augmentation. Differential privacy is ensured by clipping the per-example gradients and adding Gaussian noise before releasing the local generator update $\theta_G^{k,(t)}$ \cite{9767588}. The raw patient data stay at the originating institution. The patient data is not directly transferred by being kept $\mathcal{D}_k$ inside the institution $H_k$, but the learned parameters might leak information about individual training records. iGRAIL therefore
applies DP-SGD to the critic updates that directly access real patient
samples \cite{11391495}. For a real observation $(x_i,y_i)\in\mathcal{D}_k$, let $g_i
=
\nabla_{\theta_{\mathcal{D}}}
\mathcal{L}_{\mathcal{D}}
(x_i,y_i)$ denote its per-example gradient. Its contribution is bounded by the clipping $C$ as:
\begin{equation}
\widetilde{g}_i
=
g_i
\min
\left(
1,
\frac{C}{\|g_i\|_2}
\right)
\label{eq:clip}
\end{equation}

For mini-batch $B_k$, the privatized critic gradient is:
\begin{equation}
\bar{g}_{\mathcal{D}}
=
\frac{1}{|B_k|}
\left[
\sum_{i\in B_k}
\widetilde{g}_i
+
\mathcal{N}
\left(
0,\sigma^2C^2I
\right)
\right]
\label{eq:noise}
\end{equation}

and the critic parameters are updated according to $\theta_{\mathcal{D}}^k
\leftarrow
\theta_{\mathcal{D}}^k
-
\eta_{\mathcal{D}}
\bar{g}_{\mathcal{D}}$. The generator is subsequently optimized against the privatized critic, $\theta_{\mathcal{G}}^k
\leftarrow
\theta_{\mathcal{G}}^k
-
\eta_{\mathcal{G}}
\nabla_{\theta_{\mathcal{G}}}
\mathcal{L}_{\mathcal{G}}$ without independently claiming record-level DP from clipping gradients
with respect to random latent variables. Since the generator is
obtained through interaction with the differentially private training mechanism, subsequent operations on its released output are governed by the post-processing property of DP, provided that no additional access to private records occurs outside the privacy-accounting mechanism.

\subsubsection{Privacy Guarantee}
Let $\mathcal{D}_k$ and $\mathcal{D}'_k$ be neighboring datasets that
differ in one patient record. The randomized local training mechanism
$\mathcal{M}$ satisfies $(\epsilon,\delta)$-DP if $\Pr[
\mathcal{M}(\mathcal{D}_k)\in O
]
\leq
e^\epsilon
\Pr[
\mathcal{M}(\mathcal{D}'_k)\in O
]
+
\delta$ for every measurable output set $O$. Let
$q=|B_k|/n_k$ denote the sampling rate and let $T_{\mathrm{DP}}$
denote the number of private optimization steps. The total privacy cost is therefore: $\epsilon = \mathcal{A}\left(q,\sigma,T_{\mathrm{DP}},\delta\right)$ where $\mathcal{A}(\cdot)$ is the privacy accountant used in the implementation. Therefore, in the experiments, the privacy budget is calculated based on the sampling rate, clipping norm, number of private updates, and the target $\delta$, rather than the noise multiplier alone.

\subsection{Federated Generative Learning}
At communication round $t$, institution $H_k$ initializes its local
generator using $\theta_{\mathcal{G}}^{k,(t)}
\leftarrow \theta_{\mathcal{G}}^{(t)}$ and performs $T$ local optimization steps on $\mathcal{D}_k$. The resulting
$\theta_{\mathcal{G}}^{k,(t)}$ is protected before leaving the local environment. A submitted update does not immediately participate in the aggregation. However, blockchain-based verification yields a binary decision $v_k^{(t)} \in \{0,1\}$. The set of accepted clients is:
\begin{equation}
\mathcal{V}^{(t)}
=
\left\{
k\mid v_k^{(t)}=1
\right\}
\label{eq:verified_clients}
\end{equation}

Then the normalized contribution of a verified institution $H_k$ in sample-size-weighted aggregation is:
\begin{equation}
w_k^{(t)}
=
\frac{n_k}
{\displaystyle
\sum_{j\in\mathcal{V}^{(t)}}n_j}
\text{ where}
\sum_{k\in\mathcal{V}^{(t)}}w_k^{(t)}=1
\label{eq:aggregation_weight}
\end{equation}

The next global generator is therefore:
\begin{equation}
\theta_{\mathcal{G}}^{(t+1)}
=
\sum_{k\in\mathcal{V}^{(t)}}
w_k^{(t)}
\theta_{\mathcal{G}}^{k,(t)}
\label{eq:fed}
\end{equation}

Equation~\eqref{eq:fed} differs from conventional aggregation in the admission condition of each update. A conventional server aggregates the received client updates according to its aggregation rule, whereas iGRAIL first requires $v_k^{(t)}=1$. Thus, a submitted model with failed authorization, invalid origin authentication, or inconsistent integrity information is filtered off before it can contribute to $\theta_{\mathcal{G}}^{(t+1)}$.

\begin{algorithm}[!t]
\SetAlgoLined
\KwIn{Local datasets $\{\mathcal{D}_k\}_{k=1}^{K}$,
clipping norm $C$, noise multiplier $\sigma$, local iterations $T$,
federated rounds $R$, client key pairs, DP parameters
$(\epsilon,\delta)$}
\KwOut{Verified global generator
$\theta_{\mathcal{G}}^{(R)}$ and predictive model}

Initialize $\theta_{\mathcal{G}}^{(0)}$\;

\For{$t=0$ \KwTo $R-1$}{
    \ForEach{$H_k\in\mathcal{H}$}{
        $\theta_{\mathcal{G}}^{k,(t)}
        \leftarrow\theta_{\mathcal{G}}^{(t)}$\;

        \For{$\tau=1$ \KwTo $T$}{
            Sample $B_k\subset\mathcal{D}_k$\;

            Compute per-example critic gradients
            $g_i=\nabla_{\theta_{\mathcal{D}}}
            \mathcal{L}_{\mathcal{D}}(x_i,y_i)$\;

            $\widetilde{g}_i
            \leftarrow
            g_i\min(1,C/\|g_i\|_2)$\;

            $\bar{g}_{\mathcal{D}}
            \leftarrow
            |B_k|^{-1}
            [\sum_i\widetilde{g}_i+
            \mathcal{N}(0,\sigma^2C^2I)]$\;

            Update privatized critic
            $\theta_{\mathcal{D}}^k$\;

            Update generator
            $\theta_{\mathcal{G}}^{k,(t)}$
            against the privatized critic\;
        }

        Serialize local generator update
        $m_k^{(t)}$\;

        Derive session key
        $K_k^{(t)}=\mathrm{KDF}(\mathrm{ECDH}(\cdot))$\;

        Protect update
        $c_k^{(t)}
        =\mathrm{AEAD}_{K_k^{(t)}}(m_k^{(t)})$\;

        Compute
        $h_k^{(t)}
        =H(c_k^{(t)}\|k\|t\|\nu_k^{(t)})$\;

        Sign
        $\xi_k^{(t)}
        =\mathrm{Sign}_{sk_k}
        (h_k^{(t)}\|k\|t\|\nu_k^{(t)})$\;

        Register verification metadata on-chain\;
    }

    $\mathcal{V}^{(t)}
    \leftarrow
    \{k\mid
    a_k I_k^{(t)}S_k^{(t)}F_k^{(t)}=1\}$\;

    $\theta_{\mathcal{G}}^{(t+1)}
    \leftarrow
    \sum_{k\in\mathcal{V}^{(t)}}
    w_k^{(t)}
    \theta_{\mathcal{G}}^{k,(t)}$\;
}

Generate synthetic samples
$\mathcal{S}$ using
$\theta_{\mathcal{G}}^{(R)}$\;

Construct augmented training data
$\widetilde{\mathcal{D}}$\;

Train predictive model
$\theta_{\mathcal{F}}^{*}$ on
$\widetilde{\mathcal{D}}$\;

Evaluate $\theta_{\mathcal{F}}^{*}$
using held-out real test data\;

\Return{
$\theta_{\mathcal{G}}^{(R)},
\theta_{\mathcal{F}}^{*}$
}\;

\caption{Privacy-Preserving Generative Federated Learning and
Blockchain-Orchestrated Verification in iGRAIL}
\label{algo1}
\end{algorithm}

\subsection{Blockchain-Orchestrated Security and Access Control}
The blockchain layer coordinates verification between local training and federated aggregation. Raw patient records and complete model content are not stored on-chain. Instead, protected information remains off-chain, while its cryptographic verification information is recorded on the blockchain.

\subsubsection{Protected model exchange and integrity}
The model update $\theta_{\mathcal{G}}^{k,(t)}$ can contain a large
number of floating-point parameters and is therefore not encrypted
directly as an elliptic-curve point. iGRAIL uses ECC for shared-key
establishment and symmetric authenticated encryption for the model
payload. Let $(d_k,Q_k)$ denote the private-public key pair of
institution $H_k$ and $(d_s,Q_s)$ denote the corresponding key pair
of the receiving service. The shared secret is $Z_k = d_kQ_s = d_sQ_k$, from which a session key is derived as $K_k^{(t)}
=
\mathrm{KDF}
\left(
Z_k\parallel k\parallel t
\right)$. Let $m_k^{(t)}=\mathrm{Serialize}
(\theta_{\mathcal{G}}^{k,(t)})$. The protected payload is:
\begin{equation}
c_k^{(t)}
=
\mathrm{AEAD}_{K_k^{(t)}}
\left(
m_k^{(t)};
k,t,\nu_k^{(t)}
\right)
\label{eq:hybrid_encrypt}
\end{equation}

where $\nu_k^{(t)}$ is a unique nonce and $(k,t,\nu_k^{(t)})$ is bound to the ciphertext as authenticated metadata, and the digest is:
\begin{equation}
h_k^{(t)}
=
\mathbb{H}
\left(
c_k^{(t)}
\parallel k
\parallel t
\parallel \nu_k^{(t)}
\right)
\label{eq:model_hash}
\end{equation}

The full model payload is still off-chain. Verification metadata is registered on the blockchain layer only for auditing and integrity checking.

\subsubsection{Origin authentication and smart-contract verification}
Integrity alone does not tell you what institution produced an update. So each institution signs its submission with its signing key. Let $sk_k$ and $pk_k$ be the signing and the verification keys of $H_k$. The signature associated with $t$ is:
\begin{equation}
\xi_k^{(t)}
=
\mathrm{Sign}_{sk_k}
\left(
h_k^{(t)}
\parallel k
\parallel t
\parallel \nu_k^{(t)}
\right)
\label{eq:signature}
\end{equation}

The smart contract stores the authorization variable $a_k\in\{0,1\}$ and rejects a previously used nonce or a submission associated with an invalid federated round. We define the following:
\begin{equation}
\begin{aligned}
I_k^{(t)}
&=
\mathbb{I}
\left[
\mathbb{H}
(c_k^{(t)}\parallel k\parallel t\parallel\nu_k^{(t)})
=
h_k^{(t)}
\right],\\
S_k^{(t)}
&=
\mathbb{I}
\left[
\mathrm{Verify}_{pk_k}
(\xi_k^{(t)},h_k^{(t)},k,t,\nu_k^{(t)})
=1
\right],\\
F_k^{(t)}
&=
\mathbb{I}
\left[
t=t_{\mathrm{current}}
\land
\nu_k^{(t)}\notin\mathcal{N}_{\mathrm{used}}
\right]
\end{aligned}
\label{eq:verification_components}
\end{equation}

The final admission decision becomes:
\begin{equation}
v_k^{(t)}
=
a_k
I_k^{(t)}
S_k^{(t)}
F_k^{(t)}
\end{equation} 
Consequently, $v_k^{(t)}=1
\iff
a_k=I_k^{(t)}=S_k^{(t)}=F_k^{(t)}=1$. Here, this formulation identifies four different security properties. $a_k$ permits, $I_k^{(t)}$ checks integrity $S_k^{(t)}$ validates the source of the incoming update and $F_k^{(t)}$ confirms freshness A modified ciphertext makes $I_k^{(t)}=0$. An update without a valid client signature makes $S_k^{(t)}=0$. Reuse of a previously accepted nonce or stale round makes $F_k^{(t)}=0$. In all cases, $v_k^{(t)}=0$ and the model update is excluded from the federated aggregation.

\subsubsection{Synthetic data generation and protection}
The learned distributed generative model after $R$ federated rounds is $\theta_{\mathcal{G}}^{(R)}$. Synthetic records are made as:
\begin{equation}
\mathcal{S}
=
\left\{
s_j=
\mathcal{G}
\left(
z_j,y_j;
\theta_{\mathcal{G}}^{(R)}
\right)
\right\}_{j=1}^{N},
\qquad
z_j\sim p_z(z)
\label{eq:synth}
\end{equation}

Since $s_j$ is an elliptic-curve point rather than a healthcare feature vector, it is serialized and secured using the session key established via ECDH. Thus, $c_j^{S}
=
\mathrm{AEAD}_{K_S}
\left(
\mathrm{Serialize}(s_j);
j,\nu_j
\right)$, where $K_S$ is derived from the authorized parties' ECDH shared secret and $\nu_j$ is a unique nonce. The encrypted synthetic records remain off-chain. The blockchain stores only the metadata needed for authorization, integrity verification, and auditability.

\subsubsection{Role-based access control}
Let $\mathcal{R}=\{r_1,r_2,\ldots,r_M\}$ be the set of supported roles, $\rho(u)\in\mathcal{R}$ be the role assigned to user $u$ and $\mathcal{P}(r)$ be the permissions associated with role $r$. Let $p_{\mathrm{read}}$ denote the permission to access the protected synthetic information. Then, the authorization decision is $A(u) = \mathbb{I} \left[ p_{\mathrm{read}} \in \mathcal{P}(\rho(u)) \right]$. The smart contract response is then:
\begin{equation}
\mathcal{SC}(u,c_j^{S})
=
\begin{cases}
c_j^{S}, & A(u)=1,\\
\varnothing, & A(u)=0.
\end{cases}
\label{eq:access}
\end{equation}

Possession of the ciphertext alone does not reveal $s_j$. An authorized requester must obtain or derive the corresponding decryption key through the defined key management procedure. Therefore, blockchain-based authorization and cryptographic confidentiality remain separate security functions.

\subsection{Synthetic Augmentation and Predictive Learning}
The global generative model is not treated as the final disease predictor. Its role is to provide additional samples for predictive learning when the participating institutions contain limited or heterogeneous observations. For institution $H_k$,
let $\mathcal{S}_k$ denote the synthetic samples generated for local augmentation. The augmented training set is $\widetilde{\mathcal{D}}_k
=
\mathcal{D}_k
\cup
\mathcal{S}_k$. A predictive model $\mathcal{F}_k$ is trained by minimizing
\begin{equation}
\theta_{\mathcal{F}}^{k,*}
=
\arg\min_{\theta_{\mathcal{F}}}
\frac{1}{|\widetilde{\mathcal{D}}_k|}
\sum_{(x_i,y_i)\in\widetilde{\mathcal{D}}_k}
\ell
\left(
\mathcal{F}(x_i;\theta_{\mathcal{F}}),
y_i
\right)
\label{eq:classifier_optimization}
\end{equation}

For binary healthcare prediction,
$\mathcal{F}(x_i;\theta_{\mathcal{F}})\in[0,1]$ represents the estimated probability of the positive class. Given decision threshold $\gamma$, the predicted label is $\hat{y}_i
=
\mathbb{I}
\left[
\mathcal{F}
(x_i;\theta_{\mathcal{F}})
\geq\gamma
\right]$. The predictive evaluation is therefore performed on held-out real
test observations rather than on the synthetic training samples.
It is important because the objective of generative
augmentation is not to demonstrate that synthetic records can be
classified easily. The objective is to determine whether information learned through privacy-preserving distributed generation improves or maintains generalization on unseen real healthcare records.

\subsection{Non-IID Institutional Data Distribution}
Healthcare institutions generally do not contain identical patient
populations. To represent this heterogeneity, the distribution of
class $c$ across $K$ clients is sampled using $\boldsymbol{\pi}_{c}
=
(\pi_{1c},\ldots,\pi_{Kc})
\sim
\mathrm{Dirichlet}
(\alpha\mathbf{1}_{K})$, where $\pi_{kc}$ denotes the proportion of samples from class $c$ assigned to institution $H_k$. Consequently, $\sum_{k=1}^{K}\pi_{kc}=1,
\qquad
\pi_{kc}\geq0$. The concentration parameter $\alpha$ controls the degree of
statistical heterogeneity. A smaller $\alpha$ produces more unequal
client-level class distributions, whereas increasing $\alpha$ makes
the partition closer to an IID allocation. Therefore,
the local distributions generally satisfy $p_k(x,y)\neq p_l(x,y),
\qquad k\neq l$, which allows the evaluation to examine whether iGRAIL remains useful when participants contain different patient
distributions.

\begin{table}[!h]
\caption{Experimental Design and Evaluation Configuration}
\label{table:expdesign}
\centering
\footnotesize
\renewcommand{\arraystretch}{1.15}
\setlength{\tabcolsep}{3pt}
\begin{tabularx}{\linewidth}{@{}
>{\raggedright\arraybackslash}p{0.22\linewidth}
>{\raggedright\arraybackslash}p{0.34\linewidth}
>{\raggedright\arraybackslash}X
@{}}
\toprule
\textbf{Experiment} &
\textbf{Configuration} &
\textbf{Purpose} \\
\midrule

Healthcare datasets &
Cancer \cite{yasser2021breastcancer},
Diabetes \cite{mustafa_diabetes_prediction},
Healthcare \cite{mustafa_diabetes_prediction},
and Heart \cite{heart_disease_45} &
Evaluate datasets with different sizes, feature dimensions,
and class distributions. \\

Compared methods &
FL, BC-FL, and iGRAIL &
Compare conventional FL, blockchain-supported FL,
and the proposed privacy-preserving framework. \\

Main federated setting &
3 clients and 20 communication rounds &
Provide a common configuration for the main experiments. \\
Independent runs &
3 runs &
Measure variation across repeated predictive experiments. \\

Default non-IID setting &
Dirichlet $\alpha=0.5$ &
Represent heterogeneous client data in the main experiments. \\

Non-IID analysis &
$\alpha\in\{0.1,0.5,1.0,5.0\}$ &
Evaluate different levels of client data heterogeneity. \\

Client scalability &
$K\in\{2,3,4,5,6,7,8,9,10\}$ &
Evaluate performance as the number of participating
clients increases. \\

Round-wise analysis &
Rounds 1-20 &
Examine predictive behavior throughout federated training. \\

Runtime analysis &
$R\in\{5,10,15,20\}$ &
Measure local, verification, aggregation, and total runtime. \\

DP configuration &
$\sigma\in\{0.55,0.75,1.00,1.35\}$,
$\delta\in\{10^{-6},10^{-7},10^{-8}\}$ &
Evaluate the privacy-utility trade-off under different
privacy settings. \\

Privacy accounting &
RDP accountant &
Calculate $\epsilon$ for each evaluated DP configuration. \\

Attack experiment &
Attack active from round 6 &
Evaluate robustness against malicious model updates. \\

Security tests &
Valid, Tampered, Replay, and Unauthorized &
Evaluate integrity, freshness, and authorization checks
before aggregation. \\

Payload sizes &
16, 64, 256, 1024, and 4096 KB &
Measure cryptographic protection and verification overhead. \\

Payload repetitions &
30 measurements per payload &
Measure average cryptographic latency and its variation. \\

Visualization &
Matplotlib-generated figures &
Present predictive, privacy, security, scalability,
and runtime results. \\

\bottomrule
\end{tabularx}
\end{table}

\section{Experiments and Evaluation}
\label{sec:experiments}

This section evaluates iGRAIL under different healthcare learning
conditions. The evaluation covers predictive performance, learning
behavior, privacy, non-IID data, client scalability, security, and
runtime. These experiments are used to examine both the benefit and
the cost of the proposed framework. The experiments use four healthcare datasets. Three learning methods
are compared. FL is used as the conventional FL baseline. BC-FL adds blockchain-based verification to federated
learning. iGRAIL includes generative learning, differential privacy, protected model exchange, and blockchain-based verification. The
main configuration uses three clients and 20 communication rounds.
We repeat the main experiments three times. Table~\ref{table:expdesign} summarizes the experimental configuration used throughout the evaluation. The experiments cover predictive
performance, non-IID data, client scalability, differential privacy, attack robustness, security verification, and runtime overhead. The source code, experimental configuration,
and instructions required to reproduce the experiments are publicly
available at \url{https://github.com/debashis2124/iGRAIL}.

\begin{table*}[!tb]
\caption{Characteristics of the Evaluated Healthcare Datasets}
\label{table:dataset_stats}
\centering
\footnotesize
\renewcommand{\arraystretch}{1.15}
\setlength{\tabcolsep}{3pt}
\vspace{2mm}
\begin{tabular}{lrrrrrrrr}
\toprule
\textbf{Dataset} &
\textbf{Samples} &
\textbf{Raw Features} &
\textbf{Encoded Features} &
\textbf{Classes} &
\textbf{Positive} &
\textbf{Negative} &
\textbf{Positive (\%)} &
\textbf{Missing Values} \\
\midrule
Cancer
& 569 & 30 & 30 & 2 & 212 & 357 & 37.26 & 0 \\

Diabetes-Clinical
& 100,000 & 8 & 15 & 2 & 8,500 & 91,500 & 8.50 & 0 \\

Diabetes-Prediction
& 100,000 & 15 & 76 & 2 & 8,500 & 91,500 & 8.50 & 0 \\

Heart
& 920 & 14 & 29 & 2 & 509 & 411 & 55.33 & 1,759 \\
\bottomrule
\end{tabular}

\vspace{1mm}
\parbox{\textwidth}{\footnotesize
\textit{Note:} Raw features exclude the target variable.
Encoded features represent the input dimensions after
preprocessing. Missing values indicate missing feature
entries before imputation.}
\end{table*}

\begin{table*}[!tb]
\caption{Experimental Configuration and Reproducibility Parameters}
\label{table:reproducibility}
\centering
\footnotesize
\renewcommand{\arraystretch}{1.10}
\setlength{\tabcolsep}{5pt}
\vspace{2mm}
\begin{tabular}{p{3.1cm}p{4.3cm}p{7.3cm}}
\toprule
\textbf{Component} &
\textbf{Parameter} &
\textbf{Configuration} \\
\midrule

\multirow{5}{*}{Data preprocessing}
& Training/test split & 80\% / 20\% (stratified) \\
& Random seed & 42 (base seed) \\
& Numerical missing values & Median imputation \\
& Categorical missing values & Most-frequent imputation \\
& Feature transformation & Standardization and one-hot encoding \\

\midrule

\multirow{6}{*}{Federated learning}
& Compared methods & FL, BC-FL, iGRAIL \\
& Participating clients & 3 (default) \\
& Communication rounds & 20 \\
& Independent repetitions & 3 \\
& Local training epochs & 2 \\
& Aggregation & FedAvg (FL); trust-weighted (BC-FL, iGRAIL) \\

\midrule

\multirow{6}{*}{Predictive model}
& Architecture & MLP: Input--64--32--1 \\
& Activation & ReLU (hidden layers) \\
& Optimizer & Adam \\
& Learning rate & $10^{-3}$ \\
& Batch size & 128 \\
& Loss and threshold & Weighted BCE; threshold $=0.5$ \\

\midrule

\multirow{7}{*}{Generative learning}
& Generative architecture & Conditional WGAN-GP \\
& Latent dimension & 32 \\
& Hidden dimension & 96 \\
& Generator communication rounds & 3 \\
& Local generator epochs & 1 \\
& Generator batch size & 128 \\
& Synthetic-to-real ratio & 0.40 (minimum 20 samples) \\

\midrule

\multirow{5}{*}{Differential privacy}
& Privacy mechanism & Opacus DP-SGD \\
& Default noise multiplier & $\sigma=0.9$ \\
& Noise multiplier sweep & $\{0.55,0.75,1.0,1.35\}$ \\
& Gradient clipping norm & 1.0 \\
& Target $\delta$ & $10^{-5}$ (default);
$\{10^{-6},10^{-7},10^{-8}\}$ (sweep) \\

\midrule

\multirow{3}{*}{Non-IID evaluation}
& Default Dirichlet parameter & $\alpha=0.5$ \\
& Dirichlet sweep & $\{0.1,0.5,1.0,5.0\}$ \\
& Client scalability & $K\in\{2,\ldots,10\}$ \\

\midrule

\multirow{4}{*}{Security evaluation}
& Attack start & Round 6 \\
& Attacked client & Client 0 (zero-based indexing) \\
& FL poisoning scale & 4.0 \\
& Verification tests & Valid, tampered, replayed, unauthorized \\

\bottomrule
\end{tabular}
\end{table*}

\subsection{Datasets and Experimental Setting}
\label{subsec:dataset-setting}
We assess the proposed framework on four healthcare datasets, summarized in Table~\ref{table:dataset_stats}. The datasets vary in sample size, feature dimensions, class distributions, and missing-value characteristics. The Cancer dataset has 569 instances, 30 attributes, and the positive class percentage is 37.26\%. The Heart dataset has 920 samples and 29 encoded features. The positive class is 55.33\% of the data. Conversely, the diabetes clinical and the diabetes prediction datasets contain 100,000 samples each and have 15 and 76 encoded features, respectively. Both diabetes datasets have large class imbalance with only 8.50\% positive samples. This makes prediction for the minority class especially important for evaluation of precision, recall, F1 score, and AUPRC. The heart dataset also has missing values, and these are handled during pre-processing. These datasets together provide different conditions for studying predictive performance on small and large datasets, different feature dimensions, and imbalanced class distributions.

These differences allow the
experiments to examine the methods under small, large, balanced, and
highly imbalanced healthcare data settings. The experiments compare conventional FL, blockchain-enabled BC-FL,
and the proposed iGRAIL framework under the same general evaluation
settings. The main experiments use three federated clients and run
the learning process for 20 communication rounds. To study data
heterogeneity, non-IID client distributions are generated using
Dirichlet $\alpha \in \{0.1,0.5,1.0,5.0\}$ across all four datasets.
The scalability experiment further changes the number of participating
clients from 2 to 10 to examine performance with different federation
sizes. For the differential privacy experiments, $\delta \in
\{10^{-6},10^{-7},10^{-8}\}$ and noise multipliers $\sigma \in
\{0.55,0.75,1.0,1.35\}$ are evaluated. The corresponding privacy
budget $\epsilon$ is calculated using the RDP accountant for each
evaluated privacy setting.

Table~\ref{table:reproducibility} shows the experimental parameters used in the evaluation of FL, BC-FL, and iGRAIL. All the methods share the same predictive model with two hidden layers of 64 and 32 neurons. The model is trained with the Adam optimizer with a learning rate of $10^{-3}$, a batch size of 128, and 2 local epochs per communication round. The main federated experiments are conducted with three participating clients and 20 communication rounds.
The training data is partitioned among clients following a Dirichlet distribution with $\alpha=0.5$. Additional experiments vary $\alpha$ and the number of clients to examine data heterogeneity and scalability. For iGRAIL, a conditional WGAN-GP is trained collaboratively to synthesize artificial samples. The latent dimension of the generative model is 32, and the hidden dimension is 96. Synthetic samples are added to local training data with a nominal ratio of 40\%. For differential privacy, gradient clipping and Gaussian noise are used to train the local classifier. The default noise multiplier is 0.9 with a clipping norm of 1.0. Further experiments change the noise multiplier and target $\delta$ in an effort to investigate the privacy-utility trade-off. To test the security, one client is selected as the attack source from the 6th communication round. Conventional FL is fed a perturbed model update, while BC-FL and iGRAIL are tested with tampered protected submissions, which are rejected by their verification mechanisms.

\begin{table*}[!t]
\caption{Classification Performance Across the Evaluated Healthcare Datasets (Mean $\pm$ Standard Deviation)}
\label{table:mainresults}
\centering
\footnotesize
\renewcommand{\arraystretch}{1.3}
\setlength{\tabcolsep}{2.5pt}
\vspace{2mm}
\resizebox{\textwidth}{!}{%
\begin{tabular}{llccccccccc}
\toprule
\textbf{Dataset} & \textbf{Method} & \textbf{Accuracy} & \textbf{Precision} & \textbf{Recall} & \textbf{Spec.} & \textbf{Bal. Acc.} & \textbf{F1} & \textbf{AUROC} & \textbf{AUPRC} & \textbf{MCC} \\
\midrule
\multirow{3}{*}{Cancer} & BC-FL & $0.9737\pm0.0143$ & $0.9549\pm0.0363$ & $\mathbf{0.9762\pm0.0000}$ & $0.9722\pm0.0227$ & $0.9742\pm0.0113$ & $0.9651\pm0.0185$ & $0.9975\pm0.0008$ & $0.9960\pm0.0015$ & $0.9446\pm0.0297$ \\
 & FL & $\mathbf{0.9795\pm0.0083}$ & $\mathbf{0.9690\pm0.0219}$ & $\mathbf{0.9762\pm0.0000}$ & $\mathbf{0.9815\pm0.0131}$ & $\mathbf{0.9788\pm0.0065}$ & $\mathbf{0.9725\pm0.0110}$ & $\mathbf{0.9977\pm0.0005}$ & $\mathbf{0.9964\pm0.0008}$ & $\mathbf{0.9563\pm0.0176}$ \\
 & iGRAIL & $0.9415\pm0.0109$ & $0.9651\pm0.0116$ & $0.8730\pm0.0297$ & $\mathbf{0.9815\pm0.0065}$ & $0.9272\pm0.0146$ & $0.9164\pm0.0168$ & $0.9943\pm0.0006$ & $0.9904\pm0.0008$ & $0.8746\pm0.0232$ \\
\midrule
\multirow{3}{*}{\shortstack[l]{Diabetes--\\Clinical}} & BC-FL & $0.9123\pm0.0046$ & $0.4918\pm0.0149$ & $\mathbf{0.9006\pm0.0079}$ & $0.9134\pm0.0058$ & $\mathbf{0.9070\pm0.0013}$ & $0.6360\pm0.0104$ & $0.9770\pm0.0004$ & $\mathbf{0.8777\pm0.0022}$ & $0.6262\pm0.0084$ \\
 & FL & $0.9119\pm0.0041$ & $0.4906\pm0.0132$ & $0.9002\pm0.0089$ & $0.9130\pm0.0053$ & $0.9066\pm0.0019$ & $0.6349\pm0.0088$ & $\mathbf{0.9771\pm0.0004}$ & $0.8776\pm0.0019$ & $0.6251\pm0.0068$ \\
 & iGRAIL & $\mathbf{0.9633\pm0.0008}$ & $\mathbf{0.9159\pm0.0144}$ & $0.6255\pm0.0036$ & $\mathbf{0.9946\pm0.0010}$ & $0.8101\pm0.0015$ & $\mathbf{0.7433\pm0.0037}$ & $0.9445\pm0.0101$ & $0.8094\pm0.0079$ & $\mathbf{0.7396\pm0.0055}$ \\
\midrule
\multirow{3}{*}{\shortstack[l]{Diabetes--\\Prediction}} & BC-FL & $0.9070\pm0.0013$ & $0.4743\pm0.0041$ & $\mathbf{0.8698\pm0.0049}$ & $0.9104\pm0.0011$ & $\mathbf{0.8901\pm0.0029}$ & $0.6138\pm0.0045$ & $\mathbf{0.9698\pm0.0019}$ & $\mathbf{0.8529\pm0.0078}$ & $0.5998\pm0.0051$ \\
 & FL & $0.9090\pm0.0007$ & $0.4804\pm0.0021$ & $0.8653\pm0.0058$ & $0.9130\pm0.0012$ & $0.8892\pm0.0024$ & $0.6178\pm0.0011$ & $0.9691\pm0.0017$ & $0.8510\pm0.0061$ & $0.6028\pm0.0015$ \\
 & iGRAIL & $\mathbf{0.9579\pm0.0019}$ & $\mathbf{0.8623\pm0.0257}$ & $0.6012\pm0.0183$ & $\mathbf{0.9910\pm0.0021}$ & $0.7961\pm0.0088$ & $\mathbf{0.7080\pm0.0132}$ & $0.9479\pm0.0048$ & $0.7891\pm0.0105$ & $\mathbf{0.6993\pm0.0138}$ \\
\midrule
\multirow{3}{*}{Heart} & BC-FL & $\mathbf{0.7572\pm0.0861}$ & $\mathbf{0.7377\pm0.0955}$ & $0.9248\pm0.0533$ & $\mathbf{0.5488\pm0.2595}$ & $\mathbf{0.7368\pm0.1031}$ & $\mathbf{0.8135\pm0.0432}$ & $\mathbf{0.8921\pm0.0113}$ & $0.8826\pm0.0104$ & $\mathbf{0.5316\pm0.1416}$ \\
 & FL & $0.7101\pm0.0892$ & $0.6827\pm0.0985$ & $\mathbf{0.9608\pm0.0487}$ & $0.3984\pm0.2599$ & $0.6796\pm0.1059$ & $0.7910\pm0.0465$ & $0.8919\pm0.0122$ & $0.8748\pm0.0165$ & $0.4545\pm0.1519$ \\
 & iGRAIL & $0.7047\pm0.0916$ & $0.6972\pm0.1047$ & $0.9150\pm0.1001$ & $0.4431\pm0.3053$ & $0.6791\pm0.1116$ & $0.7787\pm0.0397$ & $0.8759\pm0.0065$ & $\mathbf{0.8860\pm0.0055}$ & $0.4230\pm0.1819$ \\
\bottomrule
\end{tabular}%
}
\end{table*}

\subsection{Predictive Performance Across Healthcare Datasets}
\label{subsec:predictive}
In the first experiment, FL, BC-FL, and iGRAIL are evaluated on four healthcare datasets to assess the predictive performance with the proposed privacy and security mechanisms. Table~\ref{table:mainresults} gives the results for each dataset and the average across all evaluated datasets. FL gets the best accuracy and AUROC for the cancer dataset. iGRAIL obtained a comparable AUROC of $0.9943 \pm 0.0006$. iGRAIL also shows the highest accuracy, precision, F1 score, and MCC in the diabetes-clinical dataset. It indicates that it performs better in classification across different metrics. AUROC is higher for FL and BC-FL. Likewise, iGRAIL outperforms the remaining methods in terms of accuracy, precision, F1 score, and MCC for the diabetes prediction dataset. BC-FL performs better than Heart on most metrics. And iGRAIL yields the highest AUPRC of $0.8860\pm0.0055$. To summarize, the results indicate that there is no single best method for all datasets and all metrics. The iGRAIL adds the proposed framework with additional privacy and security functionalities without sacrificing competitive predictive performance.

\begin{figure}[!h]
\centering
\includegraphics[width=\linewidth,keepaspectratio]{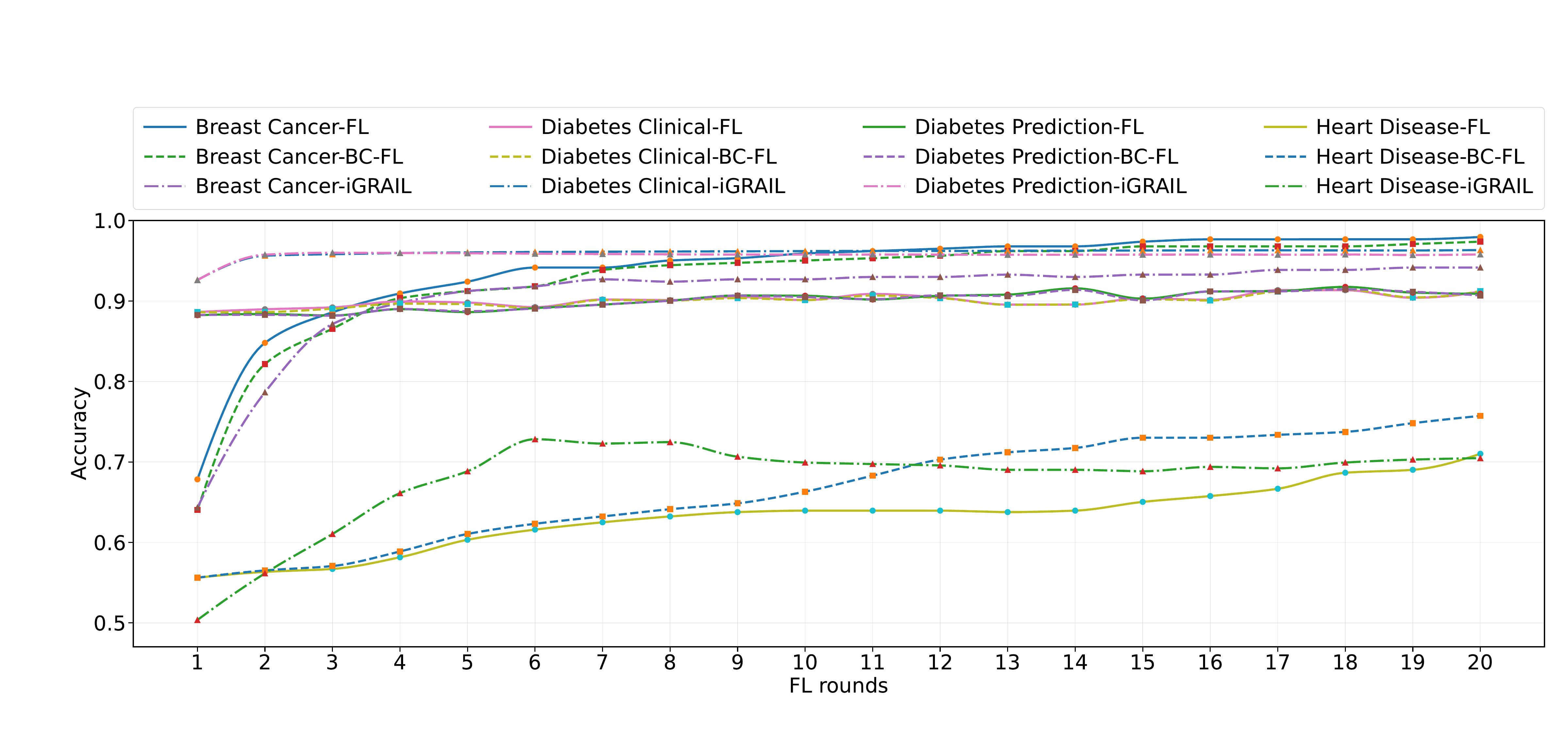}
\caption{Round-wise accuracy over 20 federated communication rounds.}
\label{fig:fig3_1}
\end{figure}

\subsubsection{Accuracy Analysis}
As shown in Figure~\ref{fig:fig3_1}, the accuracy is obtained over the communication rounds. For the cancer dataset, the three methods improve rapidly in the first rounds and then perform relatively stably. FL and BC-FL achieve slightly higher accuracies than iGRAIL, suggesting that adding privacy and security features is not necessarily correlated with improved prediction accuracy on datasets where conventional federated learning performs well. iGRAIL achieves higher accuracy than FL and BC-FL on the Diabetes Clinical and Diabetes Prediction datasets in the first learning phase. Traditional methods improve slowly but remain under iGRAIL in later rounds. This implies that the generative learning mechanism can help retain predictive performance for the data distributions examined. The Heart Disease dataset, in contrast, has lower accuracy and more variation across the methods. The results indicate that learning performance depends on dataset characteristics and that the benefits of iGRAIL are not uniform across healthcare applications.

\begin{figure}[!h]
\centering
\includegraphics[width=\linewidth,keepaspectratio]{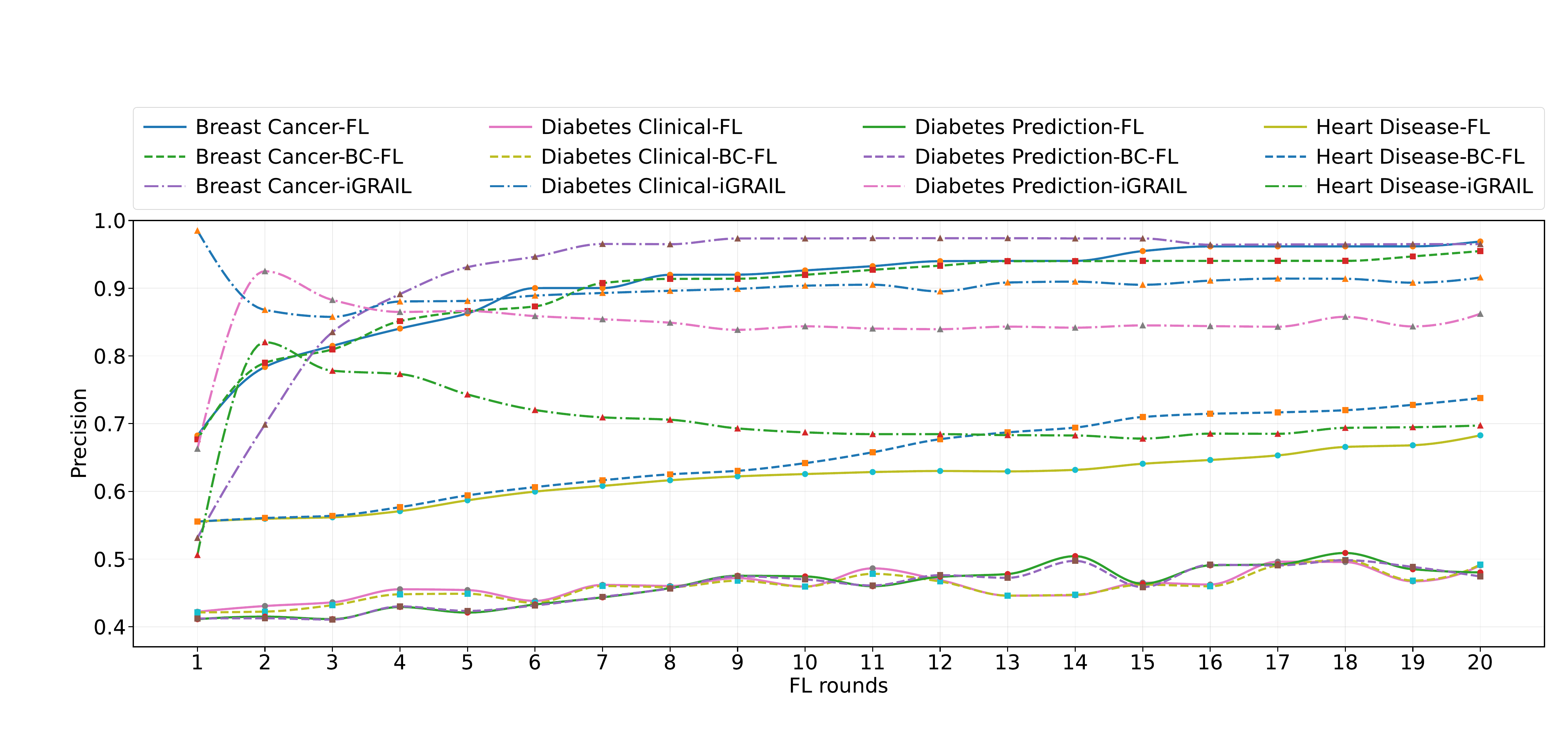}
\caption{Round-wise precision over 20 federated communication rounds.}
\label{fig:fig3_2}
\end{figure}

\subsubsection{Analysis of Precision}
Figure~\ref{fig:fig3_2} illustrates the precision of FL, BC-FL, and iGRAIL in each round over 20 communication rounds. All three methods achieve relatively high precision on the Cancer dataset with small fluctuations after convergence. In the later rounds, iGRAIL has higher precision than FL and BC-FL on both diabetes datasets, which means that a higher percentage of positive predictions made by iGRAIL are true. FL and BC-FL slowly improve in training, but still have lower precision than iGRAIL. However, precision itself cannot represent the classification effect entirely, since higher precision can be obtained by detecting fewer positive samples. Therefore, recall and F1-score are also important to assess the tradeoff between the true positive predictions and false negatives. The heart disease dataset shows lower and less stable precision for the methods. So, methods are less likely to correctly identify positive cases.

\begin{figure}[!h]
\centering
\includegraphics[width=\linewidth,keepaspectratio]{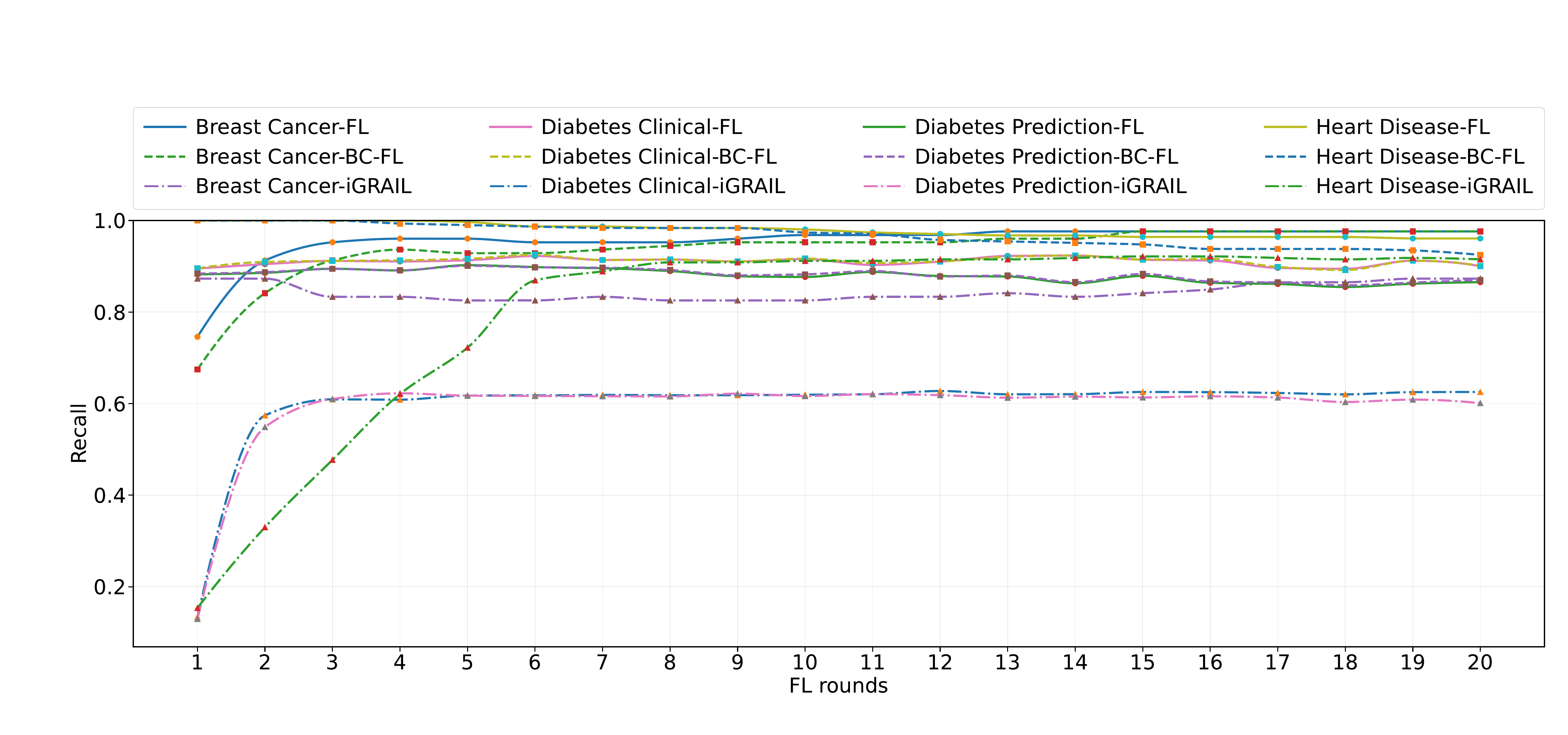}
\caption{Round-wise recall over 20 federated communication rounds.}
\label{fig:fig3_3}
\end{figure}

\subsubsection{Recall Analysis}
The recall over the communication rounds is depicted in Figure~\ref{fig:fig3_3}. Our results show a remarkable difference between iGRAIL and traditional federated approaches. For the Cancer dataset, all methods have relatively high recall, but iGRAIL is generally worse than FL and BC-FL. That is, classical methods are able to identify more true positives. For diabetes clinical data and diabetes prediction datasets, the difference is more pronounced. FL and BC-FL achieve high recall, while iGRAIL converges to much lower recall values. iGRAIL achieves better accuracy on these datasets. However, iGRAIL suffers from lower recall performance and misses more true-positive cases. This trade-off is particularly important in healthcare applications, where false negatives can be disastrous. In the heart disease dataset, we observe mostly high recall curves after the first few rounds of learning, with some variability. Overall, the results indicate that the privacy-preserving generative framework can provide useful learning performance, though its classification behavior may prefer precision to recall on some datasets. This limitation should be taken into account when defining operational limits for clinical use.

\begin{figure}[!h]
\centering
\includegraphics[width=\linewidth,keepaspectratio]{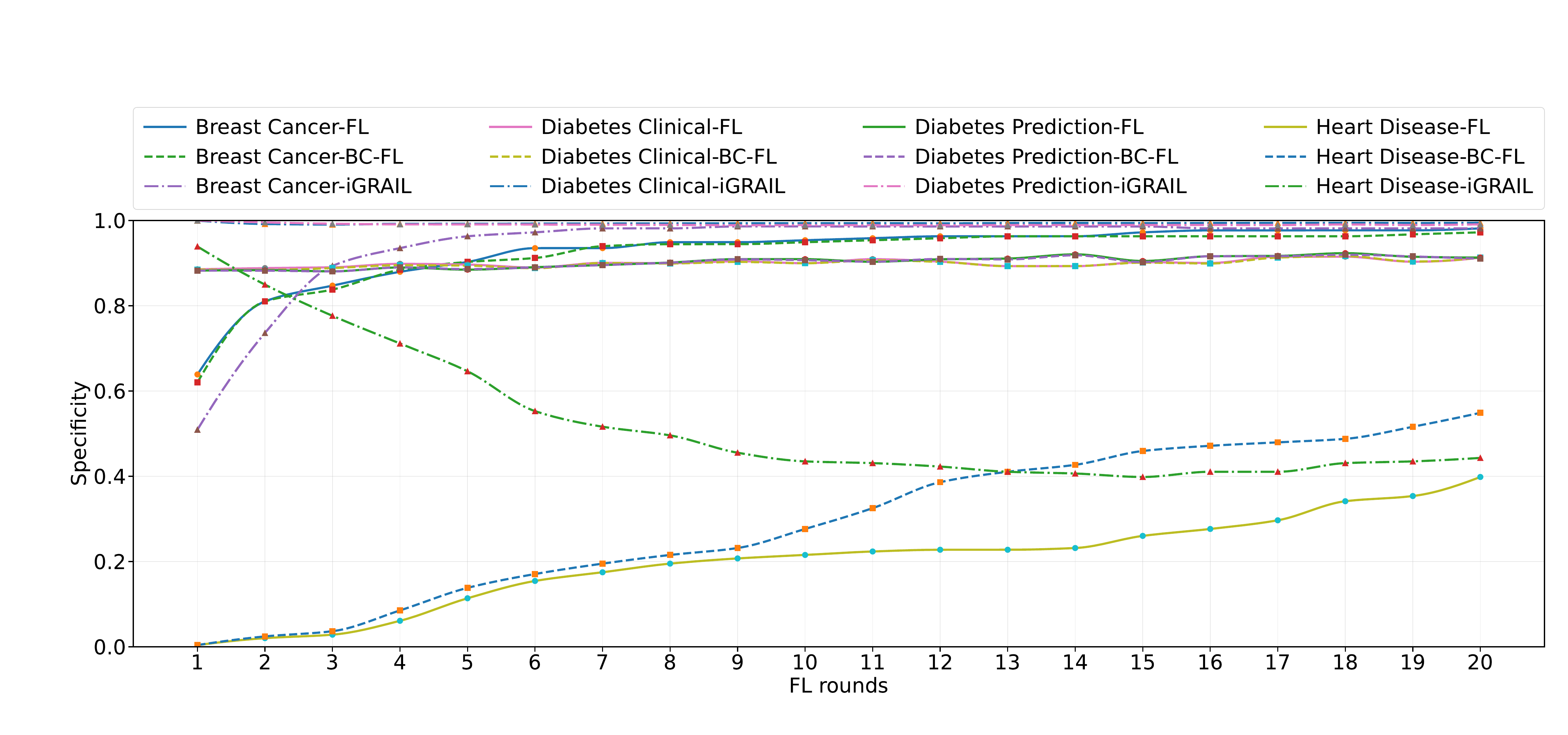}
\caption{Round-wise specificity over 20 federated communication rounds.}
\label{fig:fig3_4}
\end{figure}

\subsubsection{Specificity Analysis}
The specificities obtained by the three methods are presented in Figure~\ref{fig:fig3_4}. Specificity is the percentage of true negatives that are correctly classified by the classifier. The Cancer dataset shows high specificity for all the studied approaches, with strong identification of negative samples after the first rounds of communication. iGRAIL is more specific than FL and BC-FL for diabetes clinical and diabetes prediction datasets. This is consistent with high precision and low recall, showing iGRAIL to be more conservative in its positive classification, making fewer false positive predictions. The traditional methods are less specific, and hence the higher recall is accompanied by more false positive detections. In the Heart Disease dataset, the specificity of some methods is not very high and grows progressively during training. These results illustrate the importance of sensitivity and specificity for assessing health prediction systems. But a high recall method can still produce too many false alarms. A high specificity method can miss some positive cases.

\begin{figure}[!h]
\centering
\includegraphics[width=\linewidth,keepaspectratio]{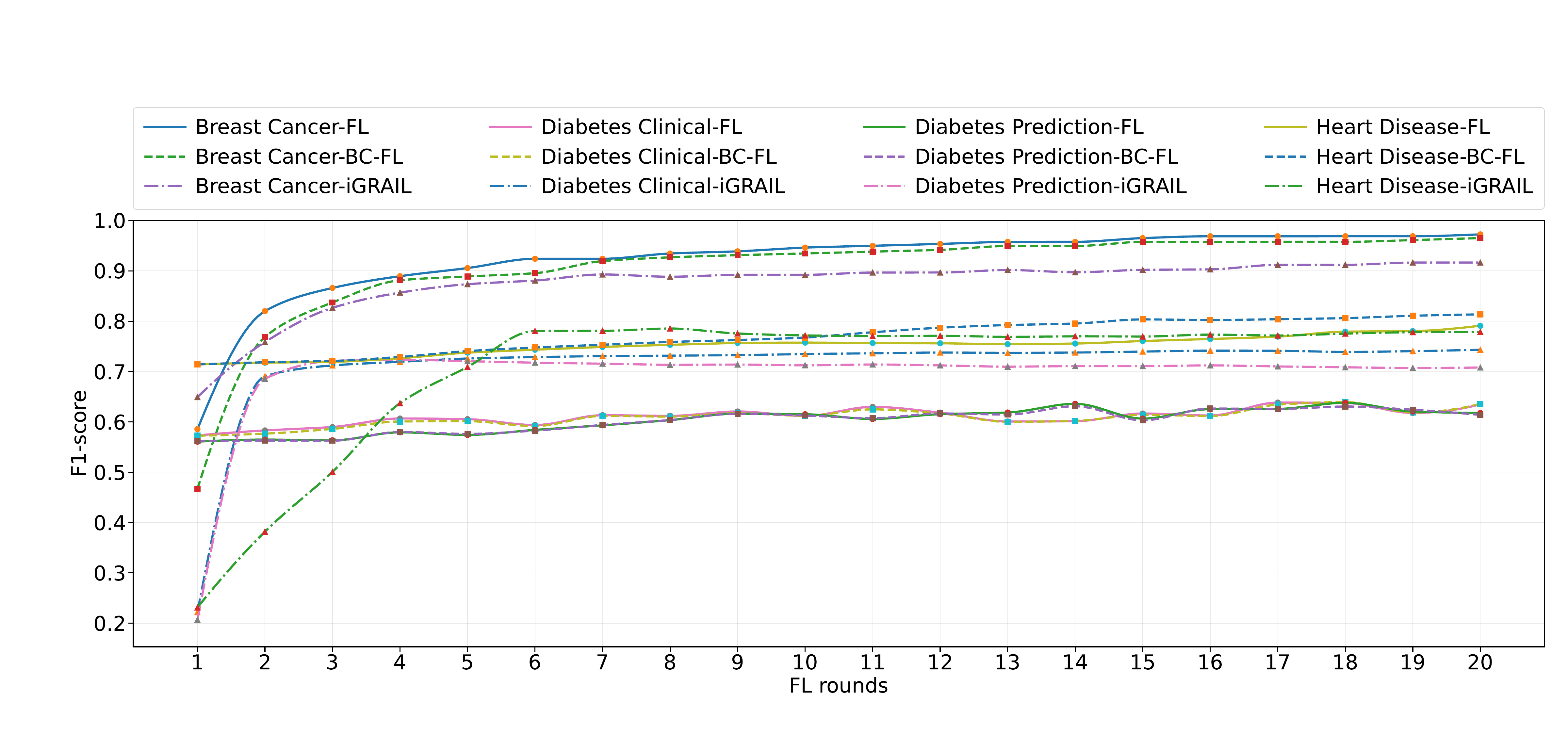}
\caption{Round-wise F1-score over 20 federated communication rounds.}
\label{fig:fig3_5}
\end{figure}

\subsubsection{F1-Score Analysis}
Figure~\ref{fig:fig3_5} shows the F1 score, which is a combined measure of precision and recall. For the Cancer dataset, FL and BC-FL achieve strong F1-scores and generally outperform iGRAIL. This result is in line with their high precision and recall on this dataset. On the Diabetes Clinical and Diabetes Prediction datasets, iGRAIL outperforms traditional methods in the later rounds of communication in terms of F1 scores. iGRAIL has lower recall but much higher precision, resulting in a better balance between the two metrics, as measured by the F1-score. This result shows that the devised learning mechanism can improve the overall precision-recall trade-off in the tested settings. For the Heart Disease dataset, the differences between the methods are smaller, and FL and BC-FL generally remain competitive in F1 score. Our results indicate that iGRAIL's performance depends on the underlying data distribution and the trade-off between false-positive and false-negative predictions. Thus, the improvements in F1-score on diabetes datasets should not be considered as a generalization of superiority on all healthcare tasks.

\begin{figure}[!h]
\centering
\includegraphics[width=\linewidth,keepaspectratio]{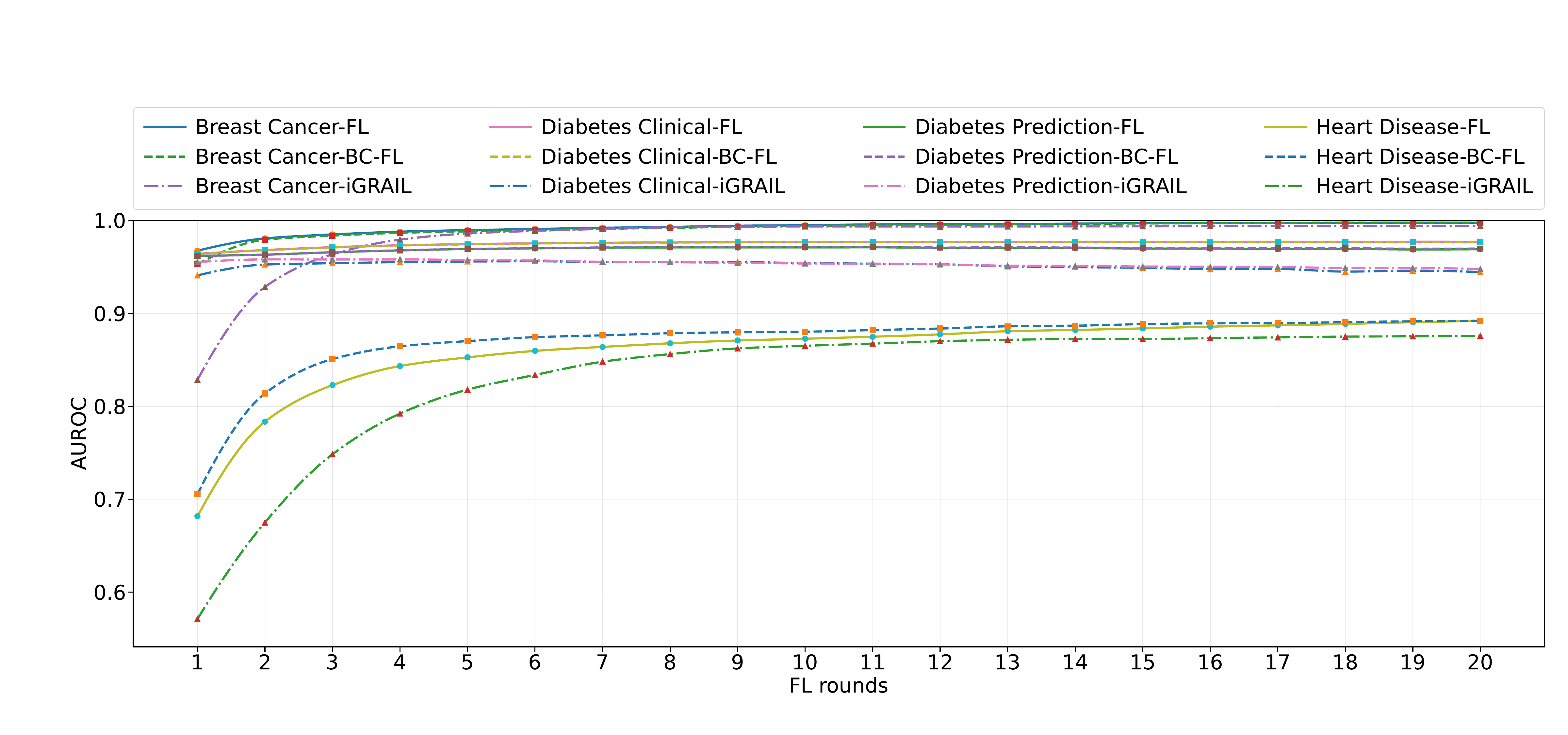}
\caption{Round-wise AUROC over 20 federated communication rounds.}
\label{fig:fig3_6}
\end{figure}

\subsubsection{AUROC Analysis}
The AUROC (area under the receiver operating characteristic curve) during federated training is shown in Figure~\ref{fig:fig3_6}. Accuracy and F1-score differ from AUROC, which measures the classifier's ability to discriminate between positive and negative cases across different classification thresholds. For the Cancer dataset, all three approaches achieve high AUROC values, with FL and BC-FL having a slight advantage over iGRAIL. FL and BC-FL generally achieve higher AUROC values than iGRAIL for the Diabetes Clinical and Diabetes Prediction datasets, although iGRAIL produces higher accuracy and F1-scores. The observed difference suggests that the improvements obtained at the chosen classification threshold do not necessarily translate to better discrimination across all possible thresholds. The AUROC curves for the heart disease dataset improve in the first few rounds of communication, and then stabilize. Even in later rounds, the differences between the methods are relatively small. The AUROC results overall suggest that iGRAIL still maintains meaningful discrimination ability, but conventional FL methods might yield improved threshold-independent performance across multiple datasets.

\begin{figure}[!h]
\centering
\includegraphics[width=\linewidth,keepaspectratio]{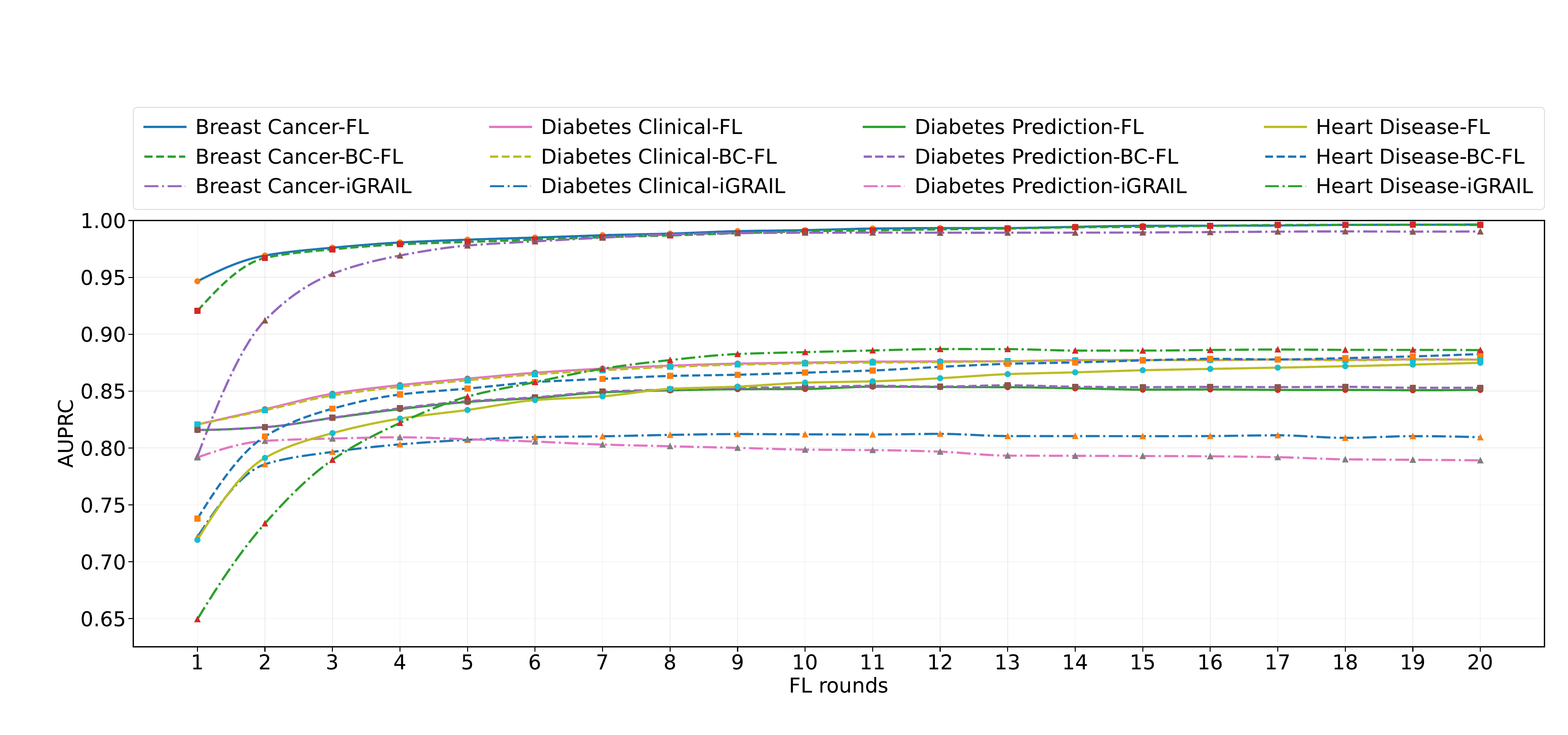}
\caption{Round-wise AUPRC over 20 federated communication rounds.}
\label{fig:fig3_7}
\end{figure}

\subsubsection{AUPRC Analysis}
Figure~\ref{fig:fig3_7} shows the AUPRC performance of FL, BC-FL, and iGRAIL over 20 communication rounds. AUPRC is especially useful for imbalanced healthcare datasets, since it evaluates precision and recall across different classification thresholds. As shown in Figure~\ref{fig:fig3_7}, all three methods achieve high AUPRC values on the Cancer dataset, where FL and BC-FL perform slightly better than iGRAIL. FL and BC-FL also achieve higher AUPRC values for the Diabetes Clinical and Diabetes Prediction datasets, whereas iGRAIL obtains better precision and F1-score at the chosen classification threshold. Thus, improvements at a fixed threshold do not necessarily translate to better performance at all thresholds. In the heart disease dataset, the three methods attain similar AUPRC values, showing minor differences during the later rounds of communication. In general, these results indicate that model performance is dependent on the evaluation metric and classification threshold. Therefore, both false-positive and false-negative predictions should be taken into account when choosing a suitable method for healthcare applications.

\begin{figure}[!h]
\centering
\includegraphics[width=\linewidth,keepaspectratio]{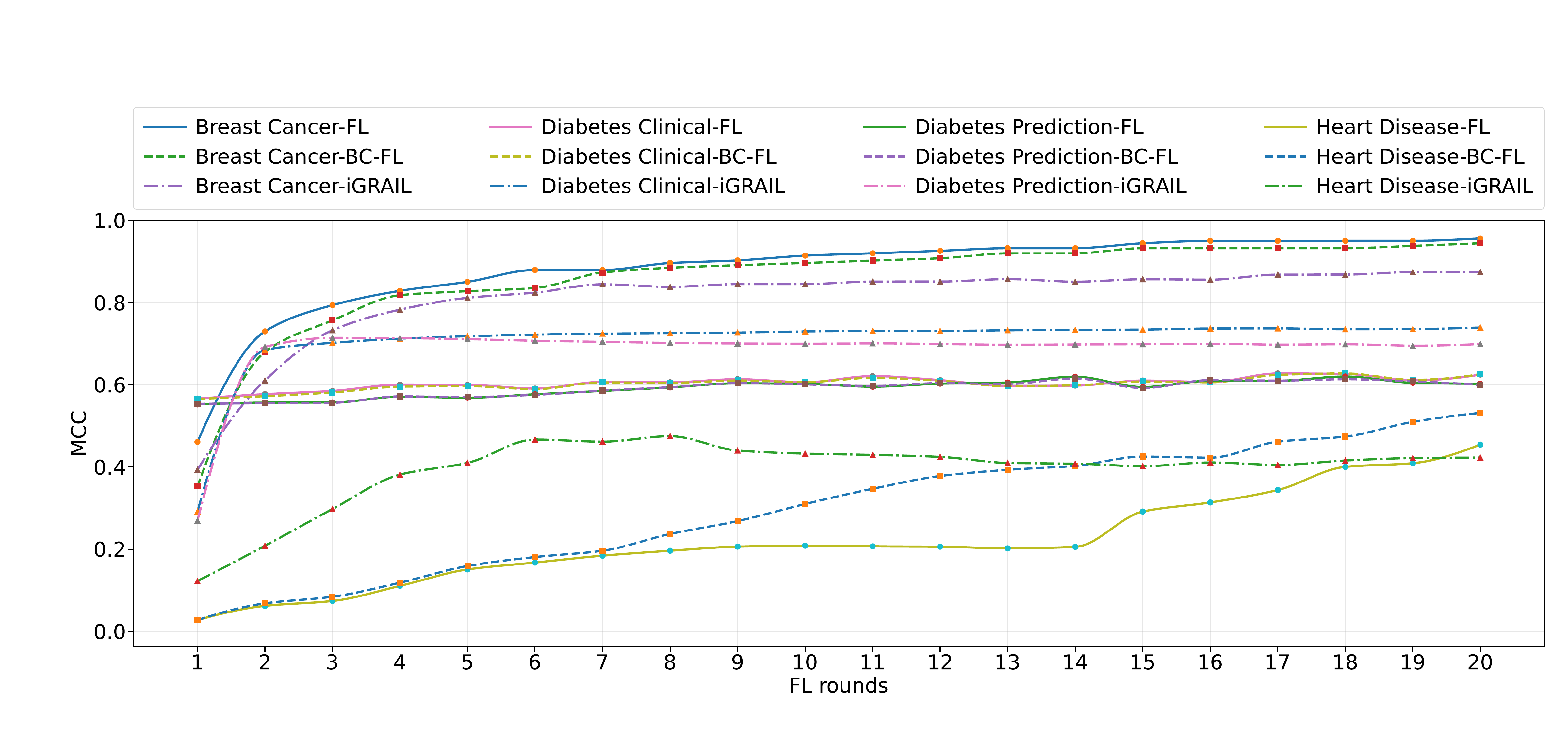}
\caption{Round-wise MCC over 20 federated communication rounds.}
\label{fig:fig3_8}
\end{figure}

\subsubsection{MCC Analysis}
The Matthews correlation coefficient (MCC) takes into account true positives, true negatives, false positives, and false negatives and is shown in Figure~\ref{fig:fig3_8}. The MCC is a good measure to evaluate the classification quality in case of imbalanced classes. For the cancer dataset, FL and BC-FL obtained high MCC values, while iGRAIL performed a bit worse. This result is consistent with the observed accuracy and F1-score. In the Diabetes Clinical and Diabetes Prediction datasets, iGRAIL shows better performance than conventional methods in terms of MCC in the later rounds of communication. This means its classification decisions are more consistent with the true labels on average for the operating conditions tested. For the Heart Disease dataset, we report lower MCC values and higher variance across rounds, indicating a more challenging classification problem. Results show that the benefits of iGRAIL are dependent on the dataset and cannot be generalized to all health prediction tasks. However, the improvements in MCC on the diabetes datasets further strengthen the claim that the proposed framework is able to achieve competitive classification performance by integrating privacy and security mechanisms.

\subsubsection{Discussion}
Three important observations can be made from the results of each round. First, all three methods show significant learning over the 20 communication rounds. A number of metrics grow fast in the first rounds and reach relatively stable values. This demonstrates that the extra mechanisms introduced in iGRAIL have no effect on the learning of the underlying federated model. Second, iGRAIL exhibits different prediction properties over the tested datasets. It obtains high accuracy, precision, specificity, F1 score, and MCC on the two diabetes datasets, while FL and BC-FL generally maintain advantages in recall, AUROC, and AUPRC. The traditional methods are better or comparable to iGRAIL in many metrics on cancer and heart disease datasets. These observations suggest that the proposed framework does not uniformly improve all evaluation metrics but instead yields dataset-dependent predictive trade-offs. Third, the differences found justify the use of multiple complementary metrics for evaluating privacy-preserving healthcare learning. Accuracy and F1-score are not enough to claim clinical reliability, especially when the cost of false-negative and false-positive predictions are not equal. Together, recall, specificity, AUROC, AUPRC, and MCC provide a more complete picture of each method's performance.

\subsection{Privacy-utility evaluation}
\label{subsec:privacy}
Privacy-Utility Experiment. We conduct a privacy-utility experiment to evaluate iGRAIL’s ability to protect sensitive local healthcare data while maintaining reliable predictive performance in federated training. DP is incorporated into the learning process to limit the amount of information that can be extracted from the learned model about individual training records. The experiments take into account four noise multipliers, $\sigma\in\{0.55,0.75,1.00,1.35\}$, and three privacy parameters, $\delta\in\{10^{-6},10^{-7},10^{-8}\}$. The corresponding privacy loss parameter $\epsilon$ is computed using a R\'enyi differential privacy (RDP) accountant. Increasing the noise multiplier for a fixed training configuration and privacy parameter $\delta$ typically decreases $\epsilon$, indicating a stronger formal privacy guarantee. However, more noise could also affect model convergence and predictive performance. The experiment therefore investigates the impact of different privacy configurations on learning behavior for the four healthcare datasets.

\begin{figure}[!h]
\centering
\includegraphics[width=\linewidth,keepaspectratio]{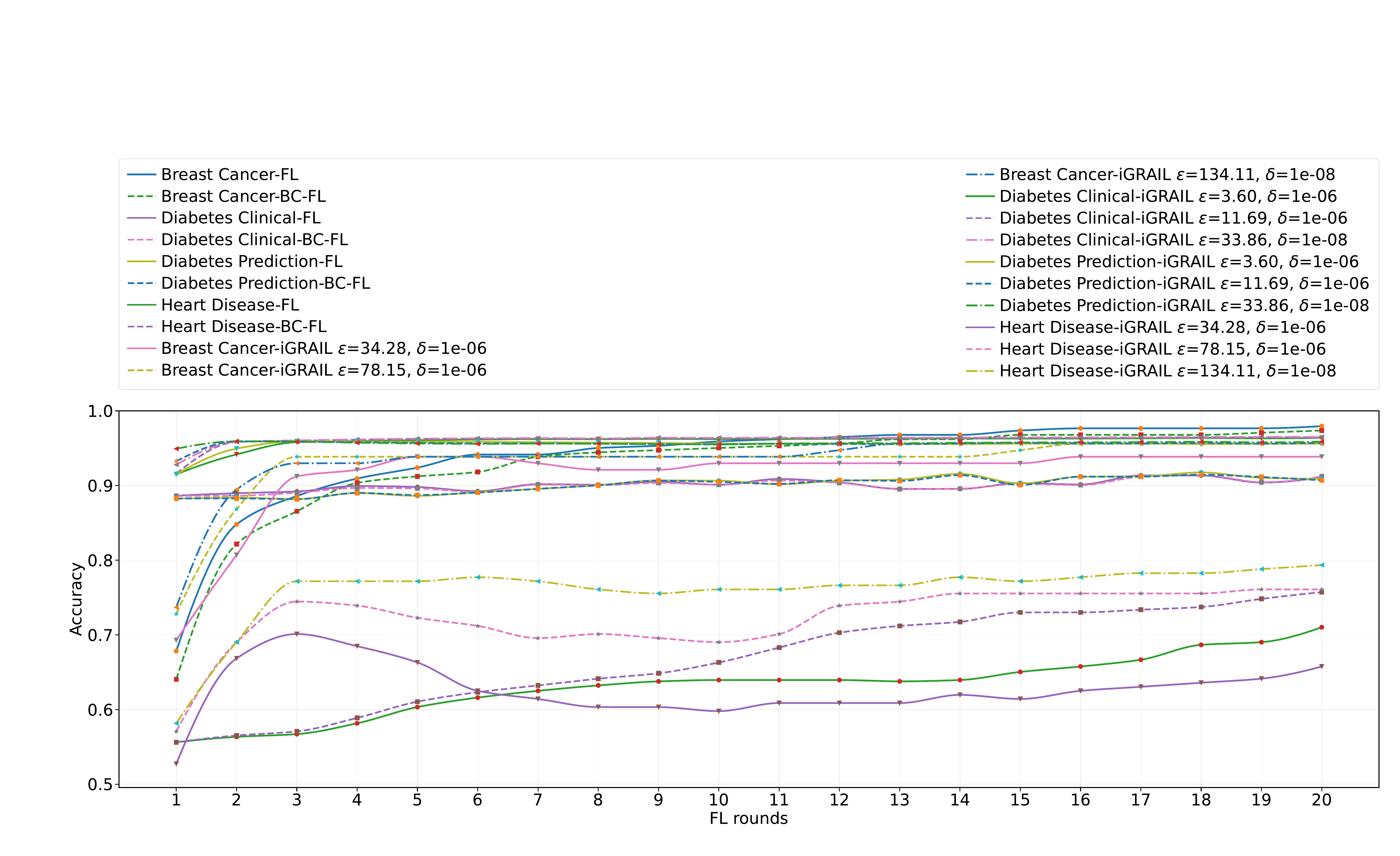}
\caption{Round-wise accuracy under differential privacy.}
\label{fig:fig4_1}
\end{figure}

\subsubsection{Accuracy Analysis}
Figure~\ref{fig:fig4_1} shows the accuracy obtained in the federated training for different privacy settings. The aim of this experiment is to see if changing the DP noise multiplier has an effect on the overall percentage of correctly classified samples. Increasing privacy introduces perturbations during learning, which can influence the parameter updates and the model’s ability to fit the underlying data distribution. The results show that iGRAIL achieves useful classification accuracy under the tested privacy settings, but the variation is affected by the characteristics of the healthcare data. This means that the model can continue to learn with the addition of privacy noise. But we should not expect the noise multiplier and accuracy to have the same relationship for different datasets. The observed performance may depend on the size of the dataset, the distribution of features, the class imbalance, and the sensitivity of the model updates. Thus, the accuracy results provide a first indication of the privacy-utility tradeoff. They show that more privacy protection does not necessarily imply a proportional loss of predictive power. However, accuracy should be considered with recall and other classification metrics to evaluate if stronger privacy affects detection of clinically relevant positive cases.

\begin{figure}[!h]
\centering
\includegraphics[width=\linewidth,keepaspectratio]{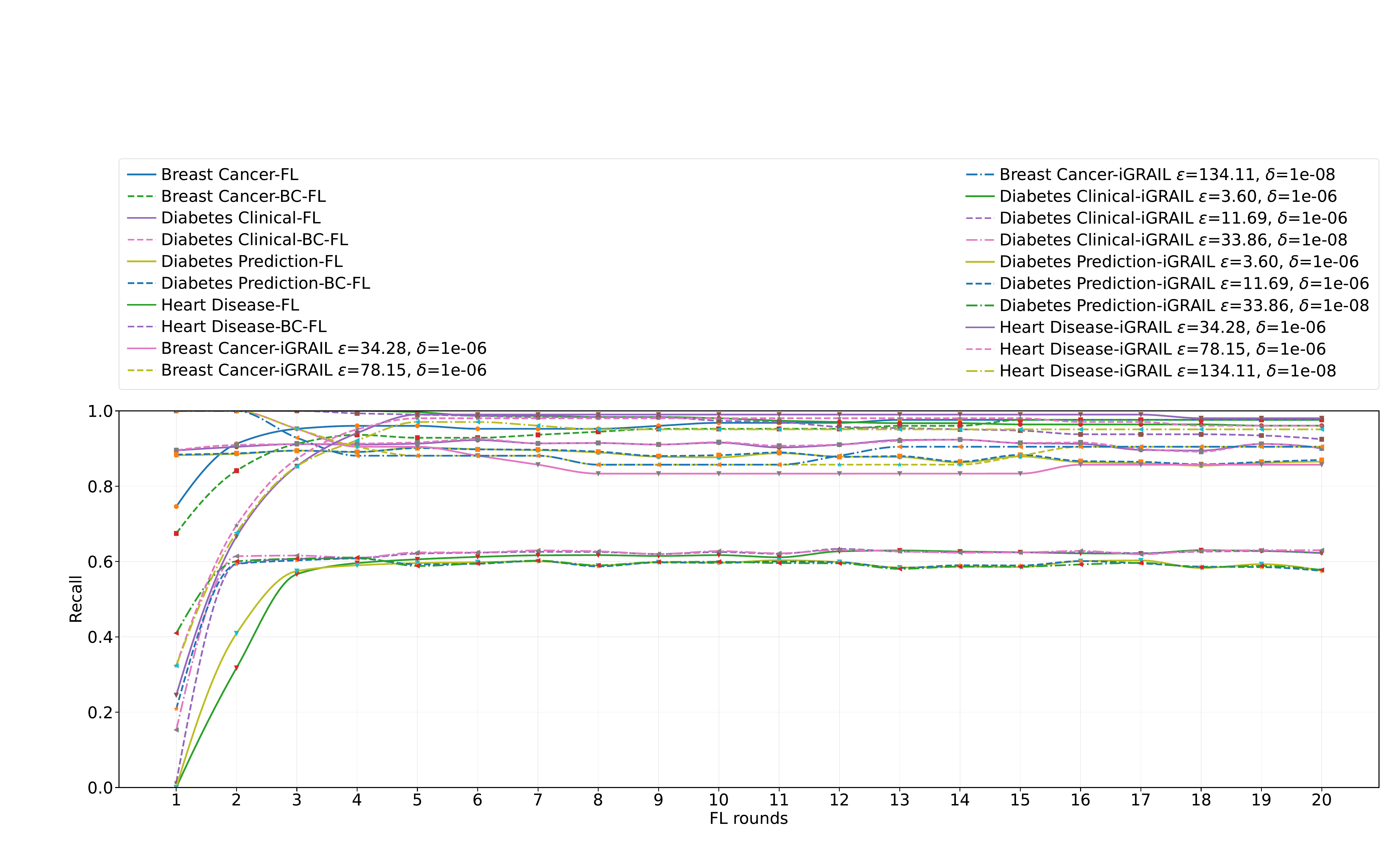}
\caption{Round-wise recall under differential privacy.}
\label{fig:fig4_2}
\end{figure}

\subsubsection{Recall Analysis}
The recall vs. communication rounds for different DP noise multipliers are depicted in Figure \ref{fig:fig4_2}. Recall is the ratio of positive instances that the model correctly predicts. This is particularly crucial in health care applications where a false-negative prediction can result in a missed diagnosis or delayed medical intervention. The recall results show that it is important to consider privacy protection in addition to the overall classification accuracy. As the noise multiplier increases, the learned decision boundary changes, and thus the number of positive samples detected by the classifier changes. The magnitude of this effect varies with the dataset and the privacy setting considered. Our results suggest that metric-dependent behavior cannot be characterized by the final privacy budget or overall accuracy alone when evaluating the performance of privacy-preserving training. The ability of a high-accuracy configuration to detect positive cases can still vary. Therefore, recall is an important complementary metric to evaluate the appropriateness of the selected privacy setting for healthcare prediction tasks where missing positive cases is costly.

\begin{figure}[!tb]
\centering
\includegraphics[width=\linewidth,keepaspectratio]{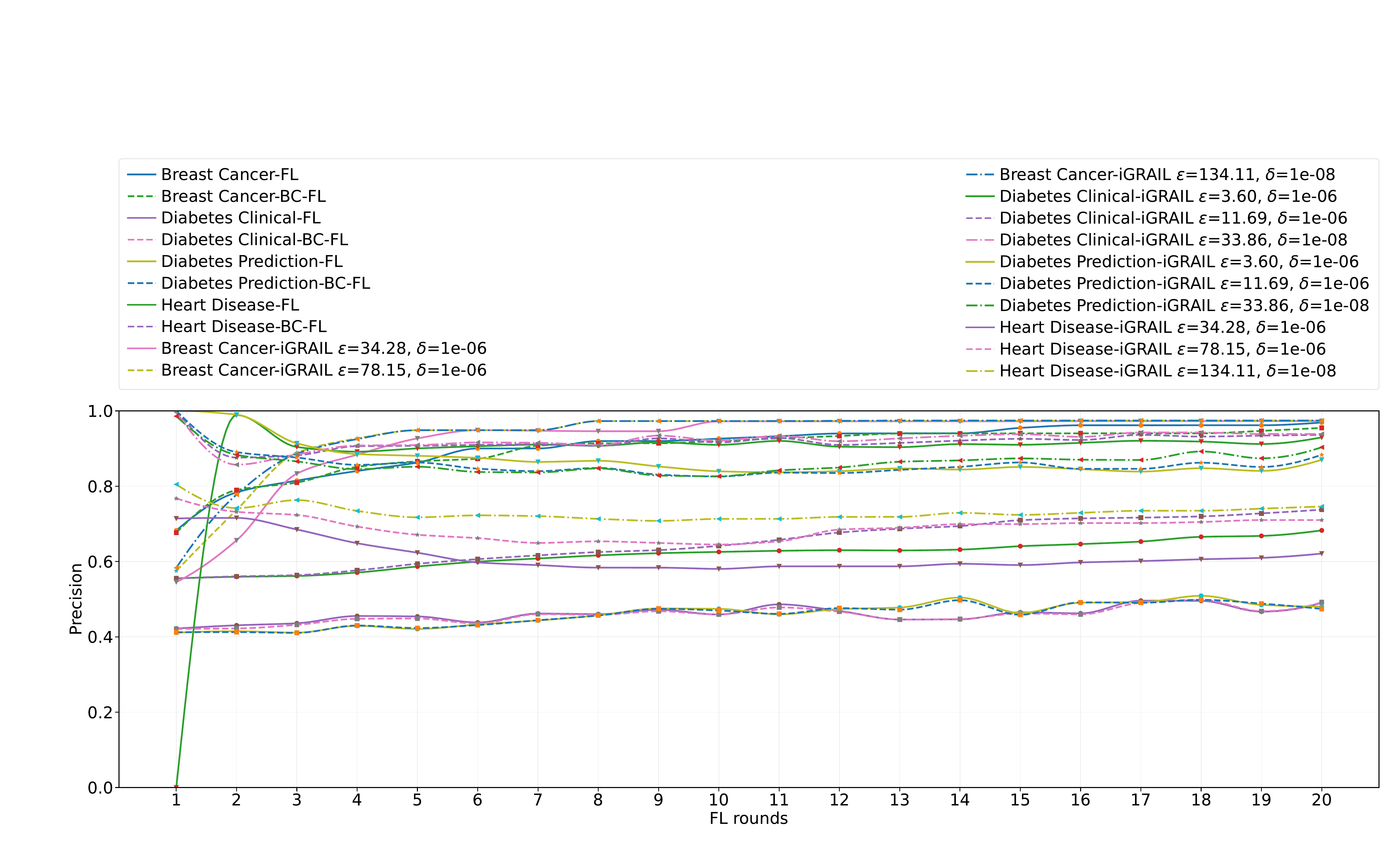}
\caption{Round-wise precision under differential privacy.}
\label{fig:privacy_precision}
\end{figure}

\subsubsection{Precision Analysis}
Figure~\ref{fig:privacy_precision} depicts the round-wise precision of FL, BC-FL, and iGRAIL with various differential privacy settings over the four healthcare datasets. The results show significant differences in precision among the datasets. For breast cancer, the precision is still relatively high, and some configurations achieve around $0.90$--$0.98$ in the first few communication rounds. The Diabetes-Prediction dataset also shows good precision in many configurations, usually between 0.85 and 0.95 in later rounds. In comparison, the Diabetes-Clinical dataset has a much lower precision, with most curves staying around 0.40–0.50. The Heart Disease dataset has moderate accuracy with values from 0.55 to 0.75. Our results also show that the precision does not increase monotonically with the number of communication rounds. Some configurations converge in the first few rounds, and others fluctuate or change slowly. However, the relative performance of iGRAIL also depends on the dataset and privacy configuration. There is no single configuration that always yields the highest precision under all circumstances. These observations are more important for the imbalanced diabetes datasets, where the false-positive predictions can significantly affect the precision.

\begin{figure}[!tb]
\centering
\includegraphics[width=\linewidth,keepaspectratio]{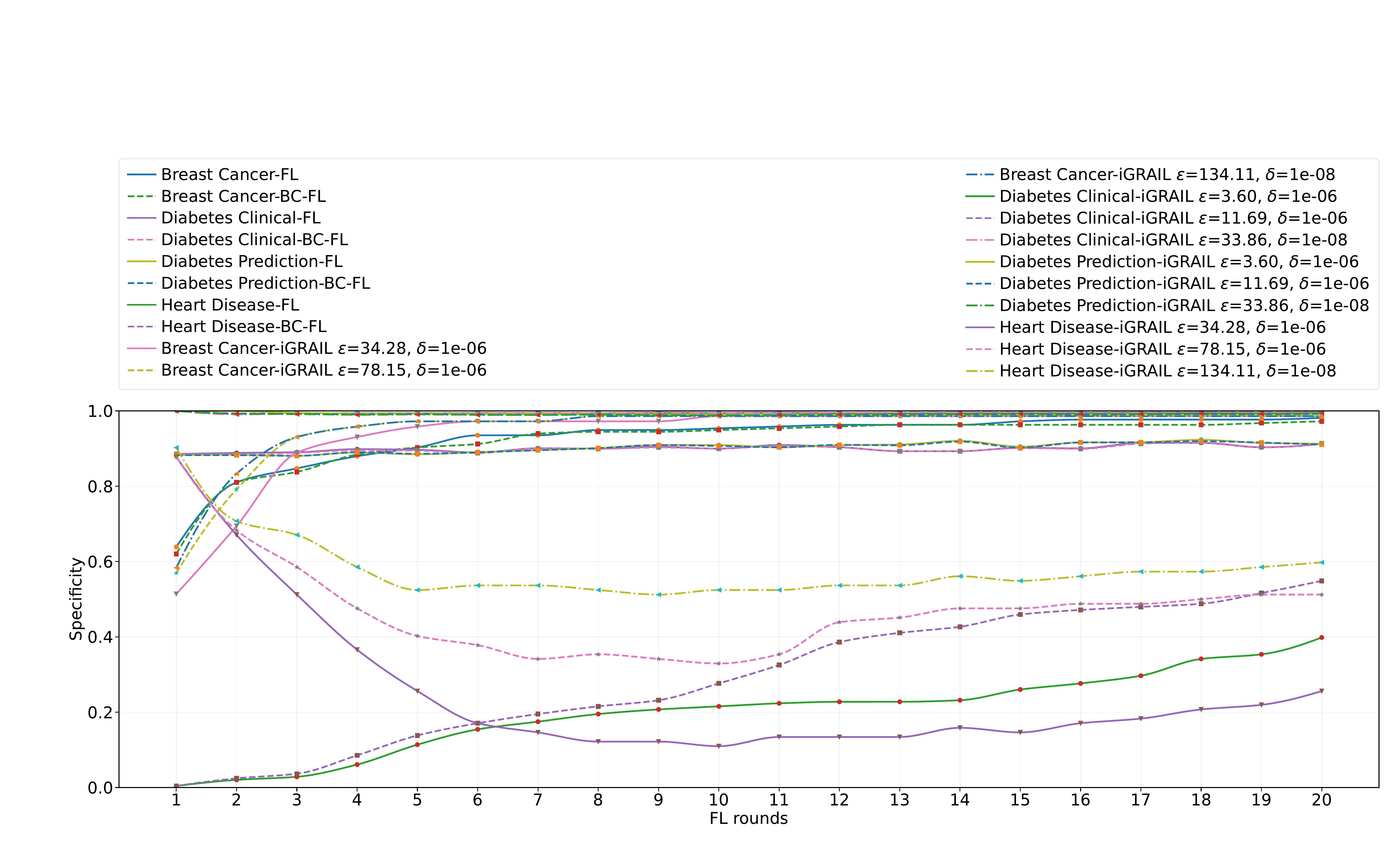}
\caption{Round-wise specificity under differential privacy.}
\label{fig:privacy_specificity}
\end{figure}

\subsubsection{Specificity Analysis}
The specificity of FL, BC-FL, and iGRAIL over 20 communication rounds with different privacy settings is shown in Figure~\ref{fig:privacy_specificity}. The results indicate that the specificity depends on the dataset and the privacy setting. Most breast cancer methods maintain high specificity, usually over 0.95 in later rounds. Similar results are observed for diabetes prediction, with specificity still ranging 0.90–0.99. However, for Diabetes-Clinical, FL and BC-FL have higher specificity, while some settings of iGRAIL have lower values between 0.33 and 0.55. That means iGRAIL will predict more false positives in those settings. In the last round, FL achieves a specificity of around 0.40, and BC-FL achieves a specificity of around 0.55 for heart disease. In contrast, some configurations of iGRAIL provide values above 0.90, showing better performance in the detection of negative cases. But this improvement does not occur uniformly across privacy settings. The results here indicate that iGRAIL does a good job in some cases, but does not always get the highest specificity. The outcomes also depend on the privacy setting you choose.

\begin{figure}[!tb]
\centering
\includegraphics[width=\linewidth,keepaspectratio]{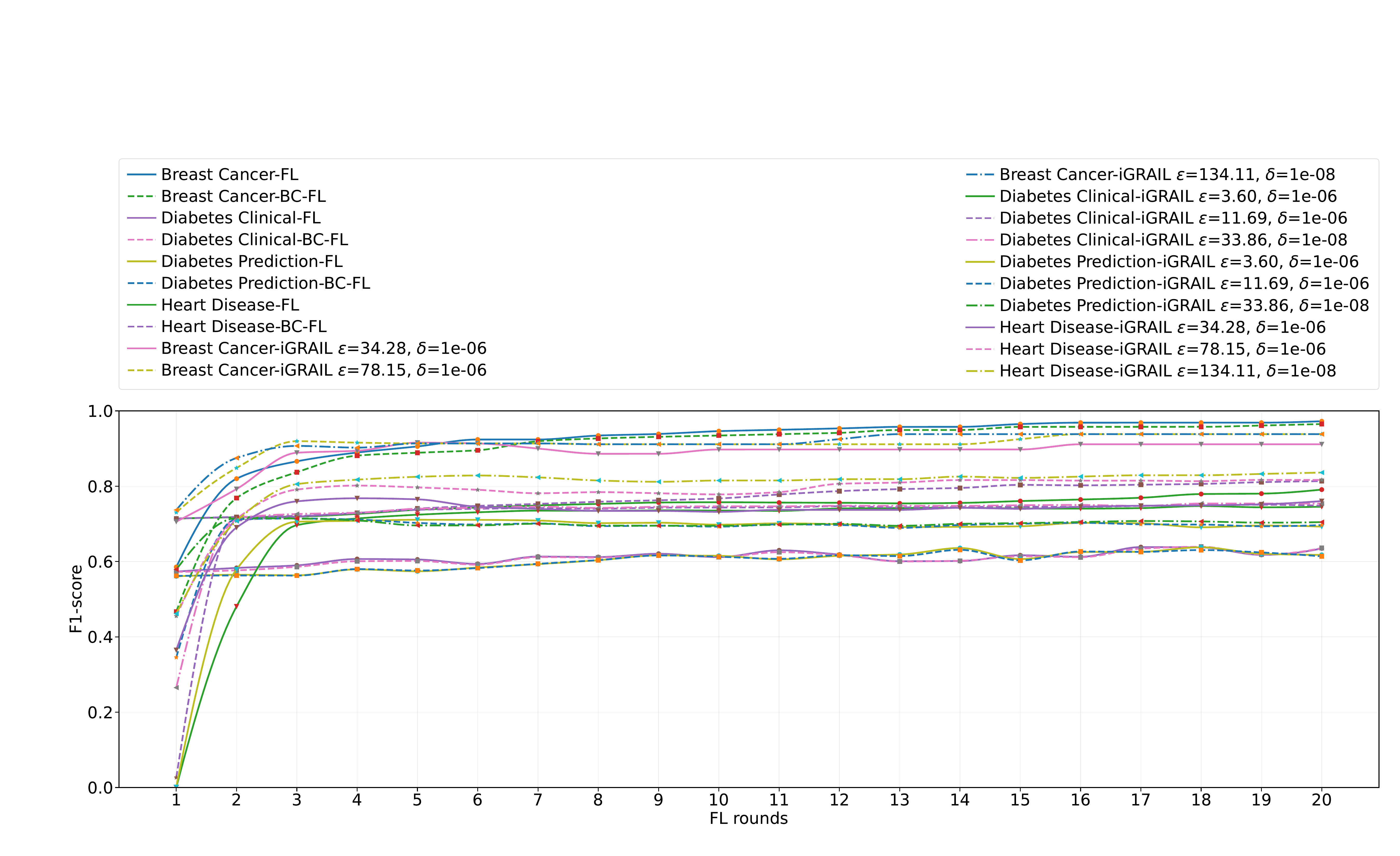}
\caption{Round-wise F1-score under differential privacy.}
\label{fig:fig4_3}
\end{figure}

\subsubsection{Analysis of F1-Score}
The F1-score achieved with different privacy settings is shown in Figure~\ref{fig:fig4_3}. The F1 score is the harmonic mean of precision and recall. It provides a more balanced view of positive-class predictions than accuracy does. The reported F1 scores remain above $0.69$ across the evaluated privacy settings on the Diabetes Clinical and Diabetes Prediction datasets. This shows that the model can maintain a useful precision-recall trade-off even as the noise multiplier increases. In particular, the results show that we can achieve better DP protection without a large drop in F1 score for these two datasets. In the Heart Disease dataset, the decrease in F1 score is more severe under stricter privacy settings. This means its predictive behavior is more sensitive to the additional perturbations injected during training. One reason might be that the properties of the data at hand are less tolerant to noisy model updates. But the exact reason should be further studied in terms of dataset size, class distribution, and model sensitivity. The results indicate that the effect of differential privacy on classification performance is dependent on the dataset. The F1 scores for the two diabetes datasets are still relatively good, but the results for the heart disease show a more pronounced privacy-utility trade-off. So the privacy setting must be selected based on the desired privacy level and the tolerable loss of forecasting performance.

\begin{figure}[!tb]
\centering
\includegraphics[width=\linewidth,keepaspectratio]{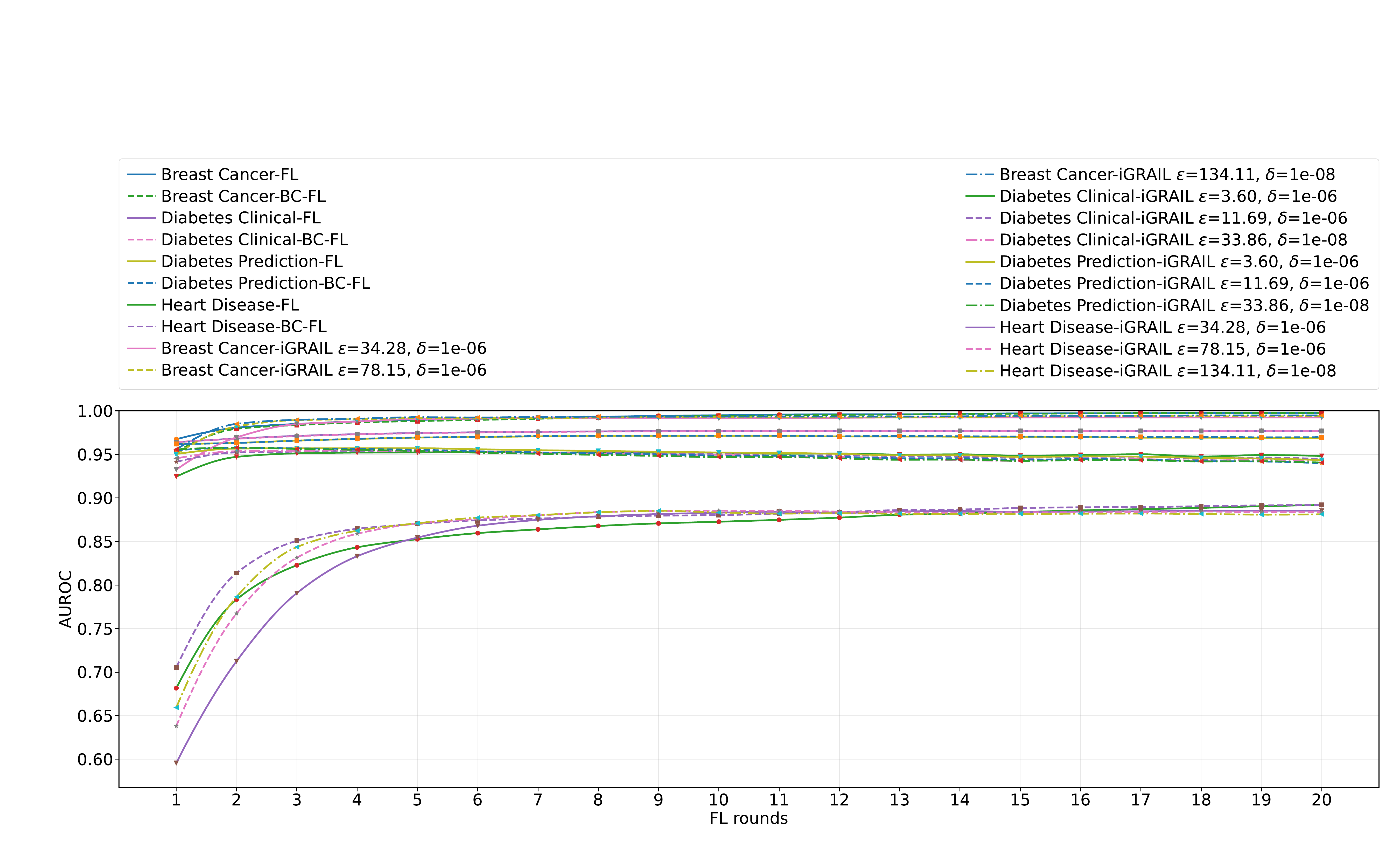}
\caption{Round-wise AUROC under differential privacy.}
\label{fig:fig4_4}
\end{figure}

\subsubsection{AUROC Analysis}
Figure~\ref{fig:fig4_4} shows the AUROC results under different DP settings. AUROC measures the model’s ability to distinguish between positive and negative classes at different classification thresholds. This metric helps to evaluate if stronger privacy impacts the total discriminative ability of the learned model. For the Cancer dataset, the reported privacy budget $\epsilon$ decreases from $124.8950$ to $34.2837$ at $\delta=10^{-6}$ by increasing the noise multiplier from $\sigma=0.55$ to $\sigma=1.35$. This is an approximately $72.5\%$ decrease in $\epsilon$. With the same configurations, AUROC only decreases from $0.9947$ to $0.9924$, an absolute reduction of $0.0023$. This result indicates that we can achieve a significant decrease in the privacy-loss parameter with a small change in AUROC for the cancer dataset. Thus, the model still has a strong capability of class discrimination under the evaluated stronger privacy configuration. However, the reported $\epsilon$ value of $34.2837$ should not be interpreted as indicating that privacy leakage is negligible. The strength of the formal DP guarantee depends on the full $(\epsilon,\delta)$ setting and the underlying threat model. The privacy budget and predictive performance for different noise multipliers, with a fixed target $\delta=10^{-6}$, are presented in Table~\ref{table:privacy_budget}.
Increasing the noise multiplier decreases the privacy-loss parameter $\epsilon$, signifying a stronger formal privacy guarantee. For the diabetes clinical and prediction datasets, $\epsilon$ decreases to 3.5981 at $\sigma=1.35$. The corresponding classification metrics show a dataset-dependent privacy-utility trade-off.

\begin{table*}[!tb]
\caption{Differential-Privacy Budget and Predictive Utility ($\delta=10^{-6}$)}
\label{table:privacy_budget}
\centering
\footnotesize
\renewcommand{\arraystretch}{1.12}
\setlength{\tabcolsep}{5pt}
\vspace{2mm}
\begin{tabular}{lccccccc}
\toprule
\textbf{Dataset} & $\boldsymbol{\sigma}$ & $\boldsymbol{\delta}$ & $\boldsymbol{\epsilon}$ & \textbf{Accuracy} & \textbf{F1} & \textbf{AUROC} & \textbf{AUPRC} \\
\midrule
Cancer & 0.55 & $10^{-6}$ & 124.8950 & 0.9561 & 0.9383 & 0.9947 & 0.9913 \\
 & 0.75 & $10^{-6}$ & 78.1506 & 0.9561 & 0.9383 & 0.9940 & 0.9902 \\
 & 1.00 & $10^{-6}$ & 51.7237 & 0.9561 & 0.9383 & 0.9934 & 0.9893 \\
 & 1.35 & $10^{-6}$ & 34.2837 & 0.9386 & 0.9114 & 0.9924 & 0.9876 \\
\midrule
Diabetes-Clinical & 0.55 & $10^{-6}$ & 28.7457 & 0.9650 & 0.7540 & 0.9432 & 0.8155 \\
 & 0.75 & $10^{-6}$ & 11.6935 & 0.9644 & 0.7481 & 0.9446 & 0.8131 \\
 & 1.00 & $10^{-6}$ & 6.0326 & 0.9639 & 0.7451 & 0.9474 & 0.8127 \\
 & 1.35 & $10^{-6}$ & 3.5981 & 0.9639 & 0.7456 & 0.9482 & 0.8105 \\
\midrule
Healthcare & 0.55 & $10^{-6}$ & 28.7457 & 0.9588 & 0.7047 & 0.9406 & 0.7845 \\
 & 0.75 & $10^{-6}$ & 11.6935 & 0.9575 & 0.6968 & 0.9399 & 0.7768 \\
 & 1.00 & $10^{-6}$ & 6.0326 & 0.9573 & 0.6967 & 0.9418 & 0.7750 \\
 & 1.35 & $10^{-6}$ & 3.5981 & 0.9566 & 0.6926 & 0.9435 & 0.7753 \\
\midrule
Heart & 0.55 & $10^{-6}$ & 124.8950 & 0.7935 & 0.8362 & 0.8813 & 0.8829 \\
 & 0.75 & $10^{-6}$ & 78.1506 & 0.7609 & 0.8167 & 0.8840 & 0.8907 \\
 & 1.00 & $10^{-6}$ & 51.7237 & 0.7391 & 0.8065 & 0.8850 & 0.8940 \\
 & 1.35 & $10^{-6}$ & 34.2837 & 0.6576 & 0.7605 & 0.8853 & 0.8949 \\
\bottomrule
\end{tabular}
\end{table*}

\begin{figure}[!tb]
\centering
\includegraphics[width=\linewidth,keepaspectratio]{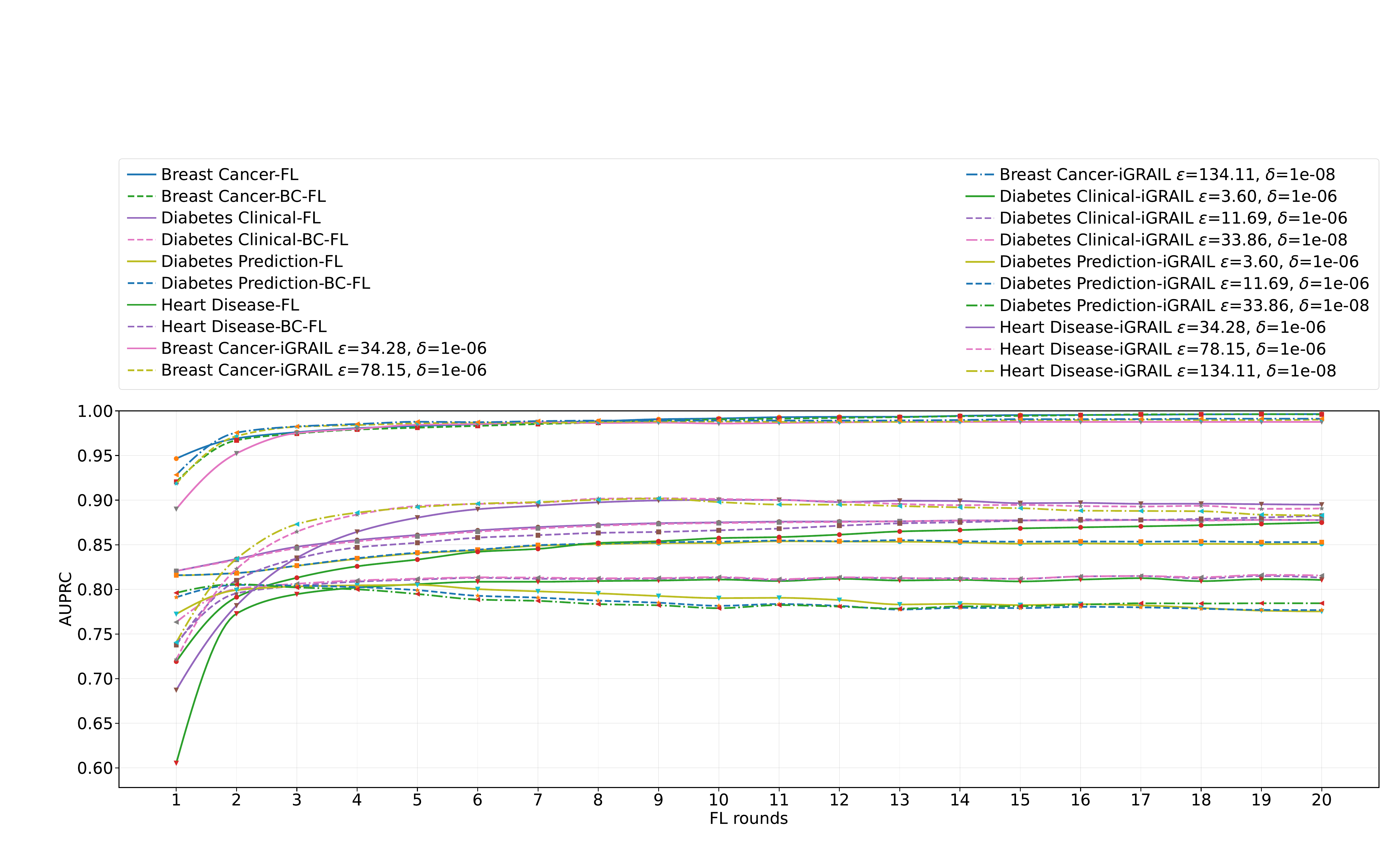}
\caption{Round-wise AUPRC under differential privacy.}
\label{fig:fig4_5}
\end{figure}

\subsubsection{Analysis of AUPRC}
Figure~\ref{fig:fig4_5} shows the AUPRC results for the evaluated privacy configurations. AUPRC measures the trade-off between precision and recall across different classification thresholds. It is especially helpful when there is an imbalance between the positive and negative classes. The AUPRC analysis offers another view of the privacy-utility trade-off because accuracy (AUROC) may not capture changes in model predictions in the same way. Privacy noise can affect the ordering of predicted scores, as well as the false-positive/false-negative relationship of predictions. These changes can affect the precision-recall curve even when overall accuracy is fairly stable. The results indicate that the impact of enhanced privacy varies among datasets and evaluation metrics. Therefore, it is important to consider AUPRC along with AUROC, especially in healthcare applications with fewer positive instances. Maintaining AUROC with stronger privacy does not necessarily mean precision-recall performance remains the same. Thus, the AUPRC results help to get a more complete picture of the impact of differential privacy on the practical usefulness of the learned classifier. The selection of the privacy configuration should be based on a reasonable trade-off between positive-case detection and false-positive control, rather than solely on overall discrimination performance.

\begin{figure}[!tb]
\centering
\includegraphics[width=\linewidth,keepaspectratio]{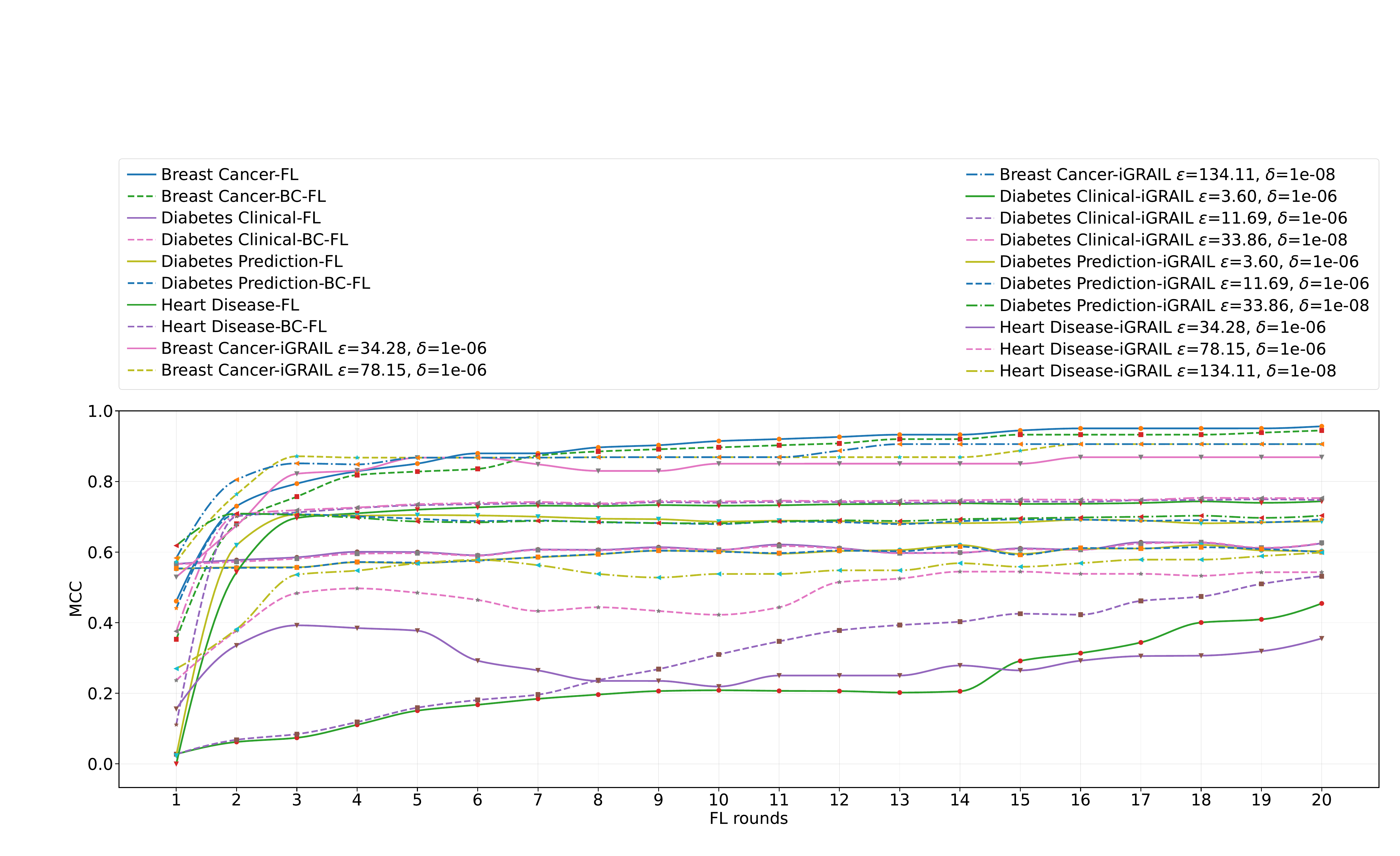}
\caption{Round-wise MCC under differential privacy.}
\label{fig:fig4_6}
\end{figure}

\subsubsection{MCC Analysis}
The MCC results are presented in Figure~\ref{fig:fig4_6} against the communication rounds and privacy settings. MCC takes into account all four elements of the confusion matrix: true positives, true negatives, false positives, and false negatives. It provides another measure of the quality of classification when class distributions are not equal. The importance of MCC analysis lies in the fact that the level of DP noise can impact the balance between positive and negative predictions. Some of these changes might not be well reflected in accuracy, especially when one class dominates the dataset. The results indicate that the effect of privacy-preserving training is not uniform across healthcare datasets. The effect of DP is based on our observations, depending on the underlying classification task and the model's sensitivity to perturbations. Thus, MCC is a complementary metric to F1-score and AUROC testing if the learned predictions are still correlated to the true class labels in stronger privacy settings. Finally, the MCC and other predictive metrics offer a more comprehensive way to measure the reliability of classification. A privacy setting should not be thought of as better simply because it results in a smaller $\epsilon$ value. The effect on the predictive quality should also be evaluated using metrics suitable for the target healthcare application.

\subsubsection{Privacy Budget vs. Utility Trade-off}
The numerical privacy results also illustrate the relationship between noise multiplier and achieved privacy guarantees. Increasing $\sigma$ from $0.55$ to $1.35$ reduces $\epsilon$ from $124.8950$ to $34.2837$ at $\delta=10^{-6}$ for the cancer data set. However, this reduction changes the AUROC by only $0.0023$. This means that the stronger privacy configuration evaluated has almost the same discrimination performance. The privacy budget decreases from $28.7457$ to $3.5981$ for diabetes clinical and diabetes prediction datasets in the reported configurations. The F1 scores are still above $0.69$ which demonstrates that it is possible to get stronger privacy while preserving useful classification performance. We find these results encouraging since they indicate that a smaller privacy-loss parameter does not imply a large decrease in forecasting utility.

In contrast, the F1-score drops more significantly for the Heart Disease dataset, whereas the AUROC stays relatively constant. This difference indicates that privacy noise can have different effects on classification threshold behavior and ranking performance. It also suggests that the same privacy setting may not cover the same utility features across different healthcare datasets. The results should also be interpreted considering the privacy parameter $\delta$. For a fixed noise multiplier and training configuration, the reported $\epsilon$ depends on the chosen $\delta$ and the accounting method. Privacy budgets should be compared under the same accounting assumption. Fixing $\delta$ and the other privacy-accounting conditions, a smaller $\epsilon$ corresponds to a stronger formal privacy guarantee.

\subsubsection{Discussion}
The privacy-utility evaluation yields three important results. First, iGRAIL achieves meaningful predictive performance for the evaluated DP configurations. For the cancer dataset, we observe that the AUROC values remain high even when the privacy-loss parameter is significantly reduced. The two diabetes datasets also have F1-scores greater than $0.69$ under the reported privacy settings. These observations suggest that privacy-preserving training can be integrated without necessarily leading to a severe drop in forecasting utility. Second, the effect of stronger privacy is dataset- and metric-dependent. The AUROC drop on the cancer dataset is small, whereas the heart disease dataset suffers a larger drop in F1 score. Moreover, stable AUROC does not mean recall, F1-score, or MCC will not change. This distinction is particularly crucial in healthcare applications, where different prediction errors have different practical implications. Third, the RDP accountant gives a quantitative basis for comparing the privacy guarantees of different noise multipliers. But the privacy budget should be considered alongside the corresponding $\delta$, the training configuration, and predictive performance. While the privacy analysis determines the formal protection offered by the assessed DP mechanism under its accounting premises, the separate security experiments investigate model-update integrity, authorization, freshness, and robustness against malicious behavior.

\subsection{Evaluation Under Non-IID Data}
\label{subsec:noniid}
\begin{figure}[!h]
\centering
\includegraphics[width=\linewidth,keepaspectratio]{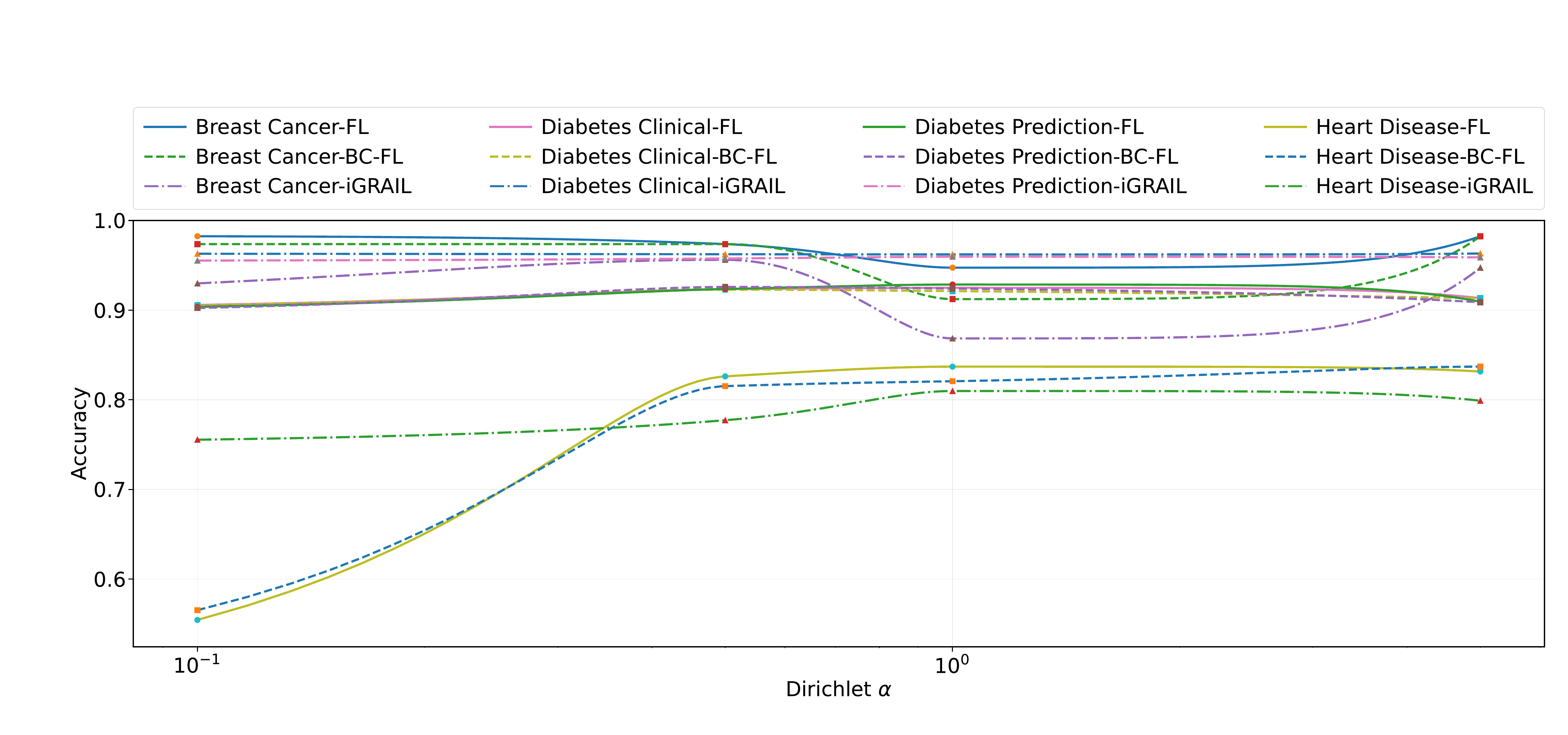}
\caption{Accuracy under different non-IID client data distributions (Dirichlet $\alpha$ values).}
\label{fig:fig5_1}
\end{figure}

\begin{figure}[!h]
\centering
\includegraphics[width=\linewidth,keepaspectratio]{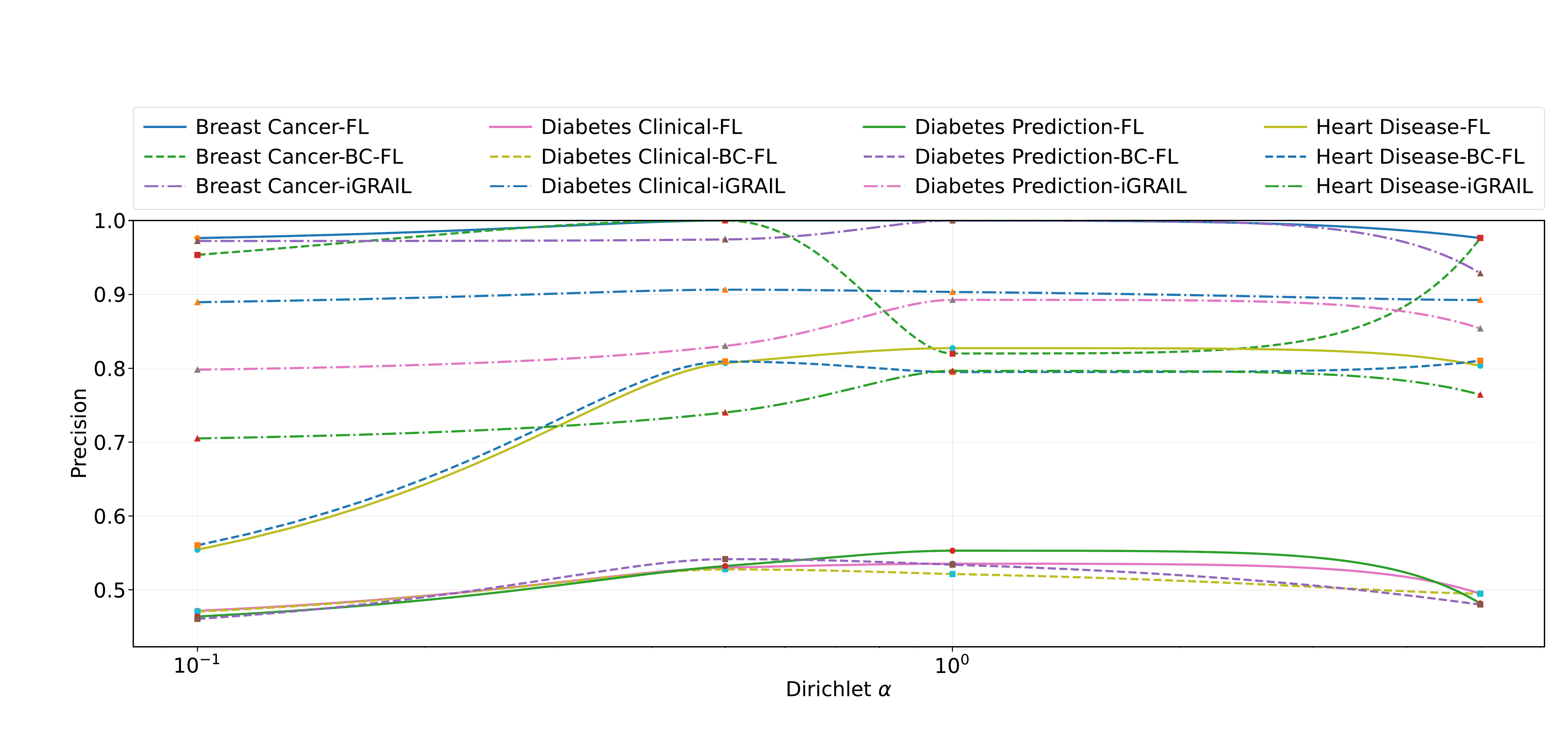}
\caption{Precision under different non-IID client data distributions (Dirichlet $\alpha$ values).}
\label{fig:fig5_2}
\end{figure}

The non-IID experiment assesses iGRAIL's performance when the local data distributions of participating federated clients differ. In real-world healthcare settings, hospitals and medical institutions may collect data from different patient populations, disease states, and clinical environments. Therefore, local datasets can have different class distributions, feature characteristics, and sample sizes. Such differences may affect local model training and lead to inconsistent updates during federated aggregation. Therefore, it is important to evaluate learning performance on heterogeneous client data to determine the applicability of iGRAIL to distributed healthcare applications. We simulate the non-IID condition by using Dirichlet-based data partitioning with concentration parameters $\alpha\in\{0.1,0.5,1.0,5.0\}$. Typically, a smaller $\alpha$ leads to more heterogeneous class distributions across the participating clients, whereas a larger $\alpha$ results in relatively balanced distributions. Specifically, $\alpha=0.1$ represents a very heterogeneous setting, whereas $\alpha=5.0$ represents a relatively less heterogeneous setting. The experiments compare the conventional FL, BC-FL, and the proposed iGRAIL framework under these partitioning configurations. We present the results for accuracy, precision, recall, specificity, F1-score, AUROC, AUPRC, and MCC in Figures~\ref{fig:fig5_1}-\ref{fig:fig5_8}, respectively. These metrics are believed to assess the effects of client data heterogeneity on overall classification performance, positive-case identification, false-positive control, class discrimination, and the reliability of model predictions.

\subsubsection{Accuracy Analysis}
The accuracy of FL, BC-FL, and iGRAIL with different Dirichlet concentration parameters is shown in Figure~\ref{fig:fig5_1}. The results show that the changes in the heterogeneity of the clients’ data affect the methods differently across the evaluated healthcare datasets. In the non-IID settings evaluated for the cancer dataset, the accuracy values are relatively high. This implies that the models are able to preserve useful global classification performance despite changes in local class distributions. However, the differences among the three methods indicate that integrating privacy-preserving generative learning with blockchain-based verification does not always yield the highest accuracy across all partitioning configurations. For the Diabetes Clinical and Diabetes Prediction datasets, iGRAIL achieves competitive accuracy for all concentration parameters considered. The relatively stable behavior suggests that predictive performance is not very sensitive to variations in the client class distributions tested. This is consistent with the stable F1-scores reported for these datasets. The variation in accuracy becomes more pronounced in the Heart Disease dataset, especially as $\alpha$ increases from $0.1$ to $1.0$. It implies that the relationship between client data heterogeneity and classification accuracy is more complex for this dataset. The more balanced distribution of local classes does not necessarily correspond to a proportional increase in prediction performance. The results of overall accuracy show that iGRAIL is still able to work under heterogeneous data conditions. However, the effect of changing $\alpha$ depends on the dataset and learning method.

\subsubsection{Precision Analysis}
The precision of FL, BC-FL, and iGRAIL under the evaluated non-IID settings is compared in Figure~\ref{fig:fig5_2}. Precision is the proportion of predicted positive samples that are correctly predicted positive. Thus, it is useful to understand how changes in local class distribution affect false-positive predictions. For the cancer dataset, the precision results indicate that the methods keep good positive prediction reliability under the tested conditions. However, the differences between the methods reveal that the impact of data heterogeneity differs between conventional and privacy-preserving federated learning. The precision values for the two diabetes datasets provide further evidence of iGRAIL's F1-score stability. The presented framework keeps useful positive-class prediction performance under variation of the Dirichlet concentration parameter. It implies that its classification behavior is relatively stable across the client distributions tested. The Heart Disease dataset is more sensitive to the changes in the non-IID setting. Precision variation indicates that changes in local class distributions can affect the proportions of false-positive predictions.

\begin{figure}[!h]
\centering
\includegraphics[width=\linewidth,keepaspectratio]{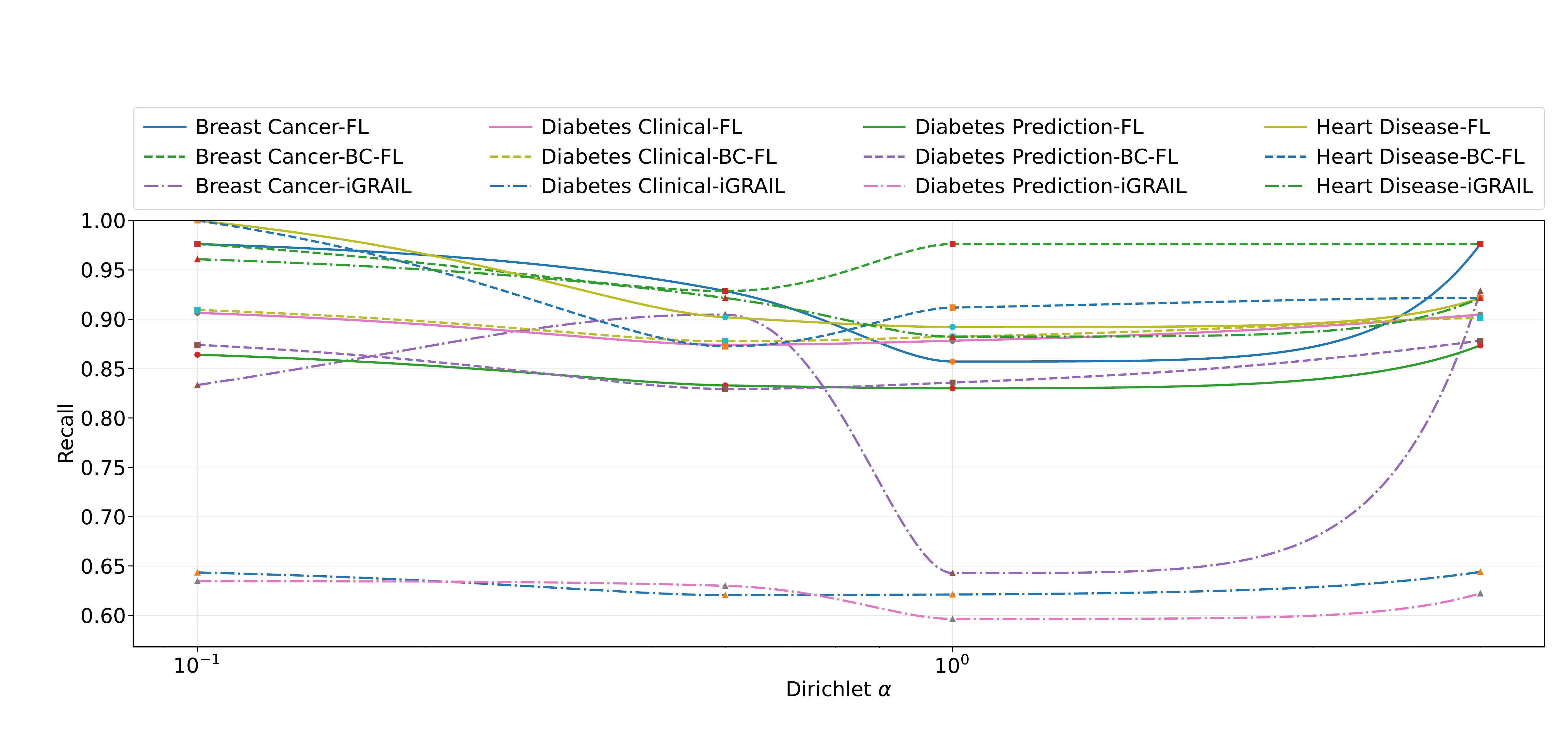}
\caption{Recall under different non-IID client data distributions (Dirichlet $\alpha$ values).}
\label{fig:fig5_3}
\end{figure}

\subsubsection{Recall Analysis}
Figure~\ref{fig:fig5_3} shows the recall obtained under different non-IID settings. Recall is the fraction of true positives predicted correctly by the classifier. In healthcare applications, recall is especially important, as a false negative prediction can have serious consequences. Results indicate that the effect of client data heterogeneity on recall depends on the dataset and the method. Under highly heterogeneous conditions, some participating clients may receive fewer positive-class samples, which may influence their local learning behavior. This might affect the global model’s ability to detect positive cases after aggregation. In the case of the two diabetes datasets, recall scores must be viewed in conjunction with the corresponding precision and F1-score. The F1-scores of iGRAIL are relatively stable, but this does not mean that the recall is the same for all concentration parameters. The changes in precision and recall may offset each other, so that the F1 scores remain similar. The recall analysis for the cancer and heart disease datasets provides this additional insight into the sensitivity of the learned models to heterogeneous client data. The results show that high accuracy or AUROC performance is not always indicative of reliable detection of positive cases. The recall scores thus indicate an important dimension of health care delivery. You can have a model with aggregate performance indicators that is stable under non-IID data but still need further false-negative assessment.

\begin{figure}[!h]
\centering
\includegraphics[width=\linewidth,keepaspectratio]{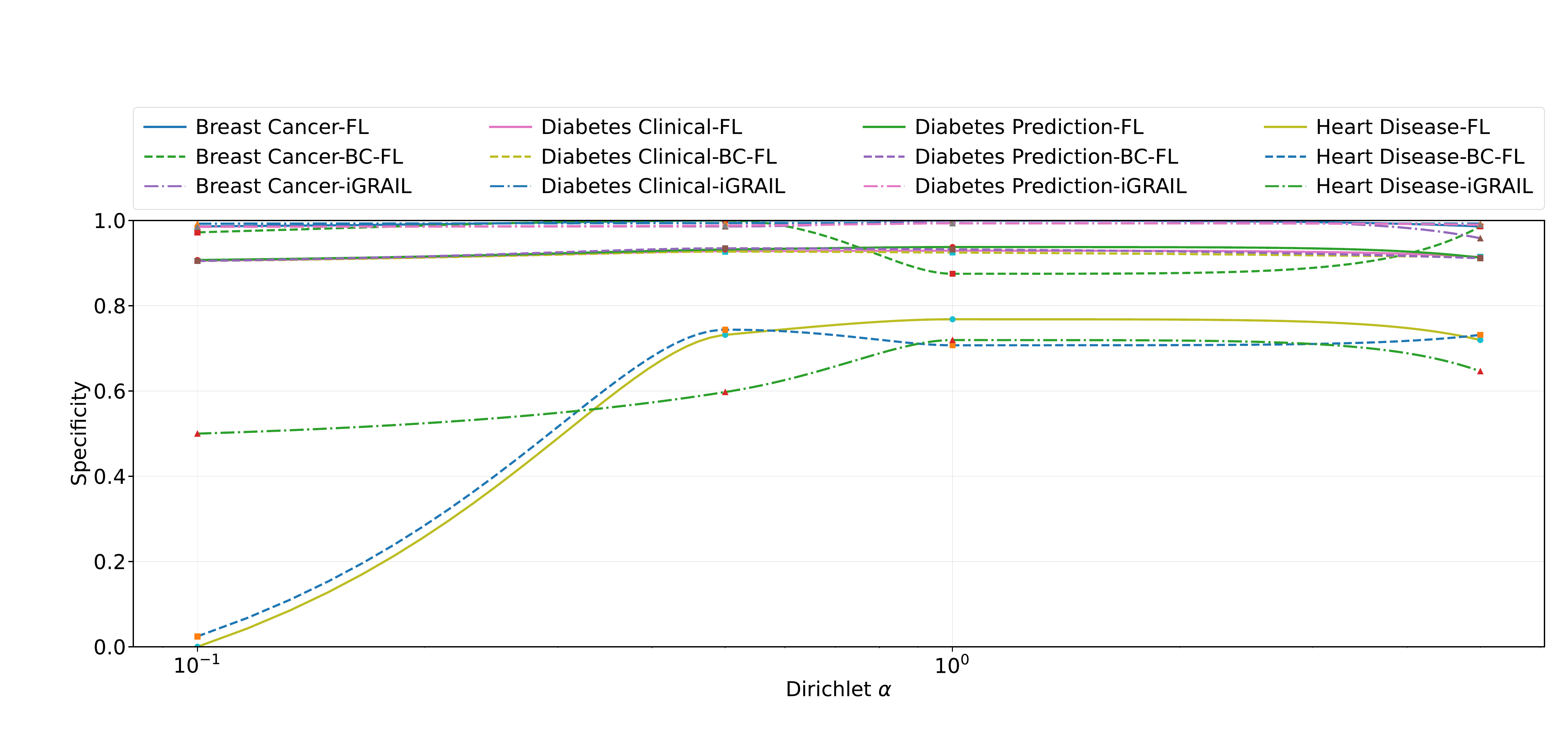}
\caption{Specificity under different non-IID client data distributions (Dirichlet $\alpha$ values).}
\label{fig:fig5_4}
\end{figure}

\subsubsection{Specificity Analysis}
In Figure~\ref{fig:fig5_4} we show the specificity of the three methods for different concentration parameters of the Dirichlet. Specificity is the proportion of true negatives that are correctly identified and is a measure of how well the test controls for false positives. The results indicate that the impact of non-IID data on specificity varies across the datasets under consideration. For the Heart Disease dataset, there are significant changes in specificity as $\alpha$ varies from $0.1$ to $1.0$. This shows that shifts in client class distributions affect the models' ability to correctly identify negative samples. The specificity results for the cancer and diabetes datasets provide complementary information on the reliability of the negative-class predictions. Differences in FL, BC-FL, and iGRAIL indicate that the methods react differently to changes in the proportions of the local classes. These observations are important because specificity and recall are two different aspects of classification behavior. A method can increase recall by labeling more samples as positive. However, specificity might decrease if there are more false positives. In contrast, lower recall may be at the expense of higher specificity. So, the joint analysis of recall and specificity helps better understand how heterogeneity in client data affects the trade-off between missing positive cases and false alarms.

\begin{figure}[!h]
\centering
\includegraphics[width=\linewidth,keepaspectratio]{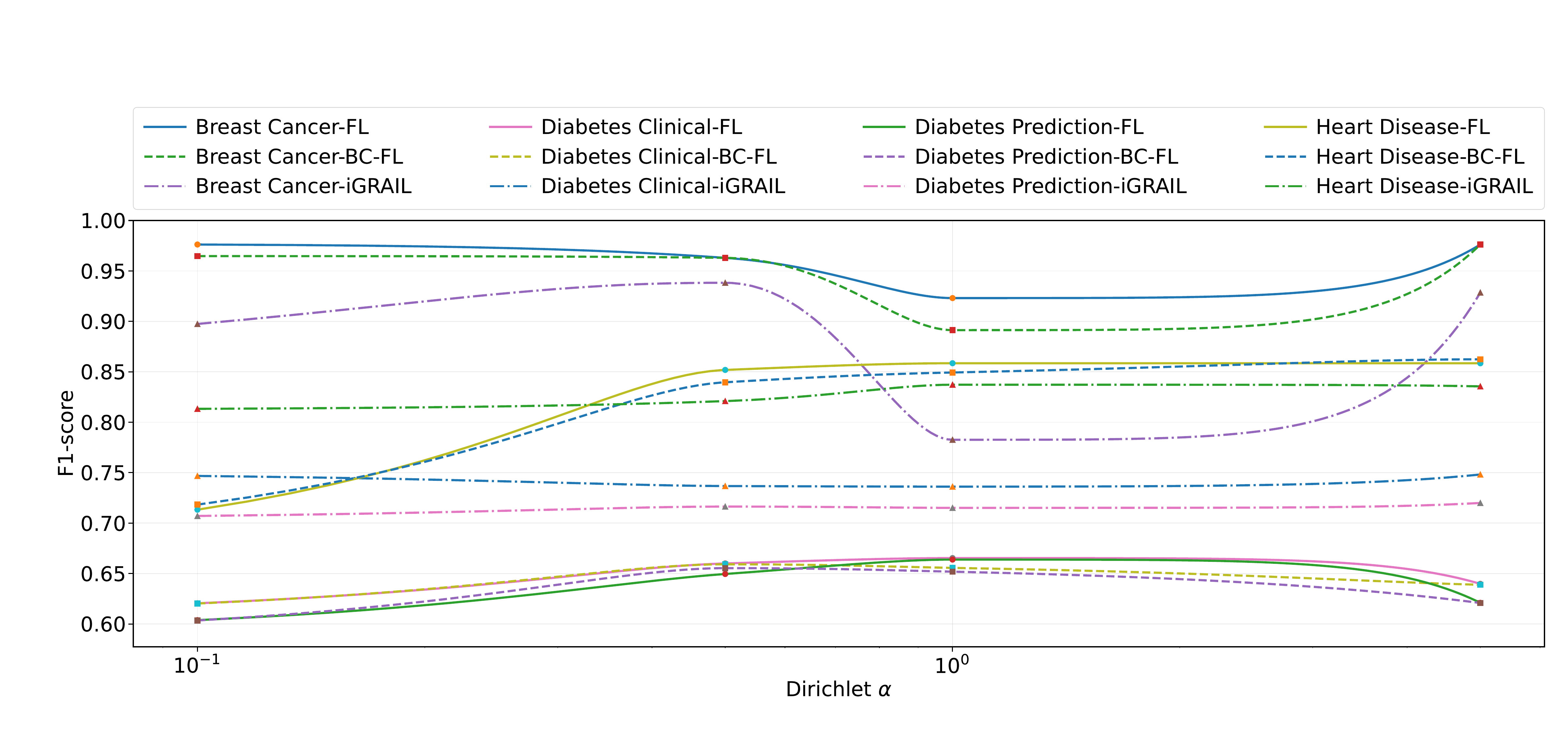}
\caption{F1-score under different non-IID client data distributions (Dirichlet $\alpha$ values).}
\label{fig:fig5_5}
\end{figure}

\subsubsection{F1 Score Analysis}
The results of the F1 score for all non-IID settings tested are shown in Figure~\ref{fig:fig5_5}. The F1-score is the harmonic mean of precision and recall and is especially useful when evaluating positive-class prediction when class distributions are unequal. iGRAIL achieves an F1 score between $0.7361$ and $0.7482$ across the four concentration parameters on the Diabetes Clinical dataset. Only the absolute difference between the maximum and minimum values is $0.0121$. This limited variation suggests that the proposed framework maintains a relatively stable precision-recall trade-off as the local client distributions change. We observe a similar trend for the Diabetes Prediction dataset, where iGRAIL achieves F1-scores ranging from $0.7071$ to $0.7200$. The corresponding deviation is $0.0129$. The result indicates that the proposed framework achieves similar F1-score performance on a highly heterogeneous and relatively balanced data partition. Another pattern is seen in the cancer dataset. The iGRAIL F1 score decreases significantly at an intermediate Dirichlet concentration parameter, particularly around $\alpha=1.0$ and rises again at a higher value. This non-monotonic behavior implies that the reduction of the client data heterogeneity does not always result in the consistent improvement of the classification performance. The constructed class partitions and their effects on the training of local models can explain the observed variation. The F1-score performance in the Heart Disease dataset is more sensitive to the particular partition setup being evaluated. Thus, for this dataset, the precision-recall trade-off of a model is more sensitive to changes in local class distributions. Overall, the F1-score results show relatively stable performance on the two diabetes datasets, while the results on cancer and heart disease are more sensitive to the heterogeneity of client data. The results suggest that the predictive advantage of iGRAIL may not be universally applicable across all healthcare settings, but rather dependent on the dataset used.

\begin{figure}[!h]
\centering
\includegraphics[width=\linewidth,keepaspectratio]{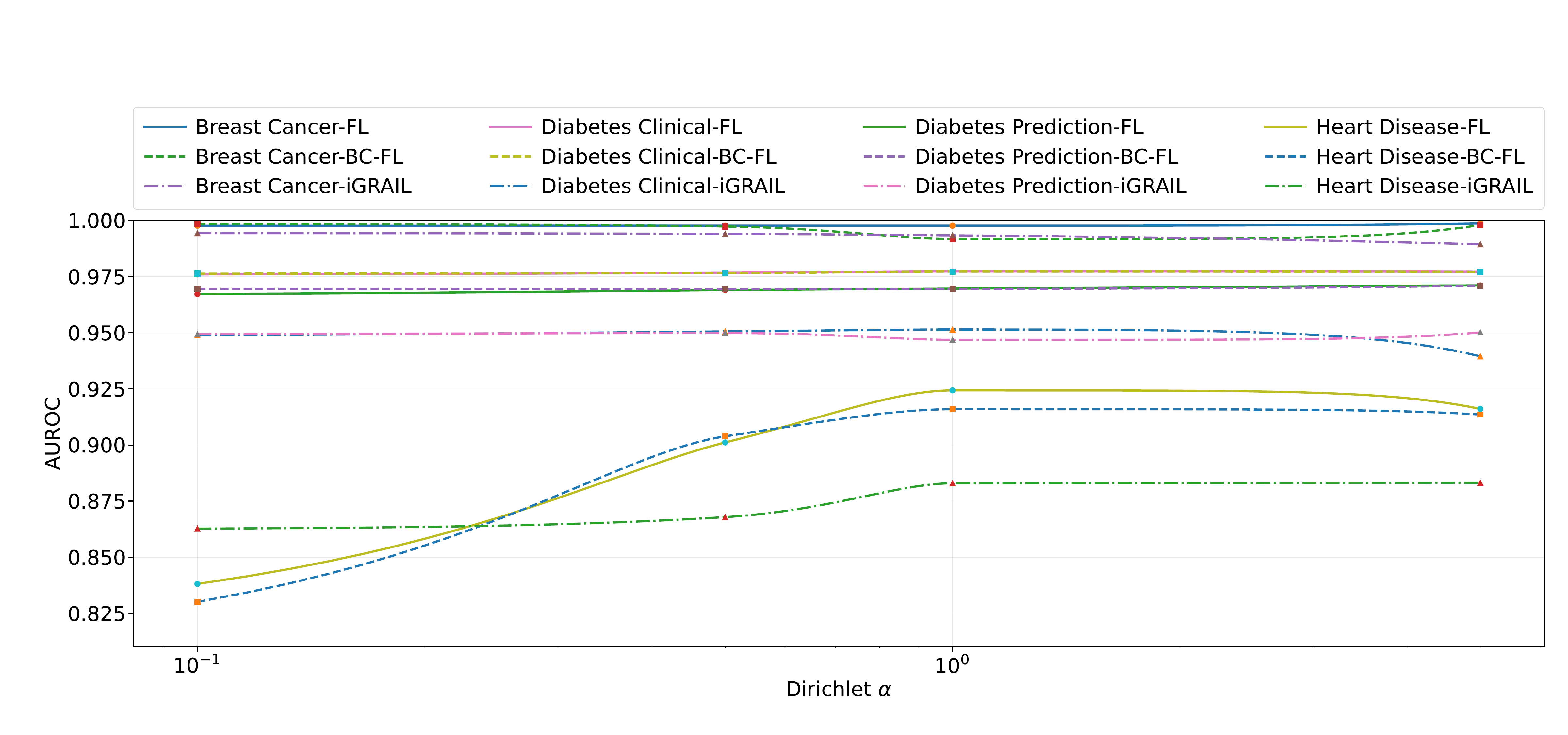}
\caption{AUROC under different non-IID client data distributions (Dirichlet $\alpha$ values).}
\label{fig:fig5_6}
\end{figure}

\subsubsection{AUROC Analysis}
Figure~\ref{fig:fig5_6} shows the AUROC results in different non-IID settings. AUROC is a classifier’s ability to discriminate between positive and negative samples with different classification thresholds. iGRAIL obtains AUROC values between $0.9894$ and $0.9944$ for the concentration parameters tested on the Cancer dataset. The absolute variation is only $0.0050$, showing that the model maintains strong discriminative performance even with changes in the distribution of client data. This result is important because the cancer F1 score shows a clear variation with the intermediate concentration parameter. The stable AUROC indicates that the model has good overall ranking ability, even when the classification performance changes at the chosen operating threshold. Thus, a change in the F1 score does not necessarily correspond to a correlated decrease in threshold-independent discrimination. In the Heart Disease dataset, iGRAIL achieves AUROC values in the range $0.8627$ to $0.8832$. The change is $0.0205$, which is larger than the change for the cancer dataset. However, the AUROC values are still above $0.86$ for all the tested concentration parameters, showing that the model maintains useful discrimination ability under the non-IID conditions tested. The AUROC results on the Diabetes Clinical and Diabetes Prediction datasets corroborate the effect of client data heterogeneity on class discrimination. These results are complementary to the F1-score analysis as they consider predictive behavior over multiple classification thresholds instead of a single threshold. To summarize, the AUROC results show that iGRAIL has good discrimination for cancer and useful discrimination for heart disease. However, the degree of performance difference is dataset-dependent, which means that the effect of non-IID data varies for classification problems.

\begin{figure}[!h]
\centering
\includegraphics[width=\linewidth,keepaspectratio]{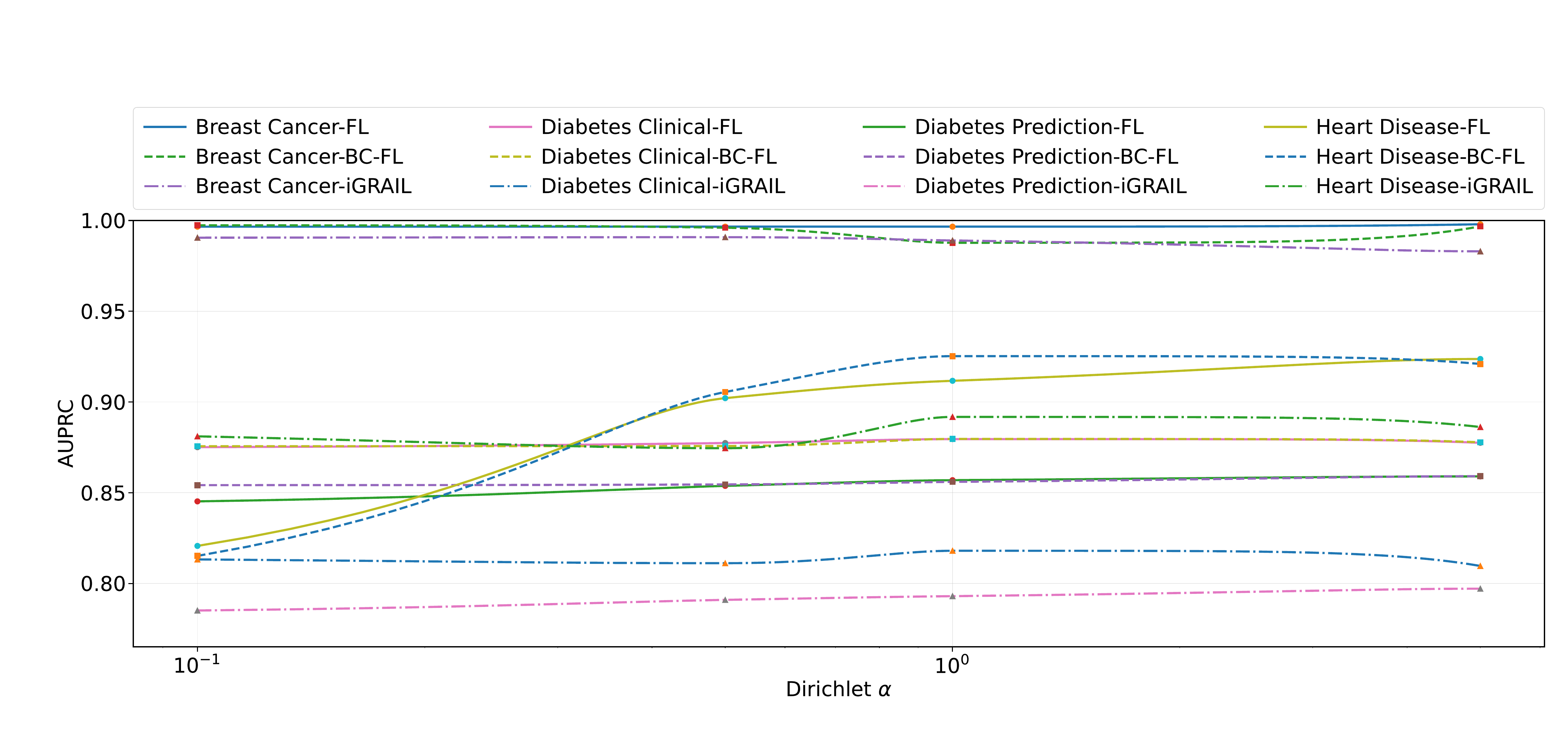}
\caption{AUPRC under different non-IID client data distributions (Dirichlet $\alpha$ values).}
\label{fig:fig5_7}
\end{figure}

\subsubsection{AUPRC Analysis}
The AUPRC results for various Dirichlet concentration parameters are shown in Figure~\ref{fig:fig5_7}. AUPRC is a measure of the tradeoff between precision and recall at various classification thresholds. It is particularly useful when the positive and negative classes are imbalanced. The results provide an additional view of the impact of client data heterogeneity, as changes in class distributions might affect positive-class ranking performance differently than overall accuracy or AUROC. The AUPRC analysis for the cancer dataset complements the high AUROC values by assessing the reliability of positive-class predictions at various thresholds. The AUPRC results for the two diabetes datasets provide further insight into whether the relatively stable F1-scores are accompanied by consistent precision-recall behavior. In the case of the Heart Disease dataset, the differences in AUPRC indicate that the trade-off between precision and recall should be investigated in depth across the different partitioning configurations. This is especially true when positive cases are rare or when the cost of missing a detection differs from the cost of a false alarm. The AUPRC results thus highlight the need to use multiple evaluation metrics when evaluating non-IID federated learning. Even if a model's AUROC is stable, its precision-recall performance can change. This is especially true if the class distributions among clients differ.

\begin{figure}[!h]
\centering
\includegraphics[width=\linewidth,keepaspectratio]{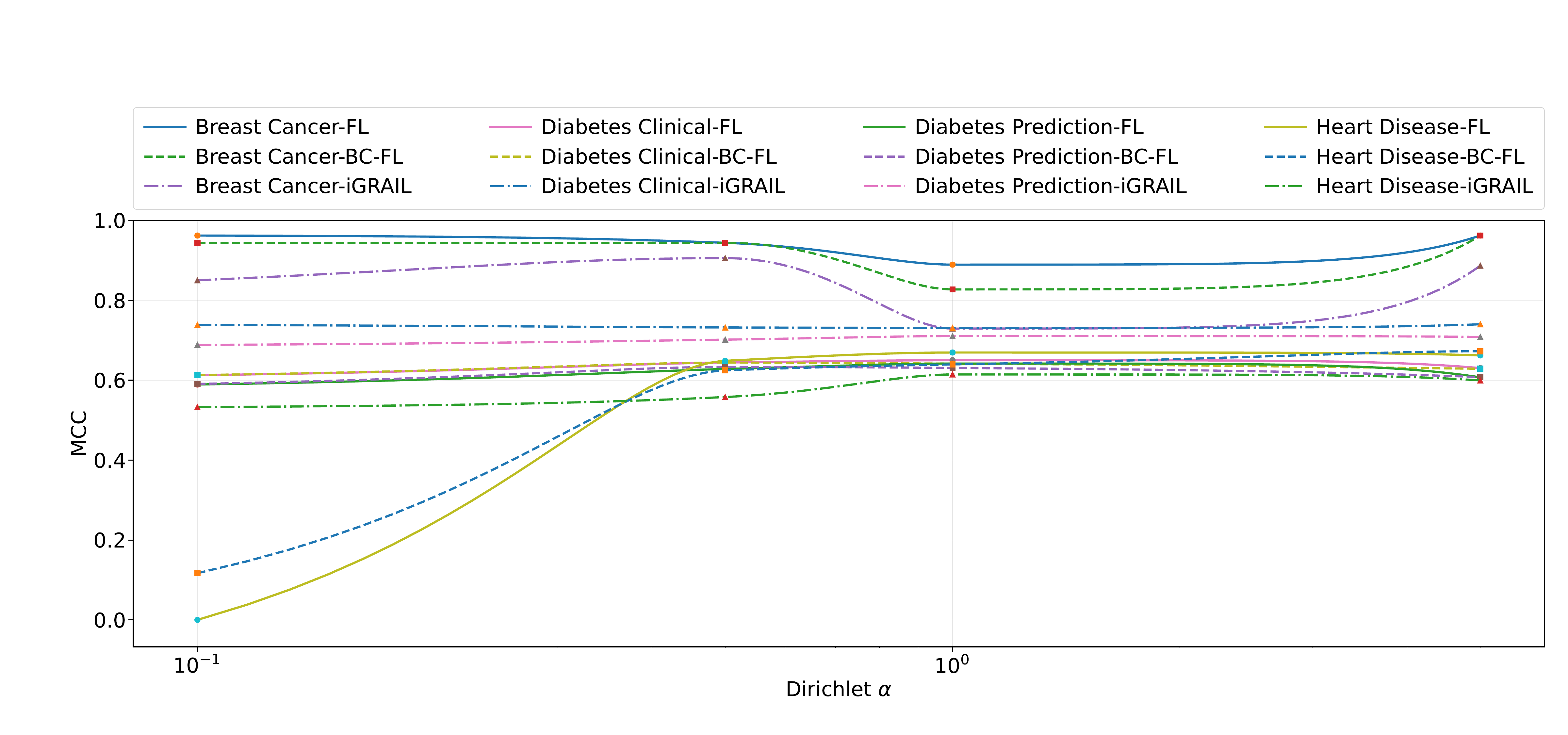}
\caption{MCC under different non-IID client data distributions (Dirichlet $\alpha$ values).}
\label{fig:fig5_8}
\end{figure}

\subsubsection{MCC Analysis}
Figure~\ref{fig:fig5_8} shows the Matthews correlation coefficient (MCC) under the non-IID conditions evaluated. MCC integrates true-positive, true-negative, false-positive, and false-negative predictions into a single measure. It is thus useful to measure the classification quality when the class distributions are unequal. The results show that heterogeneity of client data can affect the global relationship between predicted and actual class labels. This effect may differ from the trends seen in accuracy or F1 score since MCC is sensitive to both positive and negative classification outcomes. The MCC analysis for the two diabetes datasets complements the relatively stable F1-score results by providing an additional metric for classification reliability. For Cancer and Heart Disease, changes in MCC explain why a stable AUROC does not necessarily mean the same classification performance at a fixed threshold. The differences among FL, BC-FL, and iGRAIL show that the effect of heterogeneous client data depends on the learning method and dataset. The MCC results thus provide a broader measure of whether the proposed framework retains useful classification behavior across different non-IID settings.

\subsubsection{Effect of Client Data Heterogeneity}
Overall results indicate that increasing the Dirichlet concentration parameter does not necessarily result in a monotonic improvement in forecasting performance. In general, higher values of $\alpha$ lead to less heterogeneous class distributions. However, the final performance of the global model also depends on the specific client partitions and their influence on federated training. The Cancer dataset is an example of such behavior. iGRAIL achieved consistently high AUROC values, but its F1 score showed a significant drop around $\alpha = 1.0$ before recovering at a higher concentration parameter. It implies that the effect of client data heterogeneity may differ for threshold-dependent and threshold-independent evaluation metrics. Across two diabetes datasets, the changes in F1-score are small, suggesting that iGRAIL maintains a relatively stable precision-recall trade-off across the tested levels of heterogeneity. The Heart Disease dataset shows higher variance in some metrics (e.g., accuracy and specificity), indicating higher sensitivity to changes in local data partitions. The results indicate that $\alpha$ is not the only factor that influences the relationship between data heterogeneity and predictive performance. The characteristics of the dataset, class distributions, and the local training conditions that follow also influence the characteristics of the final model.

\subsubsection{Discussion}
The non-IID evaluation results in three important findings. First, iGRAIL shows relatively stable F1-score performance in the Diabetes Clinical and Diabetes Prediction datasets. The corresponding deviations of 0.0121 and 0.0129 suggest that the changes in the heterogeneity of the client data that we assessed have limited impact on the precision-recall balance for these datasets.  Secondly, the AUROC results show that iGRAIL has good performance in class discrimination for cancer and useful performance in class discrimination for heart disease. The AUROC values of the Cancer dataset are in the range $[0.9894, 0.9944]$ and the AUROC values of the Heart Disease dataset are in the range $[0.8627, 0.8832]$. However, the larger variation for heart disease indicates that the sensitivity for heterogeneous data distributions depends on the healthcare task. Third, the joint analysis of all eight metrics reveals that the effects of non-IID data are not uniform. In particular, the results for the cancer F1-score and heart disease accuracy and specificity show that reducing data heterogeneity does not always improve classification performance. Plus, just because one metric is stable doesn't mean the other metrics will be.

\subsection{Scalability with the number of clients} \label{subsec:scalability}

The scalability experiment evaluates the prediction behavior of FL, BC-FL, and iGRAIL as the number of participating federated clients increases from 2 to 10. The number of hospitals and medical institutions participating in distributed healthcare environments may vary over time. Increasing the size of the federation may affect the availability of local data, class distributions, model updates, and the resulting global prediction performance. Thus, the experiment aims to study the ability of the proposed framework to maintain useful predictive performance under the various client configurations.

\begin{figure}[!h]
\centering
\includegraphics[width=\linewidth,keepaspectratio]{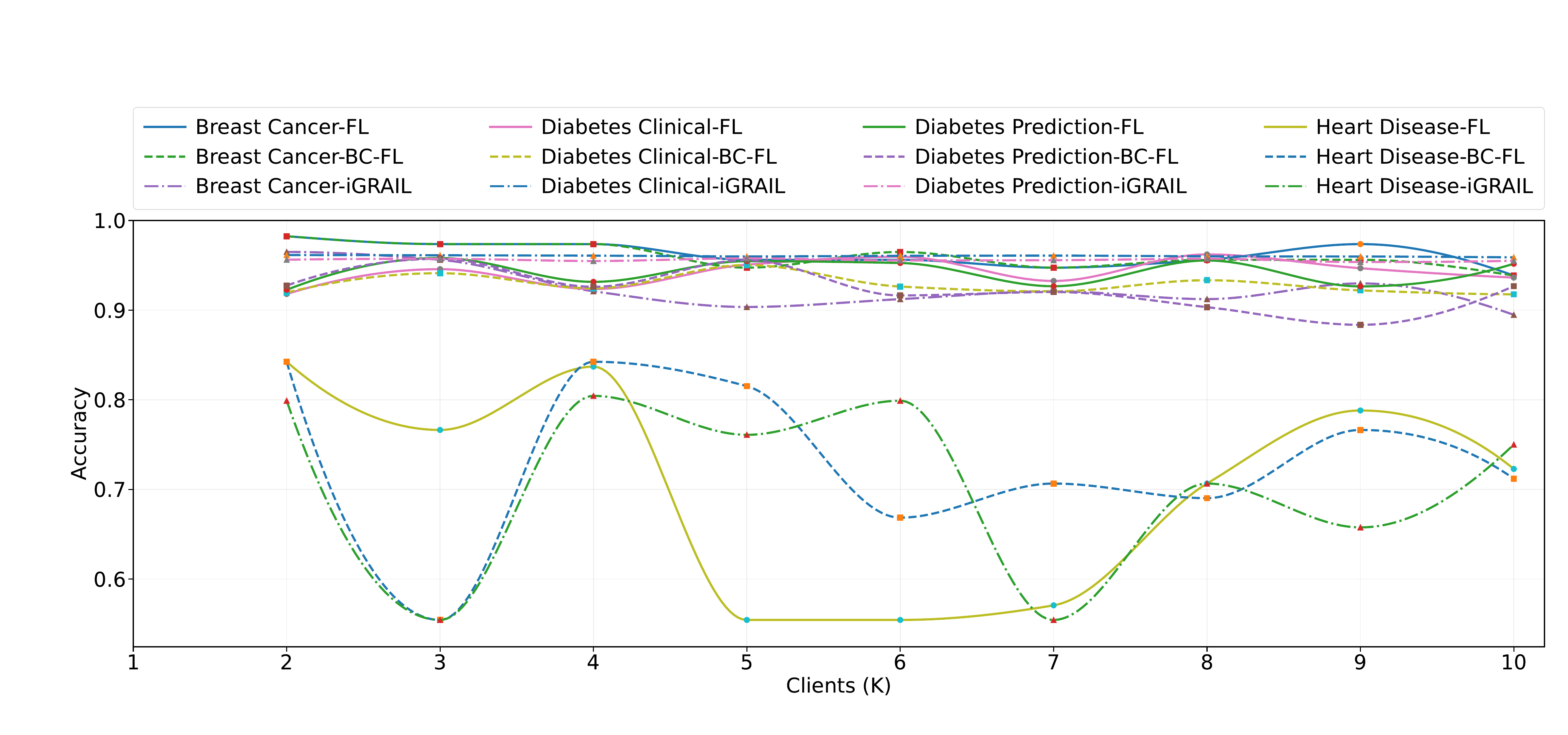}
\caption{Accuracy as the number of federated clients increases.}
\label{fig:fig6_1}
\end{figure}

\subsubsection{Accuracy Analysis}
The classification accuracy as a function of the number of clients is shown in Figure~\ref{fig:fig6_1}. The accuracy of all three methods is consistently high on the cancer dataset, with relatively small oscillations across the evaluated federation sizes. This means that the models still perform well on classification in general for cancer data partitioned across different numbers of clients. The accuracy curves of the Diabetes Clinical and Diabetes Prediction datasets are relatively stable, but there are some differences among FL, BC-FL, and iGRAIL. The results indicate that increasing the number of clients does not lead to a continuous increase or decrease in accuracy. The heart disease dataset is now behaving quite differently. Some models’ accuracy curves are very sensitive to the number of clients. For some configurations, we see significant dips at intermediate client numbers, recovering at larger federation sizes. Such non-monotonic behavior implies that the predictive performance of increasing federation size highly depends on the dataset and the resulting client data partitions.

\begin{figure}[!h]
\centering
\includegraphics[width=\linewidth,keepaspectratio]{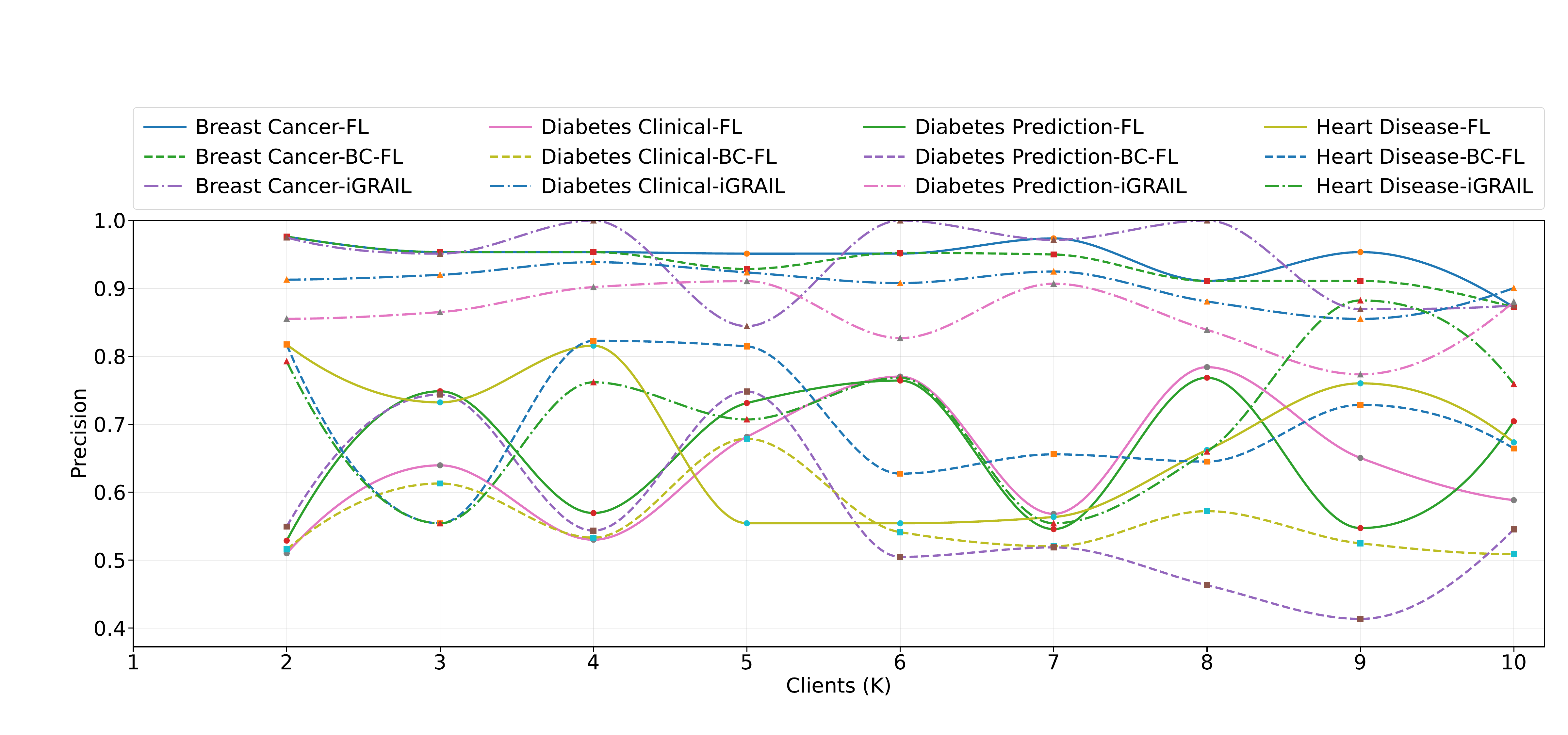}
\caption{Precision as the number of federated clients increases.}
\label{fig:fig6_2}
\end{figure}

\subsubsection{Precision Study}
Figure~\ref{fig:fig6_2} depicts the precision over different federation sizes. The Cancer dataset also has high overall precision, but the differences among the methods are noticeable for some numbers of clients. This means that the reliability of positive predictions for this dataset is still pretty high. The two diabetes datasets show different precisions for the client configurations tested. Some methods get better for a certain mid-range number of clients and then get worse for larger federation sizes. These variations mean that the fraction of correct positive predictions can be modified by distributing the available training data over more clients. The Heart Disease data set also has a high variance in precision. The observed changes show that the relation between true-positive and false-positive predictions is sensitive to the client configuration. Thus, in addition to recall, precision should be taken into account to evaluate whether changes in the detection of positive cases are associated with changes in the reliability of positive predictions.

\begin{figure}[!h]
\centering
\includegraphics[width=\linewidth,keepaspectratio]{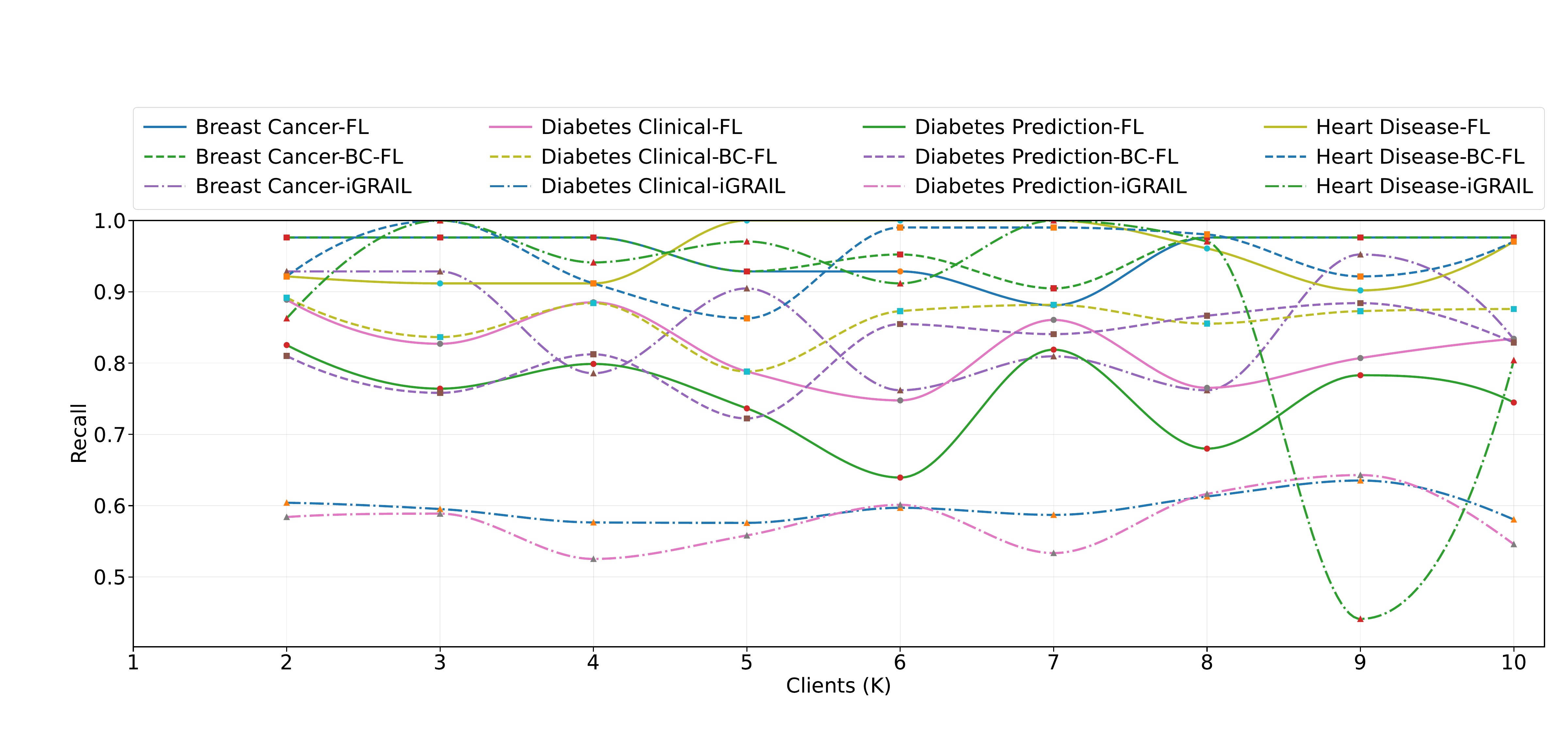}
\caption{Recall as the number of federated clients increases.}
\label{fig:fig6_3}
\end{figure}

\subsubsection{Recall analysis}
The recall for the counts of clients we evaluated is shown in Figure~\ref{fig:fig6_3}. For the cancer dataset, the recall is generally high, indicating that as the federation grows, the models keep detecting a large fraction of actual positive cases. The Diabetes Clinical and Diabetes Prediction datasets show more variation between methods. In particular, the recall curves demonstrate that the ability to retrieve positive cases is not necessarily constant with the number of clients. Some configurations exhibit fairly stable recall. Some show significant changes at intermediate-size federations. The heart disease results show more differences in detecting positive cases. The diversity of client environments could mean that the number of clients involved might affect the ratio of correct to incorrect predictions. These results are relevant for healthcare applications, as high accuracy or AUROC may not translate to stable recall. A model might have good overall discrimination but not be able to find more positive cases at a specific cut point.

\begin{figure}[!h]
\centering
\includegraphics[width=\linewidth,keepaspectratio]{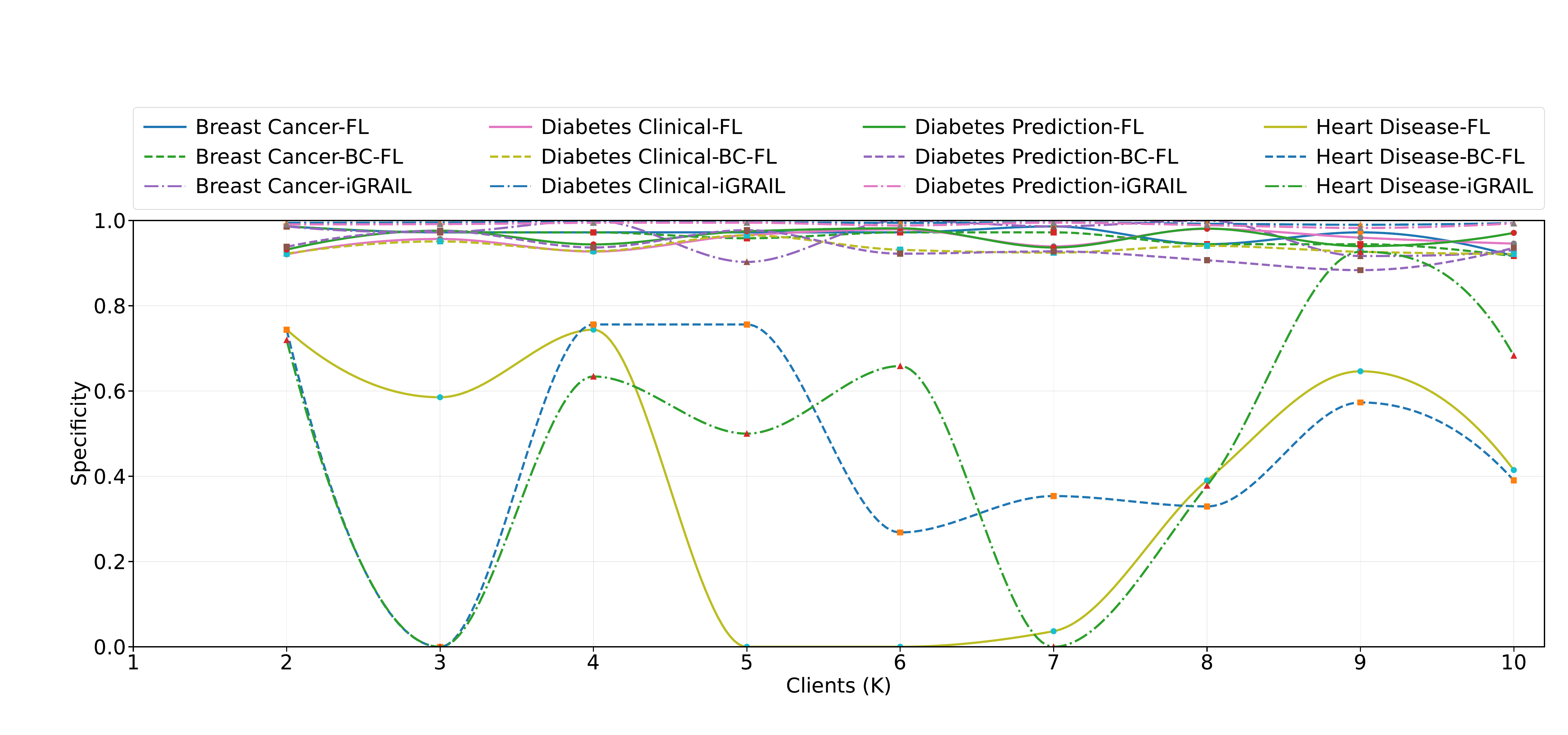}
\caption{Specificity as the number of federated clients increases.}
\label{fig:fig6_4}
\end{figure}

\subsubsection{Specificity analysis}
The specificity for different client configurations is shown in Figure~\ref{fig:fig6_4}. The cancer dataset still has a relatively high specificity for all the methods considered, meaning that the negative cases are correctly identified. In general, for diabetes clinical and diabetes prediction datasets, many configurations achieve high specificity, but differences between methods are obvious. This shows that the learning method and the size of the federation affect the capability of avoiding false positive predictions. The Heart Disease dataset has the largest variance. In some configurations of the method, the specificity values are close to zero for an intermediate number of clients, but they partially recover for larger federations. If the specificity is close to zero, most of the actual negative samples will be wrongly classified as positive for the decision threshold that is chosen. This behavior is particularly important since a model can have a high recall but still make too many false-positive predictions. Results of heart disease show that increasing the number of clients can greatly affect the sensitivity-specificity ratio. The results also demonstrate why predictive scalability cannot be measured solely by accuracy and recall.

\begin{figure}[!h]
\centering
\includegraphics[width=\linewidth,keepaspectratio]{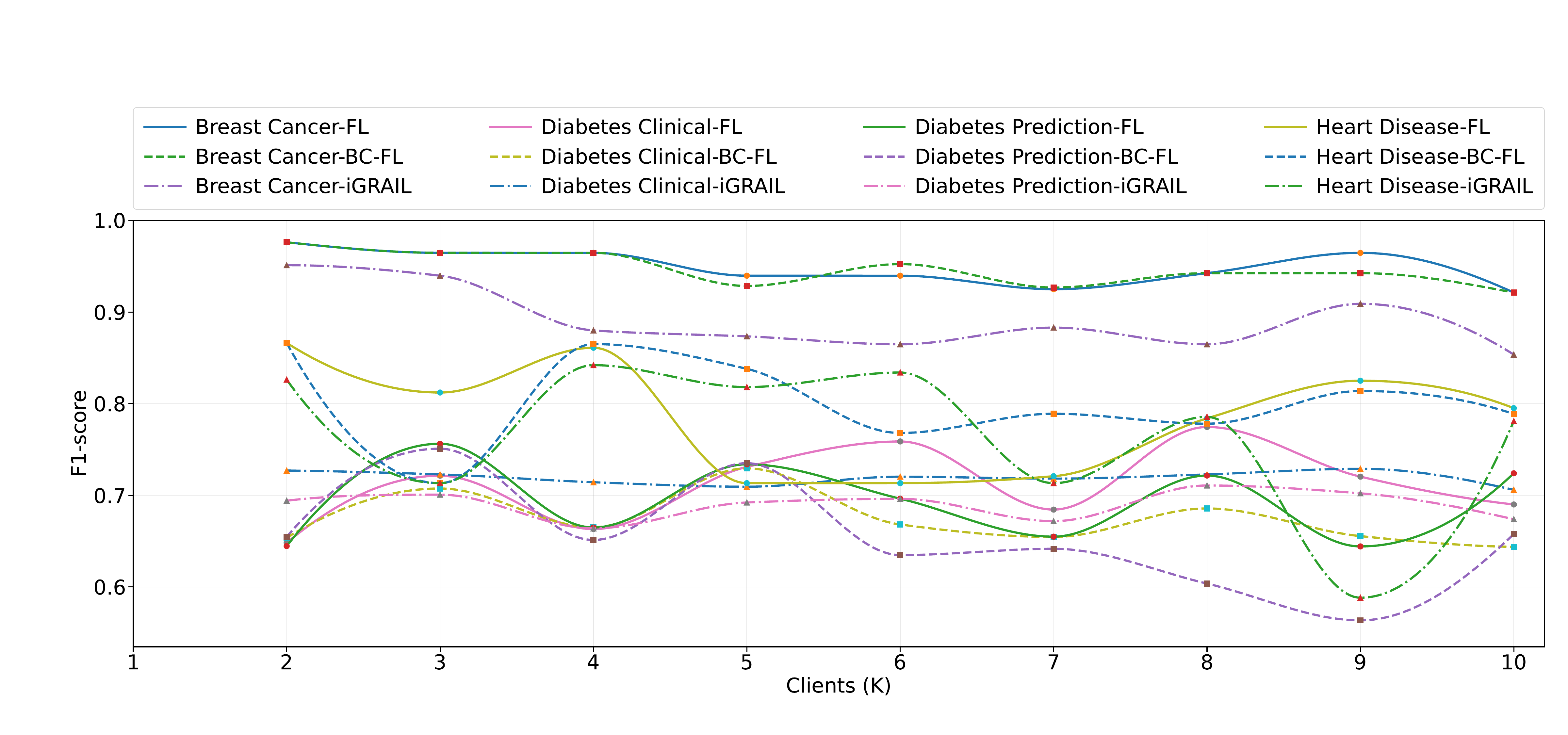}
\caption{F1-score as the number of federated clients increases.}
\label{fig:fig6_5}
\end{figure}

\subsubsection{Analysis of F1-Score}
Figure~\ref{fig:fig6_5} shows the F1-score for the tested federation sizes. For the cancer dataset, the three methods show relatively high F1-scores with small variation compared to the other datasets. This shows that the precision-recall trade-off remains relatively strong as the number of clients increases. The F1-score curves for the Diabetes Clinical and Diabetes Prediction datasets have moderate fluctuations. The curves do not show a clear improvement as federation size increases, but useful classification performance is retained across several configurations. The Heart Disease dataset shows a bigger variation in the F1 score. Some configurations show a performance dip at intermediate client counts, then improve at larger federation sizes. This is consistent with the observed changes in precision, recall, and specificity. Overall, the F1-score results indicate that the relationship between federation size and the performance on predicting the positive class is not monotonic. Increasing the number of clients does not necessarily improve the precision-recall trade-off, even if the resulting model achieves good overall accuracy.

\begin{figure}[!h]
\centering
\includegraphics[width=\linewidth,keepaspectratio]{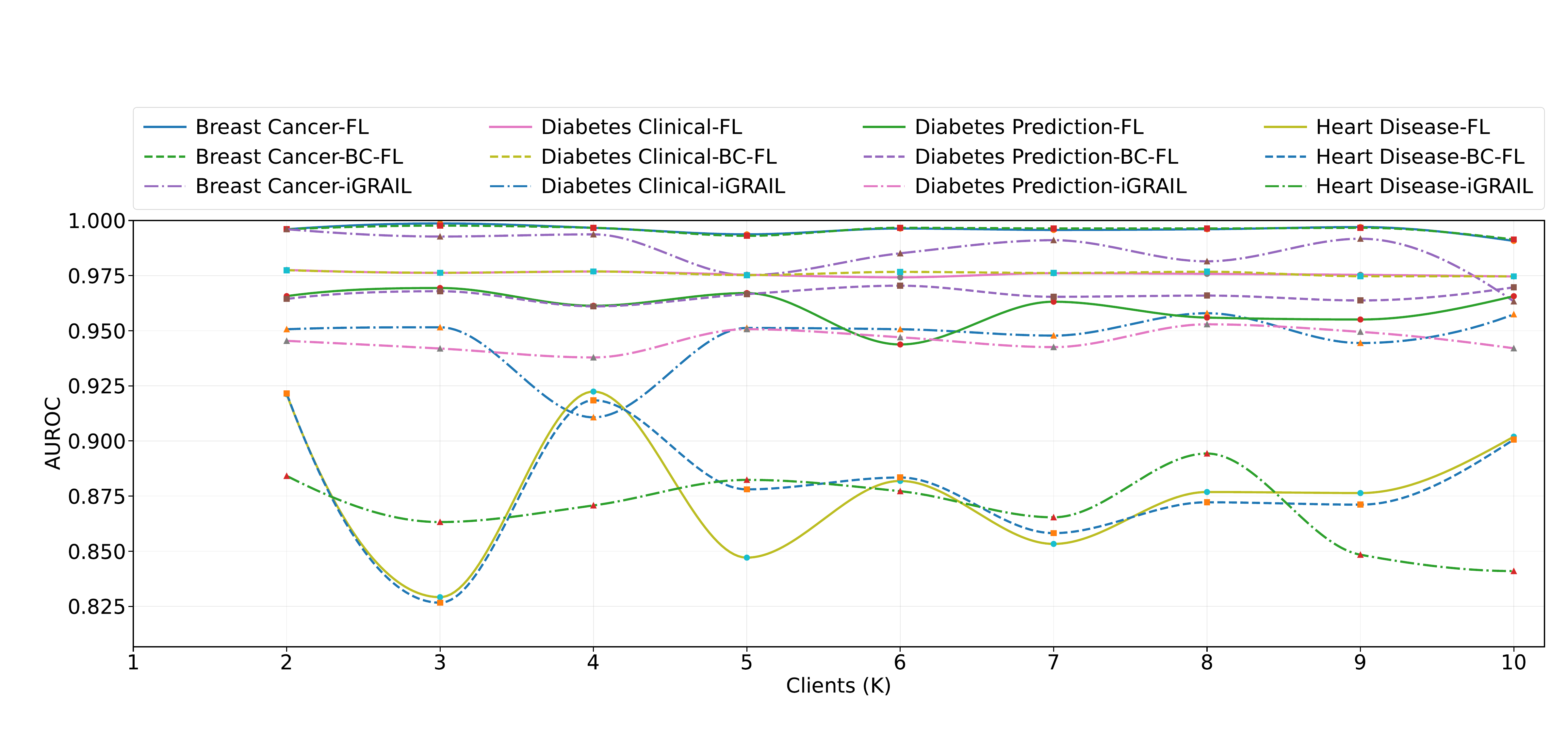}
\caption{AUROC as the number of federated clients increases.}
\label{fig:fig6_6}
\end{figure}

\subsubsection{AUROC Analysis}
Figure~\ref{fig:fig6_6} shows the AUROC as the number of clients involved increases. AUROC measures the model’s discrimination between the positive and negative classes over classification thresholds. For the Diabetes Clinical dataset, iGRAIL achieves an AUROC of $0.9507$ for 2 clients and $0.9574$ for 10 clients. The absolute increase of $0.0067$ indicates that the model still maintains good discriminative ability as the federation grows. The intermediate values are mixed, but the results at the endpoints show that increasing the number of clients does not necessarily hurt AUROC. For the Diabetes Prediction dataset, iGRAIL achieves $\text{AUROC}=0.9454$ with two clients and $\text{AUROC}=0.9420$ with ten clients. The absolute decrease is only $0.0034$, so the discrimination performance between the two endpoint configurations is relatively stable. The AUROC does not fall below $0.94$ at either endpoint.

The cancer data shows a more drastic reduction. Its AUROC decreases from 0.9960 with two clients to 0.9633 with ten clients, an absolute reduction of 0.0327. The decrease indicates that the model is more sensitive to increases in federation size than it is for the two diabetes datasets, while still performing with good discriminative power. The AUROC curves for the Heart Disease dataset, in contrast, vary more across the evaluated client configurations. The results indicate that the models' discriminative capability depends on the federation size specific to each case and is not uniformly improved by increasing the number of participants. Overall, the AUROC results suggest that iGRAIL retains useful discrimination performance across the client counts studied. But the level of stability varies between datasets. The results also show that a larger federation size does not guarantee a better discrimination performance.

\begin{figure}[!h]
\centering
\includegraphics[width=\linewidth,keepaspectratio]{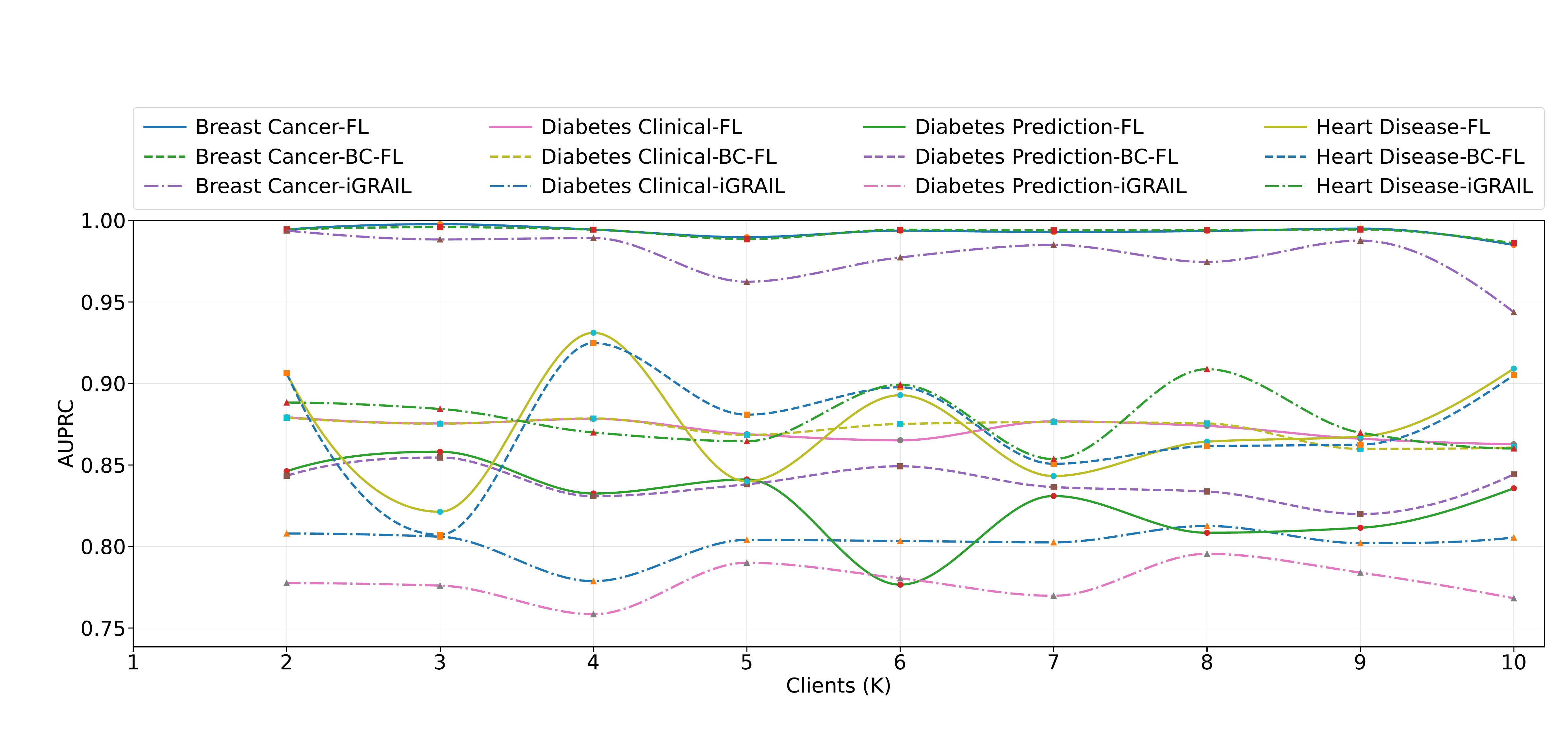}
\caption{AUPRC as the number of federated clients increases.}
\label{fig:fig6_7}
\end{figure}

\subsubsection{AUPRC Analysis}
Figure~\ref{fig:fig6_7} shows AUPRC with different federation sizes. Although differences among the three methods are evident, the Cancer dataset still maintains high AUPRC values across the evaluated configurations. The AUPRC curves for the Diabetes Clinical and Diabetes Prediction datasets differ in both the number of clients and the method. The results show that the relationship between precision and recall over classification thresholds is influenced by changes in federation size. The AUPRC fluctuations are more visible in the Heart Disease dataset at several client counts. Such behavior shows that the ranking performance for the positive class is sensitive to the created local data partitions. The AUPRC results are in line with the AUROC analysis, indicating that good overall class discrimination does not necessarily imply similarly robust precision-recall performance. This difference is particularly important in healthcare applications where the class distribution is imbalanced.

\begin{figure}[!h]
\centering
\includegraphics[width=\linewidth,keepaspectratio]{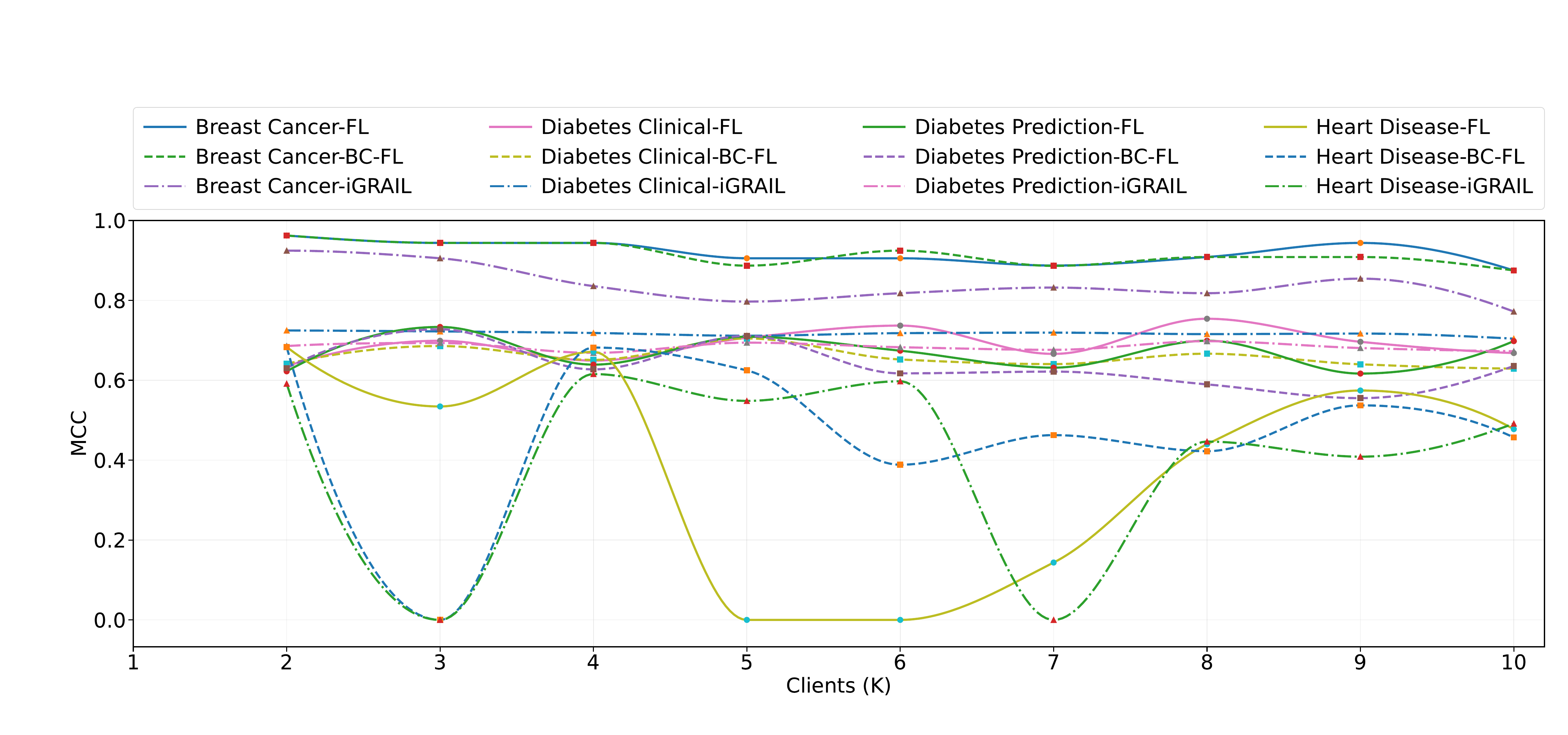}
\caption{MCC as the number of federated clients increases.}
\label{fig:fig6_8}
\end{figure}

\subsubsection{MCC Analysis}
Figure~\ref{fig:fig6_8} shows the MCC for the tested client configurations. To summarize, the cancer dataset shows high MCC values, indicating good conformity between the predicted and real class labels. The MCC for the diabetes clinical and diabetes prediction datasets varies with the methods and the size of the federation. This variability means that the number of clients involved in the test could change the proportion of true positives, true negatives, false positives, and false negatives. The maximum variation is relative to the MCC in the heart disease data set. Sometimes, the values are close to zero, which means that the predictions are weakly related to the true class labels. This is consistent with the low specificities found for some of the client counts. It implies that the MCC results may be reflective of classification constraints not well captured by accuracy or AUROC. The results show that a model can have useful discrimination ability even if at a given operating threshold it has unreliable classifications.

\subsubsection{Impact of Bigger Federation}
In general, the results indicate that the number of participating clients has different impacts on the four healthcare datasets. Cancer shows good overall classification performance, but there is a noticeable drop in iGRAIL AUROC when going from the two-client to the ten-client setup. The iGRAIL AUROC values are relatively stable for the diabetes clinical and diabetes prediction datasets. However, the precision, recall, F1 score, and MCC results indicate that the performance at a fixed classification threshold can still vary across the federation sizes. The Heart Disease dataset is more unstable for all metrics, especially for accuracy, specificity, F1 score, and MCC. The results show that classification performance does not improve with the increase in participating clients. A possible explanation is that distributing the training data over more clients changes the local availability of samples and their class representation. Such changes may impact the local optimization and the final global model. However, a more in-depth analysis of the data distributions per client, sample sizes, and training variability is required to understand the exact causes of the observed fluctuations.

\subsubsection{Discussion}
The scalability assessment yielded three main findings. First, iGRAIL achieves good AUROC performance across the evaluated federation sizes, especially for the two diabetes datasets. Increasing the number of clients from 2 to 10, the AUROC increases from $ 0.9507 $ to $ 0.9574 $ for Diabetes Clinical and from $ 0.9454 $ to $ 0.9420 $ for diabetes prediction. Second, the effect of increasing the size of the federation is dataset-dependent. Cancer has a larger drop in AUROC from $0.9960$ to $0.9633$, and the heart disease dataset has large fluctuations in several classification metrics. The results indicate that the predictive effect of increasing the number of clients is not uniform. Third, the joint analysis of all eight metrics demonstrates that stable AUROC does not necessarily imply stable classification performance at a fixed threshold. Specifically, for heart disease, the results of specificity and MCC indicate that some client configurations can result in unreliable negative class identification and weak overall classification performance. To summarize, the results show that iGRAIL can maintain a useful prediction performance for federations up to ten clients. However, the results do not show a consistent advantage over FL and BC-FL across all datasets and metrics. The experiment indicates signs of predictive scalability in the tested scope. However, further investigation of communication overhead, latency of blockchain verification, aggregation costs, and performance with larger client populations is needed to scale the system further.

\subsection{Robustness Against Malicious Model Updates}
\label{subsec:attack}
The robustness experiment measures the predictive performance of FL, BC-FL, and iGRAIL in the presence of malicious model updates during federated training. Realistic health care environments can have compromised clients, or clients who try to submit manipulated model parameters to affect the global learning process. If these updates are not correctly verified before being incorporated, they can influence the aggregated model and decrease its predictive reliability. This is particularly a concern in healthcare applications where incorrect model predictions may influence clinical decision support and patient risk assessment. To analyze this problem, we start injecting malicious model updates from communication round 6 and continue doing so for the following rounds. The experiment consists of 20 rounds of communication, and we observe the learning behavior before and after the attack. The first 5 rounds are the normal training period, and rounds 6-20 are the attack period. The experiment compares FL, BC-FL, and our proposed framework iGRAIL. The baseline is conventional FL, which does not include the update-verification mechanism explored in this experiment. In contrast, BC-FL and iGRAIL use verification mechanisms to control the acceptance of submitted model updates before aggregation. The iGRAIL framework also proposes privacy-preserving generative learning.

\begin{figure}[!h]
\centering
\includegraphics[width=\linewidth,keepaspectratio]{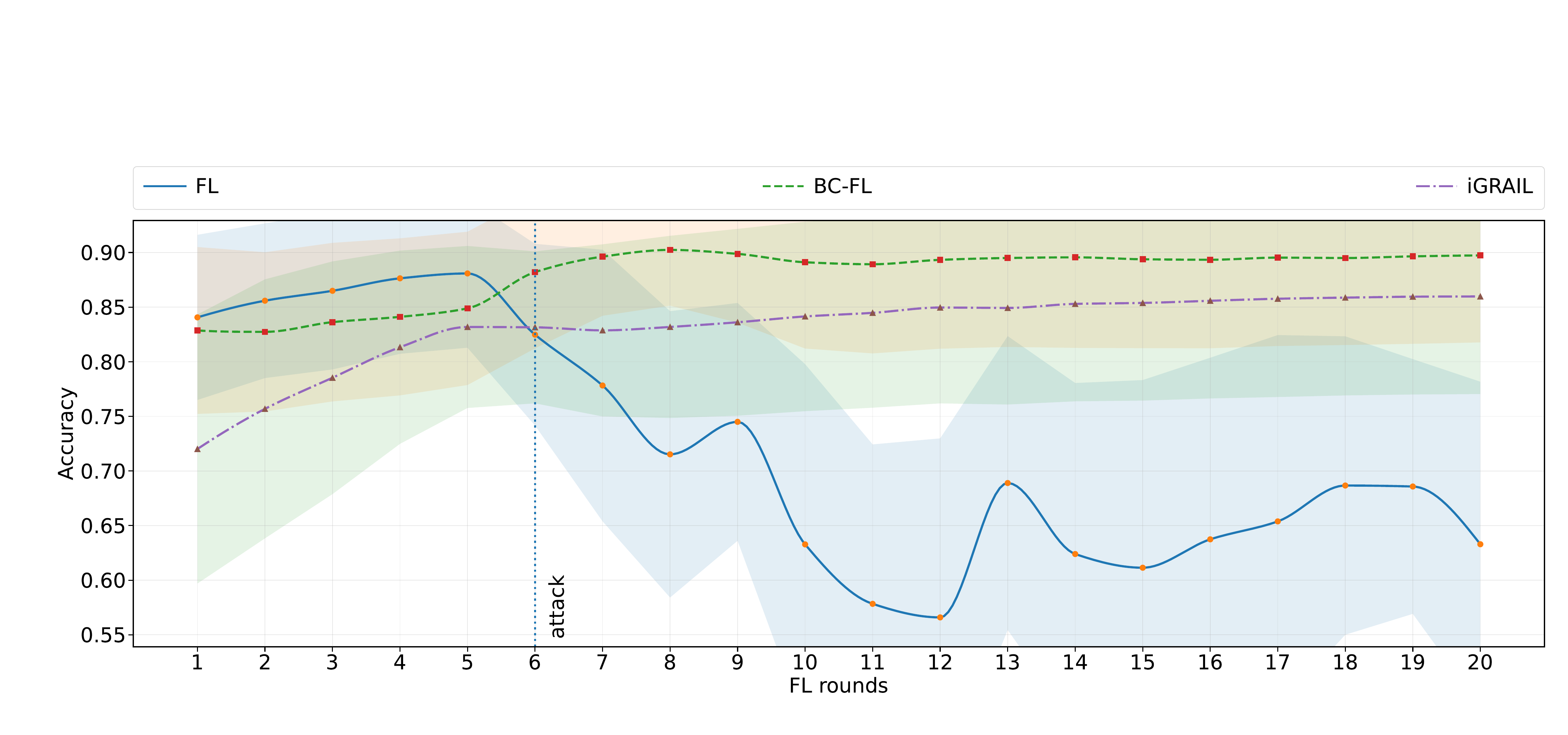}
\caption{Accuracy under malicious model updates over 20 communication rounds (attack begins at round 6).}
\label{fig:fig7_1}
\end{figure}

\subsubsection{Accuracy Analysis}
Figure~\ref{fig:fig7_1} shows the classification accuracy of FL, BC-FL, and iGRAIL in 20 communication rounds under malicious model updates. In the first 5 rounds of communication, all three methods exhibit normal learning behavior because the malicious update has not been injected.  The accuracy values reflect the predictive power of the respective learning strategies in the trained setup that was tested. The behavior of the methods changes significantly from round 6, when the attack starts. In the case of traditional FL, the accuracy is significantly reduced, i.e., the global model is negatively affected by the malicious updates. The standard FL baseline does not use the verification gate used in this experiment, so manipulated updates can influence the aggregation process. However, BC-FL has a relatively stable accuracy against rounds of attacks. This is in line with its use of update verification prior to aggregation and prevents updates that do not pass the implemented checks from affecting the global model. The proposed iGRAIL framework also tackles the severe accuracy degradation in conventional FL. After the attack is launched, the accuracy is more or less stable, which indicates that the verification mechanism is beneficial to the maintenance of the global model’s prediction performance. However, iGRAIL is not always more accurate than BC-FL across the entire attack period. These results confirm the robustness of its verification-based learning process, rather than any universal superiority in prediction.

\begin{figure}[!h]
\centering
\includegraphics[width=\linewidth,keepaspectratio]{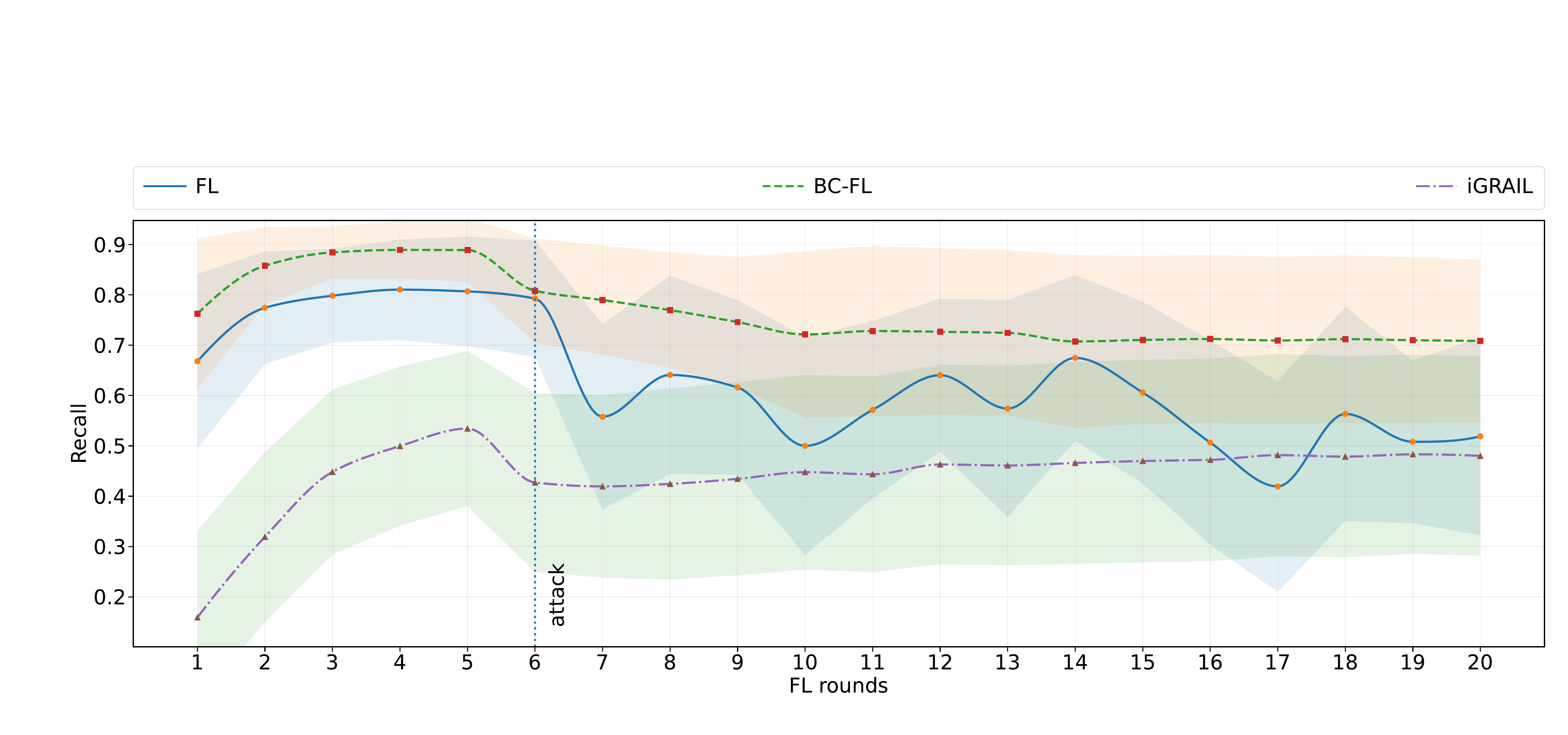}
\caption{Recall under malicious model updates over 20 communication rounds (attack begins at round 6).}
\label{fig:fig7_2}
\end{figure}

\subsubsection{Recall Analysis}
We recall the before-and-after of the malicious model update injection in Figure~\ref{fig:fig7_2}. Recall is the proportion of actual positives correctly identified by the classifier and is of particular importance in healthcare applications. In the first five rounds, the methods behave as expected to detect the positive cases. Note that the recall of the traditional FL sharply drops after injecting malicious updates at round 6. This shows that the manipulated model updates degrade the global model’s positive case detection ability. This degradation is important because the lower the recall, the higher the proportion of false-negative predictions. False negatives in healthcare settings can result in a predictive screening system failing to identify positive cases. BC-FL shows more stability in the attack rounds, benefiting from its update-verification mechanism to maintain positive-case detection. The framework mitigates the effect of malicious submissions on the evaluation by ensuring that unverified updates are not included in the aggregation. Also, iGRAIL exhibits more stable recall than traditional FL after the attack starts. This behavior shows that the proposed verification-gated aggregation process helps to sustain positive-case detection even under the presence of malicious update submissions. So the results of the recall experiment suggest that update verification can provide benefits beyond just maintaining global accuracy. In particular, it may help reduce the degradation of positive-case detection induced by the considered attack.

\begin{figure}[!h]
\centering
\includegraphics[width=\linewidth,keepaspectratio]{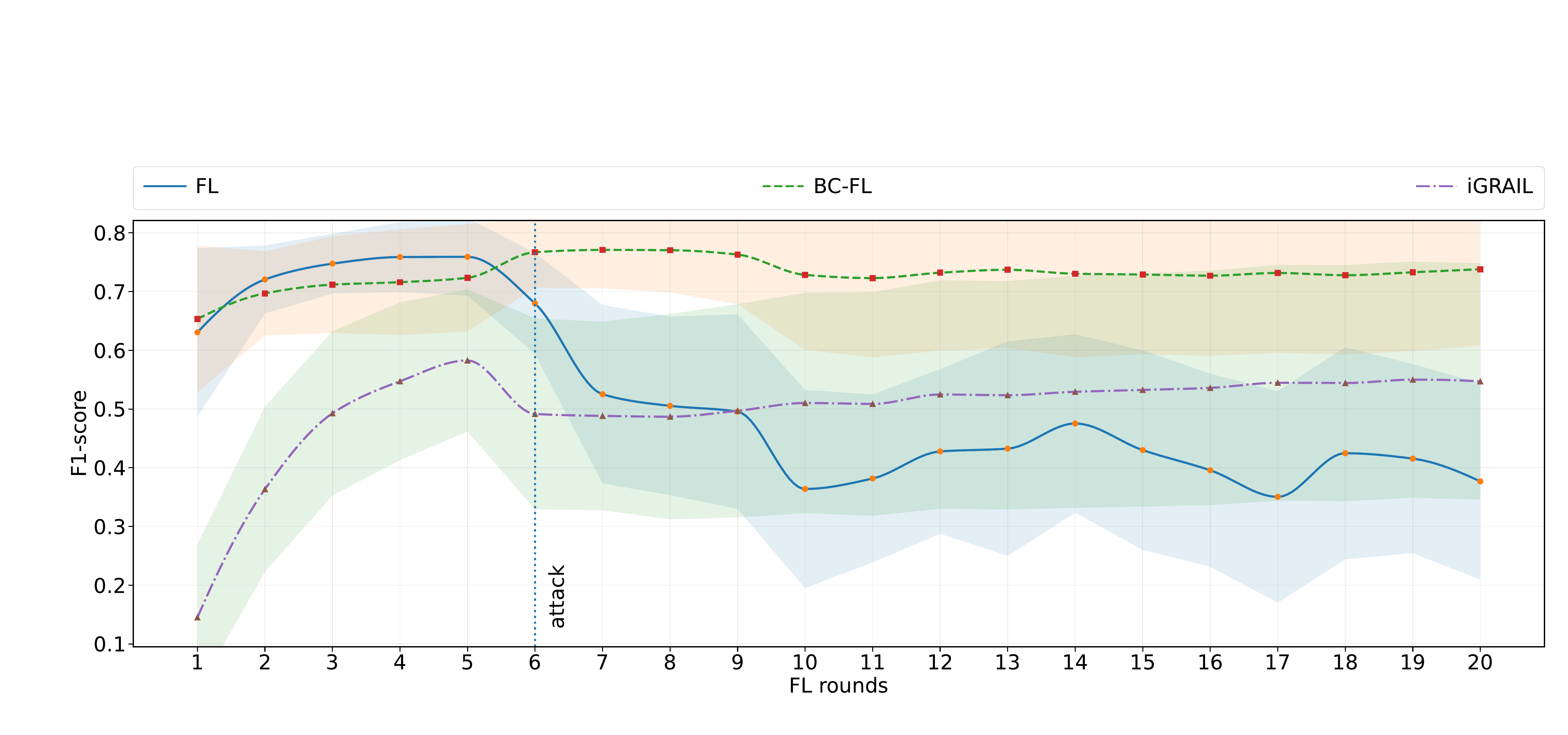}
\caption{F1-score under malicious model updates over 20 communication rounds (attack begins at round 6).}
\label{fig:fig7_3}
\end{figure}

\subsubsection{F1-Score Analysis}
Figure~\ref{fig:fig7_3} shows the F1 score during the 20 communication rounds. The F1 score is a balance between precision and recall. It gives insight into the balance between correct positive predictions and detection of positive cases. The three methods follow their learning curves normally until the attack begins. Conventional FL shows a sharp decline in F1 score after round 6, indicating that malicious updates harm its positive-class classification performance. This reduction may be caused by a change in precision, recall, or both. Hence, the F1-score provides a more comprehensive measure of the degradation in attack-related classification than recall. In contrast, the F1-score of BC-FL is relatively stable during the attacked period. This means its verification mechanism can maintain the precision-recall trade-off under the evaluated conditions for malicious updates. The proposed iGRAIL framework also achieves a more stable F1-score than conventional FL. The results show that the extra privacy-preserving learning operations do not hinder the verification mechanism in alleviating the impact of the malicious updates under evaluation. However, differences remain between BC-FL and iGRAIL in the post-attack period. These differences suggest that the extra learning components in iGRAIL do not necessarily increase the F1 score compared to the blockchain-supported baseline. To summarize, the F1-score results further provide evidence that verification-gated aggregation helps alleviate the degradation in classification observed in traditional FL.

\begin{figure}[!h]
\centering
\includegraphics[width=\linewidth,keepaspectratio]{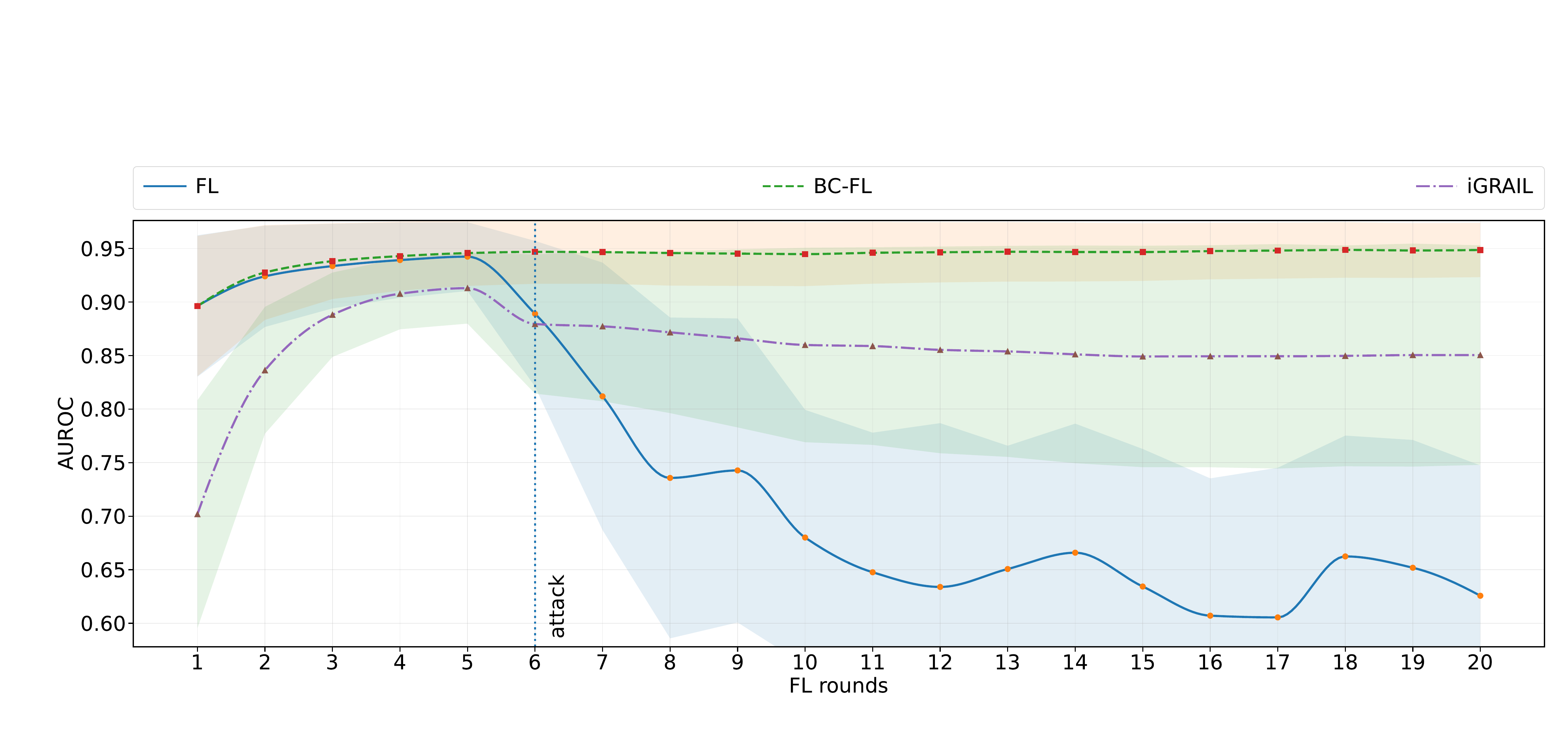}
\caption{AUROC under malicious model updates over 20 communication rounds (attack begins at round 6).}
\label{fig:fig7_4}
\end{figure}

\subsubsection{AUROC Analysis}
Figure~\ref{fig:fig7_4} shows the AUROC in 20 communication rounds. AUROC measures how well the model can discriminate between positive and negative samples across different classification thresholds. In a normal training period, the three methods exhibit their own learning behavior, without any malicious updates. Conventional FL shows a drastic drop in AUROC after the attack at round 6, indicating that the malicious updates degraded the global model's ability to distinguish classes. This is an important finding because AUROC accounts for the ranking of predictions across thresholds rather than measuring classification performance at a fixed operating threshold. We therefore observe that the degradation suggests the attack affects the discriminative ability of a model, not only the predictions at a single chosen threshold.

We observe that BC-FL achieves a higher AUROC in the attacked rounds, confirming that its verification mechanism helps the global model retain its ability to distinguish positive and negative cases. The proposed iGRAIL framework also avoids much of the AUROC degradation seen in traditional FL. Its relatively stable behavior after the attack shows that verification before aggregation helps limit the impact of the tested malicious updates on class discrimination. But BC-FL achieves stronger performance for some post-attack configurations. Thus, the security benefits of iGRAIL should be separated from its relative predictive performance. The proposed framework is robust to malicious updates but does not necessarily achieve the highest AUROC among the methods compared.

\begin{figure}[!h]
\centering
\includegraphics[width=\linewidth,keepaspectratio]{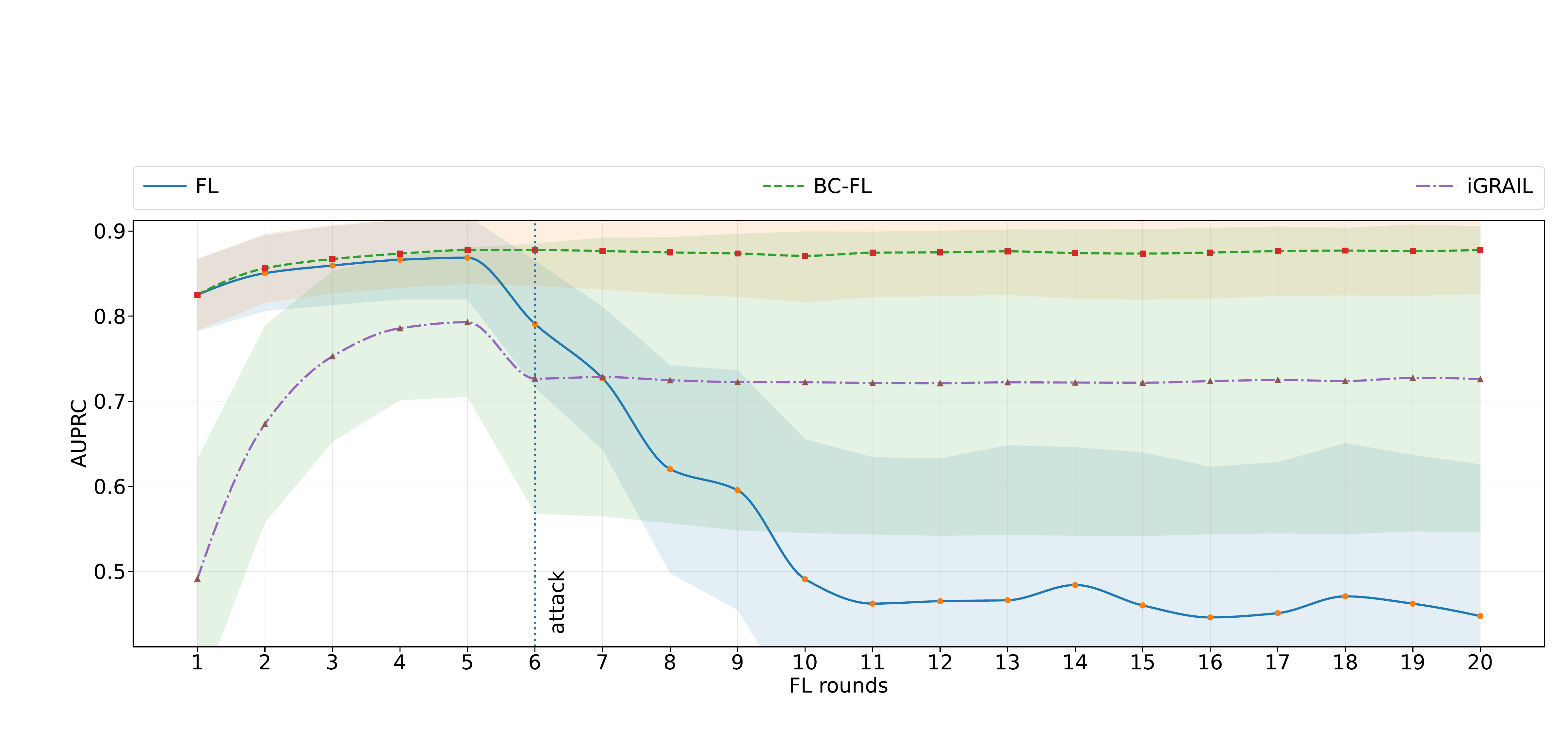}
\caption{AUPRC under malicious model updates over 20 communication rounds (attack begins at round 6).}
\label{fig:fig7_5}
\end{figure}

\subsubsection{AUPRC Analysis}
Figure~\ref{fig:fig7_5} depicts the AUPRC before and after the malicious updates. AUPRC measures the relationship between precision and recall across different classification thresholds and is particularly informative when the healthcare data evaluated has imbalanced class distributions. After the attack at round 6, traditional FL suffers a significant drop in AUPRC. This implies that the malicious updates reduce the model’s ability to produce reliable positive-class predictions across various decision thresholds. The observed degradation is substantial because an attack may compromise the identification of positive cases, even if aggregate classification measures are fairly high. So, AUPRC provides more information about the consequences of accepting manipulated updates in federated aggregation. BC-FL obtains a quite stable AUPRC during the attacked rounds. This behavior is consistent with rejecting updates that fail the implemented verification checks. iGRAIL has a more stable AUPRC compared to standard FL. The results show that its verification-gated aggregation mechanism can maintain positive-class ranking performance in the presence of malicious updates. The AUPRC results are consistent with the AUROC and F1-score results, showing that the advantages of update verification are robust across classification thresholds in terms of precision-recall behavior.

\begin{figure}[!h]
\centering
\includegraphics[width=\linewidth,keepaspectratio]{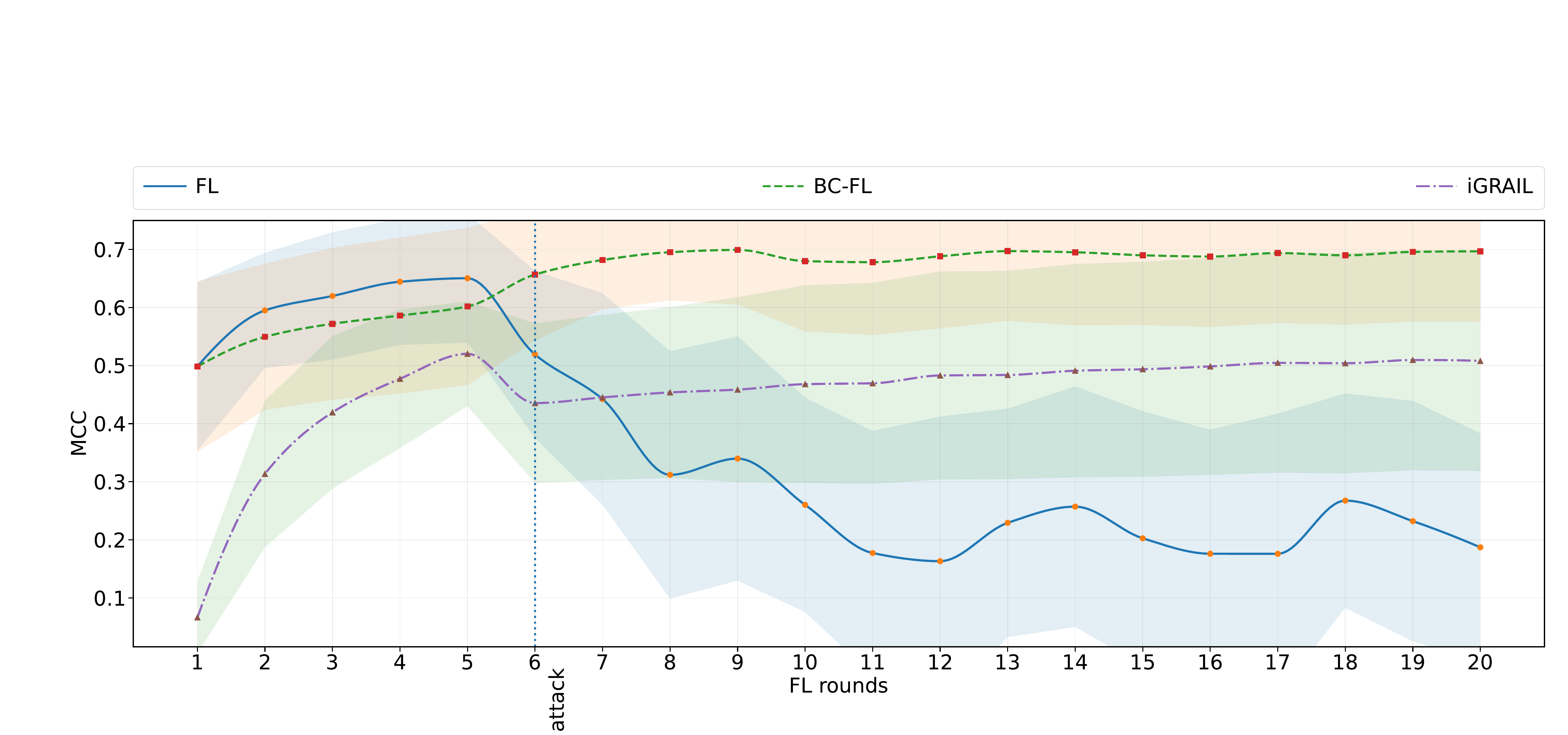}
\caption{MCC under malicious model updates over 20 communication rounds (attack begins at round 6).}
\label{fig:fig7_6}
\end{figure}

\subsubsection{MCC Analysis}
Figure~\ref{fig:fig7_6} presents the Matthews correlation coefficient throughout the 20 communication rounds. MCC considers all four elements of the confusion matrix and provides a balanced measure of the classification quality. Before the attack, the three methods show their typical classification behavior. The conventional FL exhibits a significant drop in MCC after round 6, which implies that the malicious updates degrade the relationship between the predicted and true class labels. This degradation shows that the attack affects the global classification balance, not just the predictions of a single category. As MCC considers both positive and negative classification outcomes, it provides an extra insight into the consequences of malicious model updates. BC-FL achieves relatively stable MCC during the attack rounds, indicating that verification before aggregation helps maintain the quality of the global model's predictions. The proposed iGRAIL framework also avoids much of the MCC degradation seen in conventional FL. This result further shows that the verification mechanism limits the influence of malicious submissions on the global model. The results of MCC are consistent with the observations for accuracy, F1-score, AUROC, and AUPRC. Together, these results suggest that the gain of verification-gated aggregation is reflected in several complementary predictive metrics.

\subsubsection{Intuition of Attack Robustness}
The performance differences observed after round 6 show the importance of validating submitted model updates before aggregation. Conversely, conventional FL suffers serious performance degradation against the evaluated attack, whereas BC-FL and iGRAIL maintain relatively stable predictive performance. This behavior is consistent with excluding updates that do not satisfy the verification conditions implemented. However, the robustness demonstrated is limited to the malicious-update conditions demonstrated in this experiment. Cryptographic integrity and origin authentication ensure that the submitted information is consistent with the recorded reference and originates from an authenticated participant. They alone do not prove the benignity of the model parameters. Thus, a legitimate client creating a poisoned model update can still satisfy the requirements of cryptographic verification. Thus, the present results demonstrate the benefits of verification-gated aggregation on the tested invalid submissions, but not total robustness against all model-poisoning attacks.

\subsubsection{Behavior Before and After the Attack}
The experiment can be divided into two parts: All three methods are not affected by malicious updates in rounds 1-5. Performance in this time period can be used as a baseline to assess the effect of the attack. We begin injecting malicious model updates from round 6 and continue to inject them until the end of communication rounds. The traditional FL’s performance was severely degraded at the beginning of the attack. This behavior illustrates how vulnerable the evaluated unprotected aggregation process is to manipulation by client updates. However, the prediction performance of BC-FL and iGRAIL is relatively stable in the attacked rounds. This is in line with the fact that the verification mechanisms can ignore updates from the aggregation that do not pass the implemented checks. You can see the difference between the protected and unprotected methods over a number of evaluation metrics. Our results also indicate that BC-FL dominates iGRAIL on several post-attack performance indicators. This observation is of great importance in the interpretation of the advantages of the proposed framework. In this experiment, the main contribution of iGRAIL is that it can maintain a useful predictive performance under the attack we evaluated, while introducing privacy-preserving generative learning and blockchain-based verification.

\subsubsection{Role of Verification-Gated Aggregation}
The robustness results indicate that submitted model updates should be tested before being applied to the global model. In traditional FL, the aggregation step can be vulnerable when the manipulated updates are accepted without proper security checks. The proposed iGRAIL framework introduces a verification-gated aggregation process in which submitted updates are verified before acceptance. The verification layer checks authorization, origin authentication, integrity, and freshness based on the security procedures being applied. Updates that fail these checks are discarded prior to aggregation. This minimizes the danger that a tampered, replayed, or unauthorized submission could corrupt the global model. The results in Figures~\ref{fig:fig7_1}-\ref{fig:fig7_6} demonstrate the effectiveness of this approach under the evaluated malicious-update conditions. Conventional FL deteriorates markedly, whereas verification-enabled methods maintain predictive performance more consistently. However, the obtained results do not provide protection against all potential model-poisoning strategies. In particular, an update that is properly authenticated and passes integrity and freshness checks may still carry malicious model parameters. Therefore, the protection observed must be considered in light of the low attack level and the specific verification checks used in the experiment.

\subsubsection{Attack Model and Scope of Verification}
The malicious-update experiment injects invalid model updates during communication rounds 6 to 20. As shown in Figures~\ref{fig:fig7_1}-\ref{fig:fig7_6}, conventional FL suffers from severe performance degradation, whereas BC-FL and iGRAIL show more stable performance under the various attack scenarios considered. The verification-enabled methods check the authorization, integrity, origin authentication, and freshness before aggregation. Table~\ref{table:security} shows the acceptance of valid updates and the rejection of tampered, replayed, and unauthorized submissions. The results of this provide protection against updates that do not meet the implemented verification conditions. However, the only cryptographic verification mechanism cannot detect the malicious model parameters uploaded by an authenticated client. Handling such poisoning attacks requires additional model-level validation or Byzantine-robust aggregation, which are out of scope for the present evaluation.

\subsubsection{Discussion}
The malicious-update experiment yields three important results. First, conventional FL exhibits noticeable predictive degradation after the attack begins, starting in communication round 6. The drop in multiple metrics shows that manipulated updates can hurt the aggregated global model if accepted without the verification protections in place. Second, BC-FL exhibits relatively stable performance during attack rounds and achieves the best results across several metrics. This demonstrates the effectiveness of verification-based protection under the attack conditions tested. Third, iGRAIL circumvents much of the degradation observed in conventional FL while preserving its privacy-preserving generative learning and blockchain-based verification components. The results show that these components can work together without compromising the framework’s useful predictive performance against the evaluated attack.

\begin{table}[!h]
\caption{Model-Update Verification Results}
\label{table:security}
\centering
\footnotesize
\renewcommand{\arraystretch}{1.20}
\setlength{\tabcolsep}{12pt}

\begin{tabular}{lcc}
\toprule
\textbf{Scenario} &
\textbf{Accepted} &
\textbf{Decision} \\
\midrule

Valid        & 1 & Accept \\
Tampered     & 0 & Reject \\
Replay       & 0 & Reject \\
Unauthorized & 0 & Reject \\

\bottomrule
\end{tabular}
\end{table}

\subsection{Model-Update Security Verification}
\label{subsec:security}
The following experiment tests four update conditions. The conditions are valid, tampered with, replayed, and unauthorized. The idea is to determine if the verification process gives the expected decision. The acceptance decision is shown in Table~\ref{table:security}. If accepted, the update is valid as shown in Table~\ref{table:security}. The update was corrupted and was refused. The replayed update is also dropped. Access denied; update failed. Each time, it tests a different facet of the verification process. The integrity is validated by the tampered case. The replay case proves freshness. Control of participation in unauthorized instances. The results indicate that invalid model updates can be filtered prior to aggregation. The experiment does not purport to be secure against all attacks. It simply checks the safety conditions that are implemented in the current framework.

\begin{figure}[!tb]
\centering
\includegraphics[width=0.90\linewidth,keepaspectratio]{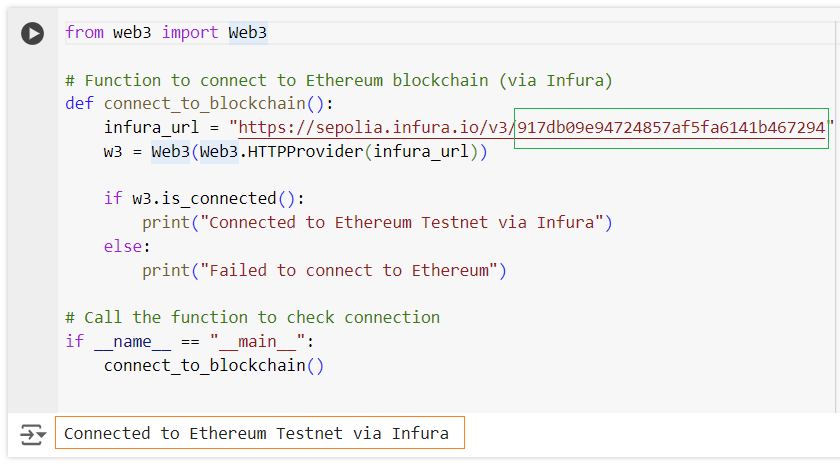}
\caption{Blockchain connectivity established for
the experimental verification environment.}
\label{fig:blockchain_connection}
\end{figure}

\subsection{Blockchain Verification and Smart-Contract Execution} \label{sec:blockchain_eval}
The experiment on blockchain verification and smart-contract execution assesses the security functionality and the associated operational overhead of the proposed iGRAIL framework. The blockchain connection setup of the experimental verification environment is shown in Figure~\ref{fig:blockchain_connection}. This connection provides the communication interface for the evaluator to access the blockchain-supported verification functions. The connectivity outcome evidences the presence of the blockchain interface in the tested configuration.
However, it does not independently establish transaction integrity, authorization correctness, or execution performance. These properties are studied separately using integrity, gas-consumption, and transaction-execution experiments.

In the classical federated learning setting, the participating clients send local model updates to a central aggregator. An incorrect verification may allow the aggregation procedure to accept an unauthorized, modified, or replayed submission, which may compromise the integrity and reliability of the resulting global model. To address these concerns, iGRAIL employs a blockchain-based verification layer and a smart contract-based participation control before accepting model updates into aggregation. The experiment uses 10 representative model update transactions to study the mechanism of verification and transaction processing. We evaluate four aspects: model-update integrity verification, gas consumption of blockchain transactions, transaction execution time, and communication overhead for encrypted model information. The aim of this experiment is to verify whether the verification mechanism is capable of detecting inconsistencies between the submitted model information and the corresponding integrity reference stored on the blockchain. For every transaction under consideration, a cryptographic hash is generated from the relevant model update information. The hash is then compared with the corresponding reference stored on the blockchain. If the two hashes are identical, then the comparison indicates that the information examined is identical to the stored reference. A mismatch, however, means the submitted data does not correspond to the data linked to the recorded hash. The mechanism is an integrity check before a model update is accepted in the aggregation process. 

\begin{figure}[!h]
\centering
\includegraphics[width=\linewidth,keepaspectratio]{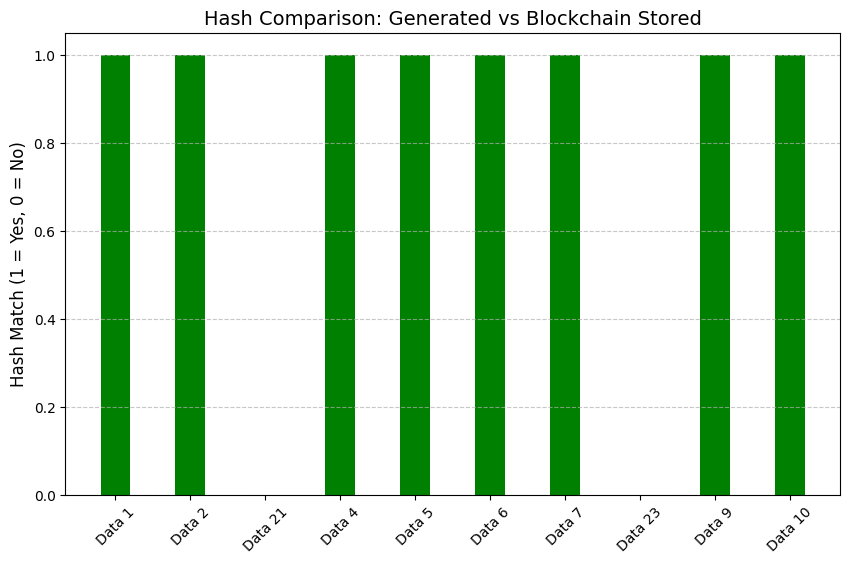}
\caption{Blockchain integrity verification using generated and blockchain-recorded hashes for ten representative model-update transactions.}
\label{fig:fig_blockchain_eval_1}
\end{figure}

The hash comparison that follows can reveal the inconsistency as long as the hash function and reference-recording process remain secure, if the model information changes after the reference hash has been created. The results depicted in Figure~\ref{fig:fig_blockchain_eval_1} show that the integrity references stored on the blockchain can be used to verify the integrity of the model updates. The function provides the ability for the blockchain to not only log transactions but also log information that can support verification decisions. Hash consistency alone does not prove that an update to the model is trustworthy and that there are no injected malicious parameters. It only verifies consistency with the recorded reference. Integrity verification must therefore be integrated with authorization, origin authentication, freshness validation, and any other model-level checks required by the threat model.

\begin{figure}[!h]
\centering
\includegraphics[width=\linewidth,keepaspectratio]{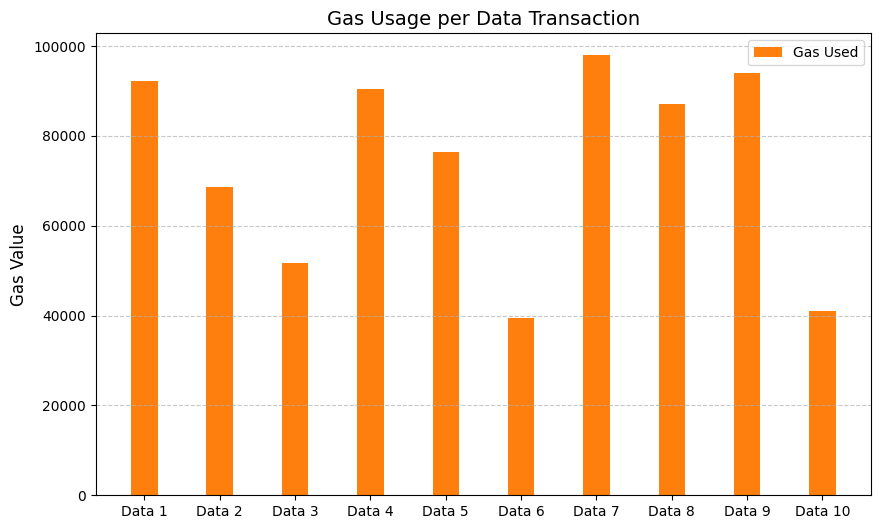}
\caption{Blockchain transaction gas consumption for ten representative model-update transactions.}
\label{fig:fig_blockchain_eval_2}
\end{figure}

\subsubsection{Blockchain Transaction Gas Consumption}
Figure~\ref{fig:fig_blockchain_eval_2} shows the gas consumed by the 10 representative blockchain transactions. Gas is a unit of account for the computational resources required to perform transaction processing and smart contract operations in an Ethereum-compatible execution environment. The proposed framework is based on intelligent contracts to support the verification and authorization workflow governing the admission and participation in model updates. Depending on the implemented contract functions, transaction processing may consist of checking the submitted information, validating participation conditions, and recording the resulting verification state. The gas consumption results give an insight into the computational cost of these blockchain operations. The measured gas usage for the 10 transactions evaluated is limited. This implies that the tested transaction processing operations have finite, measurable execution costs in the experimental setup. The amount of gas consumed can vary depending on the contract instructions, storage operations, or the transaction processing paths being executed. Querying existing information may require less gas than an operation that changes the blockchain state. The results show that the blockchain-supported verification workflow incurs additional computational cost, which can be measured on a per-transaction basis. This price includes the cost of verifiable participation controls and integrity checks in the federated learning process. Gas usage is not the same as fees for monetary transactions and should be treated accordingly. It is a measure of the amount of computational resources used by operations executed. The actual cost in money depends on the gas price and the blockchain network. The gas measurements reported are therefore indicative of execution overhead, not of a fixed monetary cost.

\begin{figure}[!htbp]
\centering
\includegraphics[width=\linewidth,keepaspectratio]{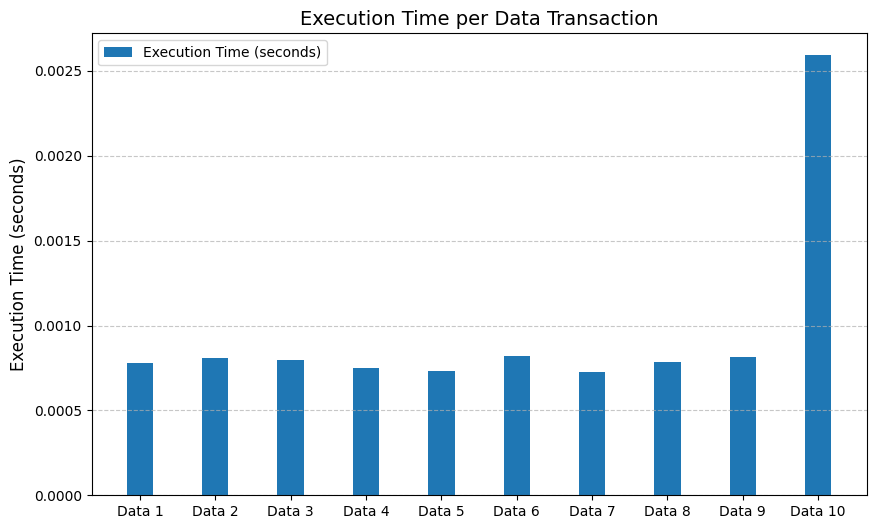}
\caption{Blockchain transaction execution time for ten representative model-update transactions.}
\label{fig:fig_blockchain_eval_3}
\end{figure}

\subsubsection{Execution Time of Smart-Contract and Transactions}
The execution time measured for the 10 representative blockchain transactions is shown in Figure~\ref{fig:fig_blockchain_eval_3}. This experiment measures the time overhead of processing blockchain-supported verification operations. iGRAIL only allows a model update to be accepted after the relevant verification and authorization processes are complete. Hence, blockchain transaction processing may introduce additional latency compared to traditional aggregation workflows, which do not perform such checks. The results show that the execution times measured are bounded for the evaluated transactions. This implies that the tested blockchain operations could be completed within a certain execution interval under the experimental conditions. This may be due to different contract execution paths, varying transaction processing conditions, or different underlying execution environments. The reported measurements provide a first-order characterization of the operational cost of the blockchain-aided model-update verification. The runtime results are especially important for federated learning, as repeated verification operations can increase the total time of a communication round. If multiple clients submit updates, the overall verification delay depends on the number of submissions and on whether the corresponding operations are executed sequentially or in parallel. However, the measured transaction execution time should not be automatically interpreted as the end-to-end blockchain confirmation latency. In a deployed blockchain environment, network propagation, block inclusion, finality, and transaction queuing may introduce additional delays. The results therefore describe the measured transaction-processing overhead of the evaluated setup, rather than full deployment latency.

\begin{figure}[!htbp]
\centering
\includegraphics[width=\linewidth,keepaspectratio]{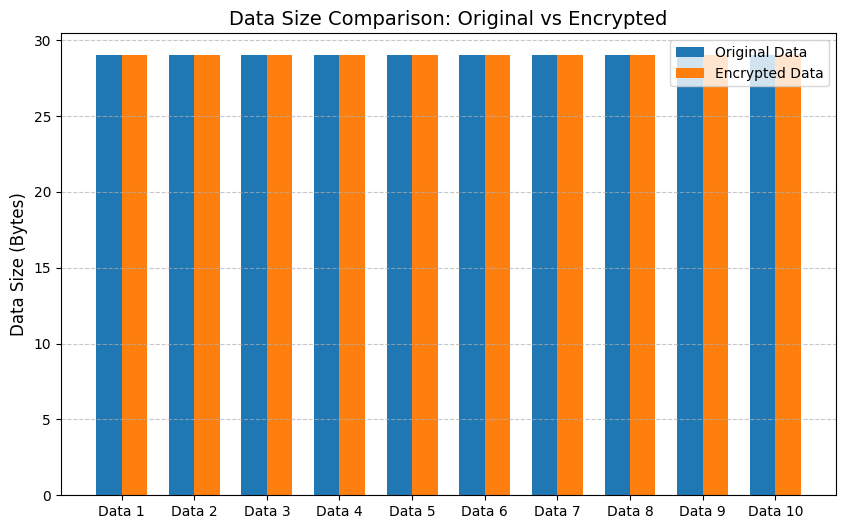}
\caption{Communication overhead from original and encrypted payload sizes for ten representative model-update transactions.}
\label{fig:fig_blockchain_eval_4}
\end{figure}

\subsubsection{Overhead of Encrypted Model-Payload Communication}
Figure~\ref{fig:fig_blockchain_eval_4} compares the sizes of the original and encrypted payloads for the 10 representative model-update transactions. The goal is to measure the additional communication costs incurred to protect model information before it is transmitted or exchanged. In the proposed framework, encryption is used to protect the model information during the exchange process. Encryption can improve confidentiality by making data unavailable to unauthorized parties without the necessary decryption credentials, but it can also increase the payload overhead due to cryptographic metadata, initialization values, authentication tags, or other protocol-specific data. The results show that the sizes of the encrypted payloads are close to those of the corresponding original payloads for the analyzed transactions. This implies that the implemented protection mechanism limits payload expansion under the tested configurations.

The observed behavior is important for federated learning, as model updates are repeatedly communicated over communication rounds. However, small increases in payload size can add up if the federation comprises many participants or many training rounds. However, the payload-size comparison quantifies only the expansion due to the encryption process under consideration. It does not account for the communication costs of the proposed framework independently, such as blockchain transaction metadata, additional verification messages, network protocol overhead, and repeated transmissions. Also, small payload expansion does not necessarily mean negligible encryption computation time. The latency of the encryption and decryption processes needs to be measured separately to assess their impact on the overall execution cost.

\subsubsection{Smart Contract-Based Model-Update Admission}
The blockchain and smart-contract elements are working in tandem to enable controlled participation in the federated learning process. Instead of directly accepting all submitted model updates for aggregation, the proposed framework checks the verification and authorization conditions that can be applied for acceptance. The verification workflow checks the integrity of the submitted information, the participating client's authorization status, and the applicable authentication and freshness conditions. The associated participation and verification decisions are enforced or recorded by intelligent contracts in accordance with the implemented protocol. Once the required conditions are met, the model update is eligible for aggregation. A submission that fails the relevant checks is rejected or excluded from the aggregation process. This approach separates the submission of model updates from their acceptance. This distinction is important because the mere submission of an update does not automatically imply that the corresponding client is authorized or that the submitted information is consistent with the recorded integrity reference. The blockchain layer also provides an auditable record of transaction information and verification decisions stored on-chain. This allows for later examination of participation and update-admission activities. However, the completeness of the audit trail depends on which events and metadata the deployed smart contract actually records.

\subsubsection{Security and Operational Overhead Trade-Off}
Taken together, the results illustrate the trade-off between verification functionality and operational overhead in iGRAIL. As shown in Figure~\ref{fig:fig_blockchain_eval_1}, the hash references in the blockchain can be used to indicate inconsistencies in the model-update information. This supports the integrity verification component of the proposed framework. Figure~\ref{fig:fig_blockchain_eval_2} presents the gas consumption of the evaluated blockchain transactions, and Figure~\ref{fig:fig_blockchain_eval_3} shows the respective execution-time overhead. In sum, these measurements show that blockchain-based verification adds processing costs that can be quantified and analyzed. Figure~\ref{fig:fig_blockchain_eval_4} also demonstrates that the implemented encryption mechanism leads to a limited payload expansion for the considered model-update transactions. This shows that the overhead of additional transmitted data related to encryption is relatively small under the tested conditions. These results suggest that the security mechanisms can be integrated into the model-update exchange and admission workflow with measurable transaction-processing and communication costs. However, the experiment is limited to 10 representative transactions and does not establish the scaling of the overhead for substantially larger transaction volumes, concurrent client submissions, or varying blockchain network conditions.

\subsubsection{Discussion}
The experiment of blockchain verification and smart-contract execution yields four important observations. First, the comparison between generated and blockchain-recorded hashes demonstrates the use of cryptographic integrity references to identify inconsistencies in model-update information before aggregation. This helps strengthen blockchain's use as a tool for verification, not just a record of transactions. Second, the gas consumption measurements indicate that the evaluated smart-contract and blockchain operations impose measurable computational overhead. The observed costs are bounded across the 10 representative transactions, but their behavior at larger transaction volumes warrants further investigation. Third, the results at execution time characterize the processing overhead associated with the evaluated blockchain transactions. The measurements provide a first estimate of the time required for the verification workflow, but they should not be considered full end-to-end blockchain confirmation latency if network and finality delays are not accounted for. Fourth, the comparison between the original and encrypted payload shows that the implemented encryption mechanism results in a limited payload expansion under the conditions evaluated. This supports the feasibility of protecting model information without significant increases in measured payload.

\begin{table}[!h]
\caption{Cryptographic Protection and Verification Overhead}
\label{table:crypto}
\centering
\footnotesize
\renewcommand{\arraystretch}{1.20}
\setlength{\tabcolsep}{12pt}

\begin{tabular}{rcc}
\toprule
\textbf{Payload (KB)} &
\textbf{Protect (ms)} &
\textbf{Verify (ms)} \\
\midrule

16   & $0.2417\pm0.0075$ & $0.3470\pm0.0027$ \\
64   & $0.2671\pm0.0024$ & $0.3750\pm0.0057$ \\
256  & $0.6121\pm0.2345$ & $0.7952\pm0.3595$ \\
1024 & $0.7439\pm0.0069$ & $0.8610\pm0.0097$ \\
4096 & $2.3649\pm0.1359$ & $2.4732\pm0.0164$ \\

\bottomrule
\end{tabular}
\end{table}

\subsection{Cryptographic Communication Overhead}
\label{subsec:crypto}
The security mechanisms also introduce processing costs. This cost
must be measured because protected model exchange is performed during
the federated process. The payload experiment uses sizes of 16, 64, 256, 1024, and
4096 KB. Each payload size is tested 30 times. The change in protection and verification time, as well as the exact mean and standard deviation, are given in Table~\ref{table:crypto}. For the 16-KB payload, the mean protection time is $0.2417$ ms.
The average time to verify is $0.3470$ms. 1024 KB, these values increase to $0.7439$ ms and $0.8610$ ms. For the largest 4096-KB payload, the protection time is $2.3649$ ms and the verification time is $2.4732$ ms. The result shows that the cost increases with payload size. However, the measured cryptographic latency for the tested payload sizes is still small. The 256-KB result has a larger standard deviation than the other
payload sizes. This variation is visible in
Table~\ref{table:crypto}. It should be retained rather than hidden
because it reflects the measured experiment.

\subsection{Runtime Analysis}
\label{subsec:runtime}
The next experiment investigates the runtime of FL, BC-FL, and iGRAIL at various communication rounds. The runtime is divided into local processing, verification, and aggregation. This comparison shows the additional computational cost incurred by iGRAIL's privacy and security operations. Table~\ref{table:runtime} shows the runtime results at the 5th, 10th, 15th, and 20th communication rounds. The total execution time generally increases with the number of rounds. FL incurs the lowest computational cost, and BC-FL incurs a small verification cost. iGRAIL needs more time due to the additional local generative and privacy-preserving operations. The runtime components also show where this extra cost is paid. For iGRAIL on Diabetes-clinical at 20 rounds, local processing requires $37.0532$ s, while verification requires only $0.0450$ s and aggregation requires $0.0069$ s. On the diabetes-prediction dataset, the corresponding
times are $48.7324$ s, $0.0528$ s, and $0.0138$ s. Therefore, the main runtime cost of iGRAIL comes from local processing rather than verification or aggregation. Overall, Table~\ref{table:runtime} shows a clear computational
trade-off. iGRAIL has a higher runtime than FL and BC-FL, but the
verification and aggregation overhead remains small compared with
local computation. The additional runtime is associated mainly with
the local operations required by the proposed privacy-preserving
framework.

\begingroup
\footnotesize
\renewcommand{\arraystretch}{1.2}
\setlength{\tabcolsep}{9pt}
\setlength{\LTcapwidth}{0.9\textwidth}
\begin{longtable}{@{}llrrrrr@{}}
\caption{Runtime Comparison of FL, BC-FL, and iGRAIL Across Communication Rounds}
\label{table:runtime}\\

\toprule
\textbf{Dataset} &
\textbf{Method} &
\textbf{Rounds} &
\textbf{Local (s)} &
\textbf{Verify (s)} &
\textbf{Aggregate (s)} &
\textbf{Total (s)} \\
\midrule
\endfirsthead

\multicolumn{7}{c}{%
\textit{Table \thetable{} (continued)}}\\
\toprule
\textbf{Dataset} &
\textbf{Method} &
\textbf{Rounds} &
\textbf{Local (s)} &
\textbf{Verify (s)} &
\textbf{Aggregate (s)} &
\textbf{Total (s)} \\
\midrule
\endhead

\midrule
\multicolumn{7}{r}{\textit{Continued on next page}}\\
\endfoot

\bottomrule
\endlastfoot

\multirow{12}{*}{Cancer \cite{yasser2021breastcancer}}
& FL     & $5$  & $0.0194$ & $0.0000$ & $0.0003$ & $0.0197$ \\
& BC-FL  & $5$  & $0.0187$ & $0.0095$ & $0.0016$ & $0.0299$ \\
& iGRAIL & $5$  & $0.0759$ & $0.0109$ & $0.0023$ & $0.0891$ \\
\cmidrule(lr){2-7}

& FL     & $10$ & $0.0418$ & $0.0000$ & $0.0005$ & $0.0423$ \\
& BC-FL  & $10$ & $0.0483$ & $0.0202$ & $0.0033$ & $0.0718$ \\
& iGRAIL & $10$ & $0.1385$ & $0.0207$ & $0.0043$ & $0.1635$ \\
\cmidrule(lr){2-7}

& FL     & $15$ & $0.0704$ & $0.0000$ & $0.0016$ & $0.0720$ \\
& BC-FL  & $15$ & $0.0616$ & $0.0284$ & $0.0047$ & $0.0947$ \\
& iGRAIL & $15$ & $0.2207$ & $0.0318$ & $0.0056$ & $0.2581$ \\
\cmidrule(lr){2-7}

& FL     & $20$ & $0.0824$ & $0.0000$ & $0.0011$ & $0.0835$ \\
& BC-FL  & $20$ & $0.0899$ & $0.0403$ & $0.0063$ & $0.1365$ \\
& iGRAIL & $20$ & $0.3369$ & $0.0428$ & $0.0075$ & $0.3873$ \\

\midrule

\multirow{12}{*}{%
\shortstack[l]{Diabetes-\\Clinical \cite{mustafa_diabetes_prediction}}}
& FL     & $5$  & $2.3684$ & $0.0000$ & $0.0004$ & $2.3688$ \\
& BC-FL  & $5$  & $2.3481$ & $0.0116$ & $0.0017$ & $2.3615$ \\
& iGRAIL & $5$  & $9.2794$ & $0.0118$ & $0.0018$ & $9.2930$ \\
\cmidrule(lr){2-7}

& FL     & $10$ & $4.7013$ & $0.0000$ & $0.0008$ & $4.7021$ \\
& BC-FL  & $10$ & $4.6747$ & $0.0223$ & $0.0041$ & $4.7012$ \\
& iGRAIL & $10$ & $18.2875$ & $0.0223$ & $0.0041$ & $18.3139$ \\
\cmidrule(lr){2-7}

& FL     & $15$ & $7.0394$ & $0.0000$ & $0.0011$ & $7.0406$ \\
& BC-FL  & $15$ & $7.0599$ & $0.0350$ & $0.0052$ & $7.1001$ \\
& iGRAIL & $15$ & $27.4470$ & $0.0362$ & $0.0052$ & $27.4884$ \\
\cmidrule(lr){2-7}

& FL     & $20$ & $9.4149$ & $0.0000$ & $0.0018$ & $9.4167$ \\
& BC-FL  & $20$ & $9.3674$ & $0.0441$ & $0.0074$ & $9.4188$ \\
& iGRAIL & $20$ & $37.0532$ & $0.0450$ & $0.0069$ & $37.1051$ \\

\midrule

\multirow{12}{*}{%
\shortstack[l]{Diabetes-\\Prediction \cite{mustafa_diabetes_prediction}}}
& FL     & $5$  & $2.6421$ & $0.0000$ & $0.0004$ & $2.6425$ \\
& BC-FL  & $5$  & $2.5347$ & $0.0117$ & $0.0026$ & $2.5490$ \\
& iGRAIL & $5$  & $12.1324$ & $0.0130$ & $0.0031$ & $12.1484$ \\
\cmidrule(lr){2-7}

& FL     & $10$ & $5.0987$ & $0.0000$ & $0.0008$ & $5.0995$ \\
& BC-FL  & $10$ & $5.0665$ & $0.0220$ & $0.0056$ & $5.0940$ \\
& iGRAIL & $10$ & $23.0823$ & $0.0251$ & $0.0063$ & $23.1137$ \\
\cmidrule(lr){2-7}

& FL     & $15$ & $7.8515$ & $0.0000$ & $0.0020$ & $7.8535$ \\
& BC-FL  & $15$ & $8.1700$ & $0.0419$ & $0.0088$ & $8.2206$ \\
& iGRAIL & $15$ & $37.6932$ & $0.0419$ & $0.0096$ & $37.7447$ \\
\cmidrule(lr){2-7}

& FL     & $20$ & $10.7918$ & $0.0000$ & $0.0020$ & $10.7937$ \\
& BC-FL  & $20$ & $10.7869$ & $0.0492$ & $0.0124$ & $10.8485$ \\
& iGRAIL & $20$ & $48.7324$ & $0.0528$ & $0.0138$ & $48.7990$ \\

\midrule

\multirow{12}{*}{Heart \cite{heart_disease_45}}
& FL     & $5$  & $0.0320$ & $0.0000$ & $0.0003$ & $0.0323$ \\
& BC-FL  & $5$  & $0.0303$ & $0.0096$ & $0.0015$ & $0.0414$ \\
& iGRAIL & $5$  & $0.1096$ & $0.0112$ & $0.0020$ & $0.1228$ \\
\cmidrule(lr){2-7}

& FL     & $10$ & $0.0604$ & $0.0000$ & $0.0006$ & $0.0610$ \\
& BC-FL  & $10$ & $0.0642$ & $0.0217$ & $0.0033$ & $0.0892$ \\
& iGRAIL & $10$ & $0.2262$ & $0.0235$ & $0.0043$ & $0.2540$ \\
\cmidrule(lr){2-7}

& FL     & $15$ & $0.0938$ & $0.0000$ & $0.0008$ & $0.0946$ \\
& BC-FL  & $15$ & $0.0920$ & $0.0309$ & $0.0053$ & $0.1282$ \\
& iGRAIL & $15$ & $0.3231$ & $0.0333$ & $0.0060$ & $0.3623$ \\
\cmidrule(lr){2-7}

& FL     & $20$ & $0.1227$ & $0.0000$ & $0.0012$ & $0.1240$ \\
& BC-FL  & $20$ & $0.1186$ & $0.0381$ & $0.0066$ & $0.1633$ \\
& iGRAIL & $20$ & $0.4573$ & $0.0438$ & $0.0085$ & $0.5096$ \\

\end{longtable}
\endgroup

\section{Conclusion} \label{conclu}
The findings of this study indicate that privacy and security in federated healthcare learning should be considered throughout the
learning process rather than as isolated operations. Protecting local data alone is insufficient when model updates must also be exchanged, validated, and aggregated across multiple institutions. The evaluation also shows that these protections involve practical trade-offs between predictive utility, privacy, and computational cost. Such trade-offs are especially important when data distributions and the number of involved institutions are changing. The current evaluation is limited to public health datasets and simulated federated environments. Real healthcare deployments may involve changing client availability, larger institutional networks, different computational resources, and more complex adversarial behavior. Future work will therefore study dynamic participation, larger federations, stronger defenses against coordinated and Byzantine attacks, and heterogeneous computing environments. Further work will also investigate methods to reduce local computational overhead and improve the privacy-utility balance. Evaluation with realistic multi-institutional healthcare settings will be an important step toward assessing the practical applicability of this approach.

\bibliographystyle{unsrtnat}
\bibliography{ref}

\clearpage
\section*{Author Biographies}


\par\noindent
\begin{wrapfigure}{l}{0.21\textwidth}
    \centering
    \vspace{-8pt}
    \includegraphics[
        width=0.21\textwidth,
        height=1.25in,
        keepaspectratio
    ]{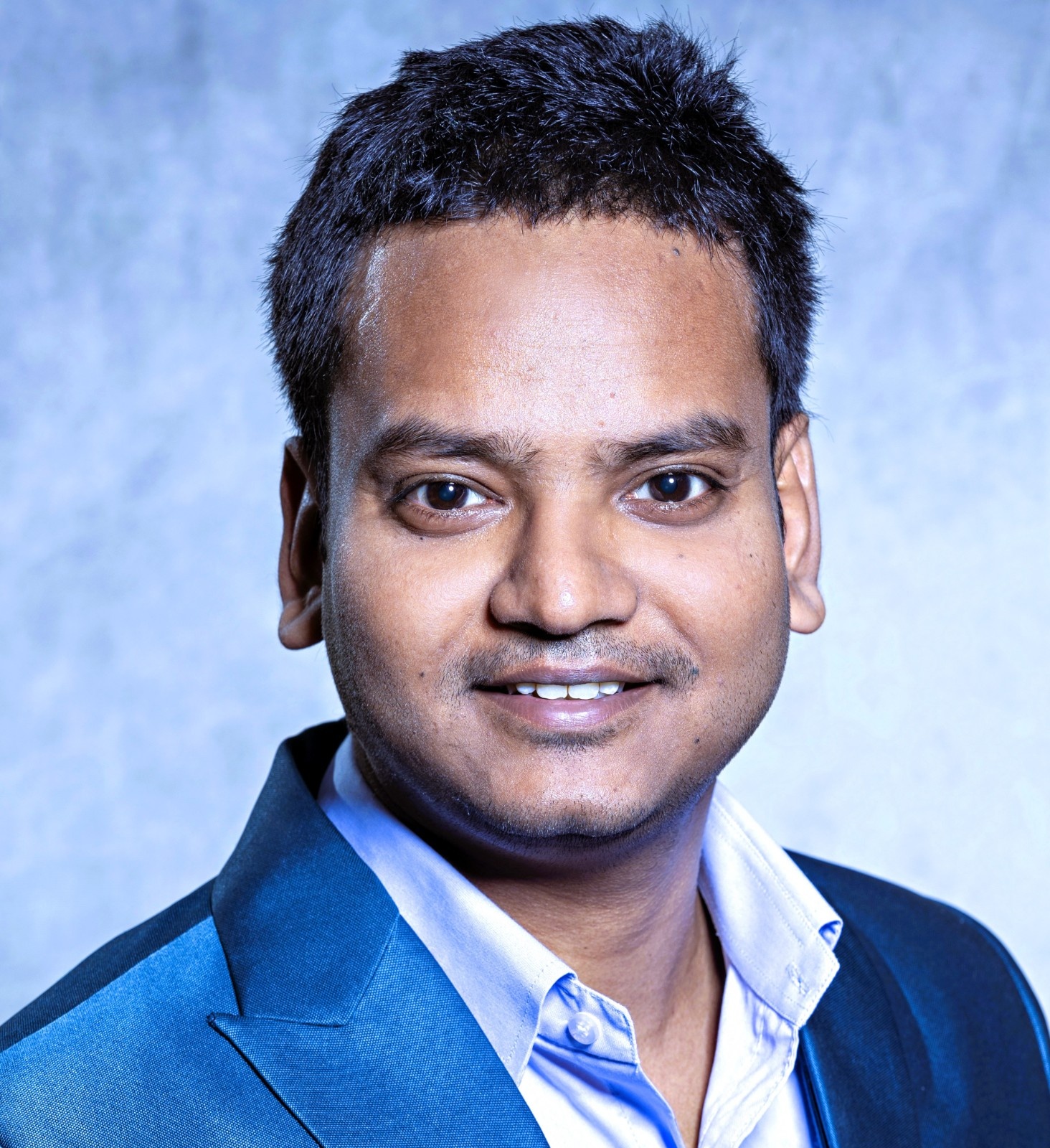}
    \vspace{-12pt}
\end{wrapfigure}
\noindent
\textbf{Debashis Das (Senior Member, IEEE)}
received his Ph.D. degree in Computer Science and Engineering from the University of Kalyani, India, in 2023, and his master's and undergraduate degrees in Computer Science and Engineering in 2018 and 2015, respectively. He is currently an assistant professor of computer science in the School of Science and Engineering at Benedict College, Columbia, South Carolina, USA. Previously, he served as a postdoctoral research associate at Meharry Medical College in the USA and as an assistant professor at Narula Institute of Technology in India. He has published over 60 peer-reviewed research articles, including several papers in IEEE Transactions and SCIE-indexed journals. His research interests include artificial intelligence, cybersecurity, privacy, blockchain technology, federated learning, and trustworthy intelligent systems. He serves as a reviewer for several international journals, including the IEEE Journal of Biomedical and Health Informatics, Future Generation Computer Systems, IEEE Transactions on Vehicular Technology, IEEE Internet of Things Journal, IEEE Transactions on Intelligent Transportation Systems, Peer-to-Peer Networking and Applications, and The Journal of Supercomputing. He has also served as an invited member of the Technical Program Committee (TPC) and as chair for several international conferences, including CICBA, BCCA, CCGRID, ICDCN, IEEE STP-CPS, ICSPIS, ISORC, and IoST. He has been an IEEE member since 2023.


\par\noindent
\begin{wrapfigure}{l}{0.20\textwidth}
    \centering
    \vspace{-8pt}
    \includegraphics[
        width=0.20\textwidth,
        height=1.25in,
        keepaspectratio
    ]{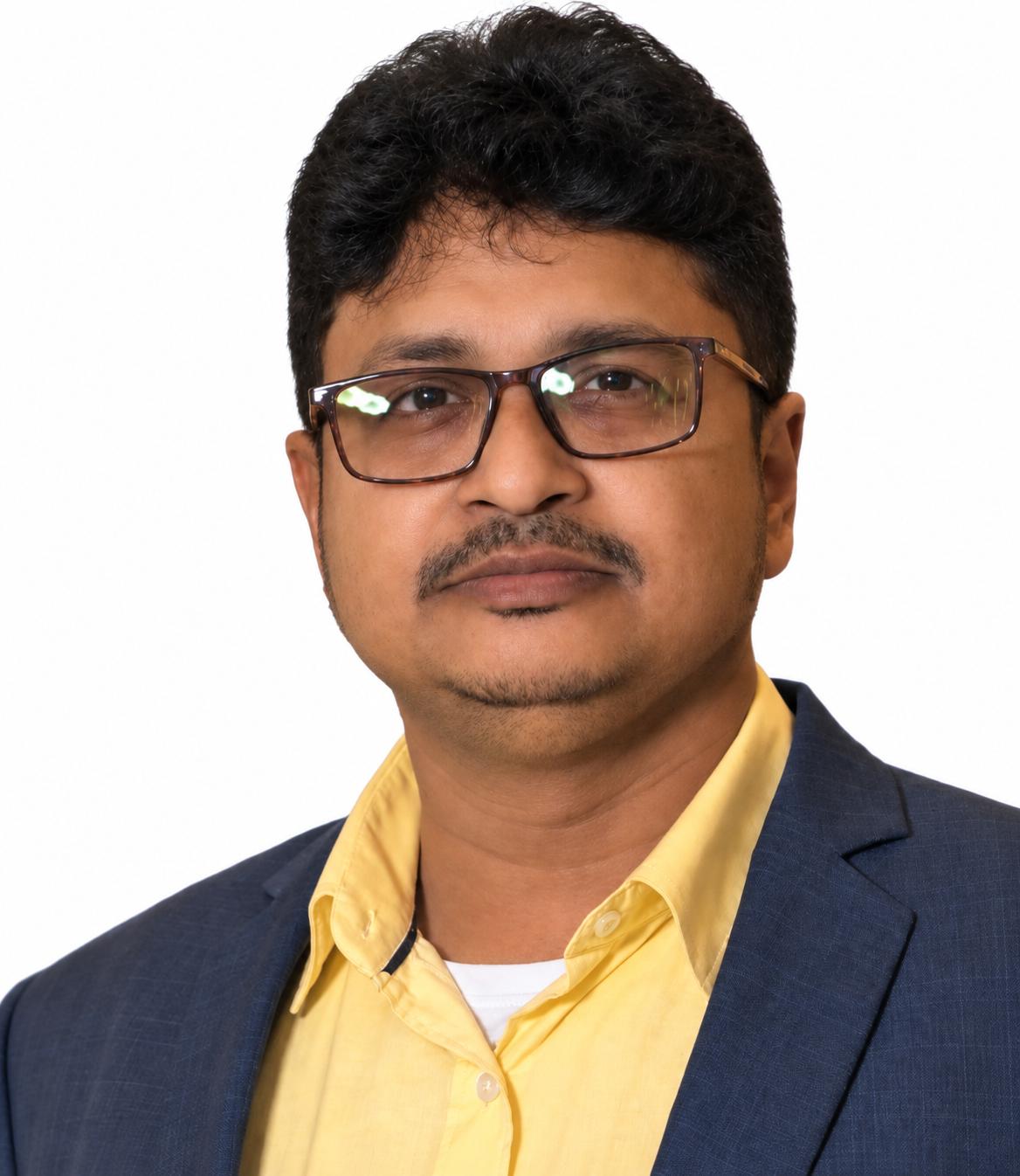}
    \vspace{-12pt}
\end{wrapfigure}
\noindent
\textbf{Sourav Banerjee (Senior Member, IEEE)}
received a Ph.D. degree in Computer Science and Engineering from the University of Kalyani in 2018. He completed his B.E. in Computer Science and Engineering and M.Tech. in Computer Science and Engineering. He is a faculty member in the Department of Computer Science and Engineering of Alipurduar Government Engineering and Management College, Village Chepani, P.O. Bakla School Danga, Block Alipurduar-II, District Alipurduar, West Bengal, Pin-736206. He has authored a number of reputable journal articles, books, book chapters, and international conference papers. He has edited a book on Green Cloud Computing. He is handling many special issues in reputed journals, such as Computers and Electrical Engineering Journal, IEEE Transactions on Consumer Electronics, Sustainability, and Electronics. His research interests include big data, blockchain, cloud computing, green cloud computing, green AI, 6G, cloud robotics, distributed computing, mobile communications, IoT, smart city, and global warming.

\par\noindent
\begin{wrapfigure}{l}{0.20\textwidth}
    \centering
    \vspace{-8pt}
    \includegraphics[
        width=0.20\textwidth,
        height=1.25in,
        keepaspectratio
    ]{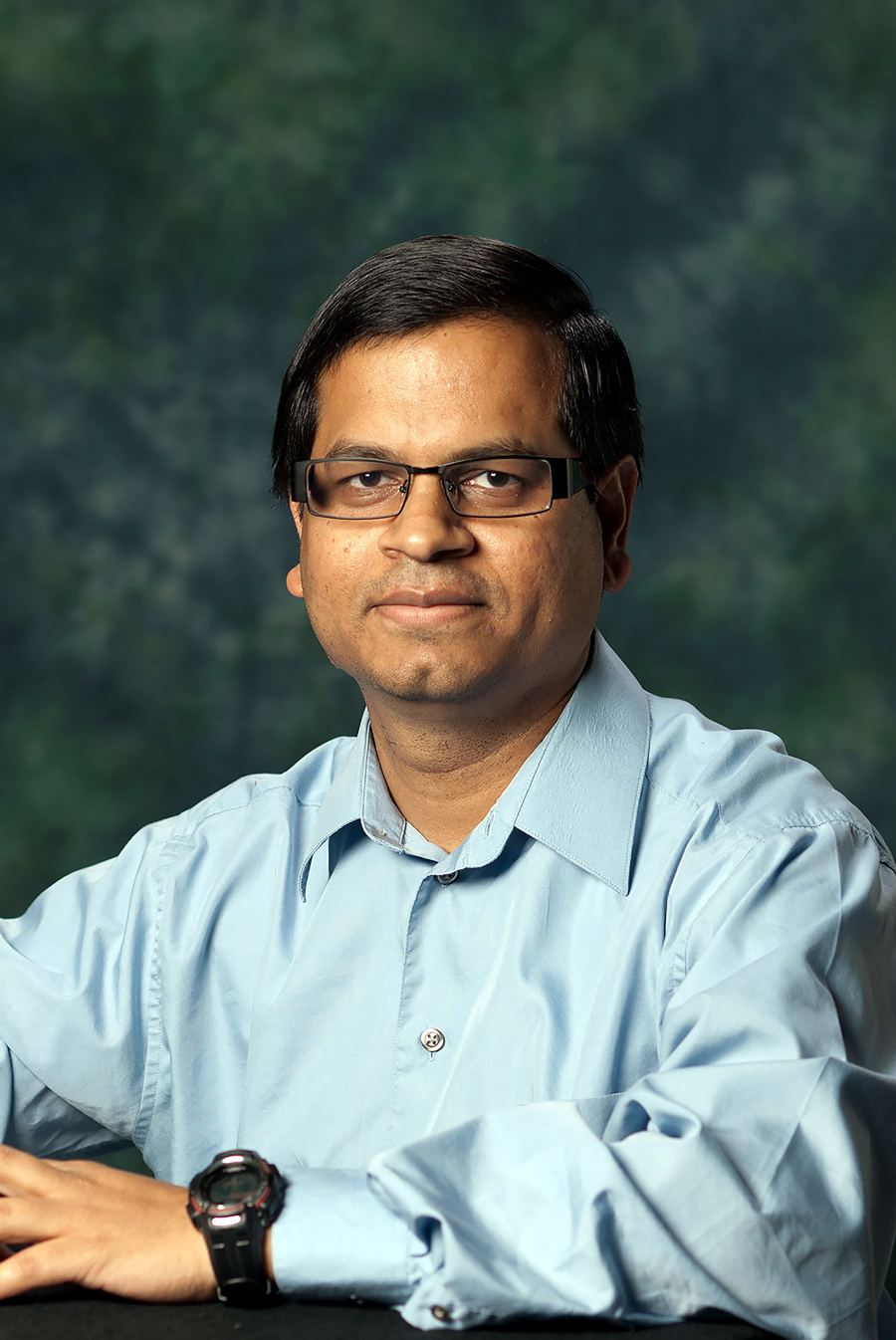}
    \vspace{-12pt}
\end{wrapfigure}
\noindent
\textbf{Saraju P. Mohanty (Senior Member, IEEE)}
received the bachelor's degree (Honors) in electrical engineering from the Orissa University of Agriculture and Technology, Bhubaneswar, in 1995; the master's degree in systems science and automation from the Indian Institute of Science, Bengaluru, in 1999; and the Ph.D. degree in computer science and engineering from the University of South Florida, Tampa, in 2003. He is a professor at the University of North Texas. His research is in ``Smart Electronic Systems,'' which has been funded by the National Science Foundation (NSF), the Semiconductor Research Corporation (SRC), the U.S. Air Force, NIDILRR, IUSSTF, and Mission Innovation. He has authored 600+ research articles, 5 books, and 6 granted patents. His Google Scholar h-index is 69, and his i10-index is 345, with 20,000+ citations. He is regarded as a visionary researcher in smart cities technology, with research on security- and energy-aware, AI/ML-integrated smart components. He introduced the Secure Digital Camera (SDC) in 2004 with built-in security features designed using Hardware-Assisted Security (HAS) or Security by Design (SbD) principles. He is widely credited with designing the first digital watermarking chip in 2004 and the first low-power digital watermarking chip in 2006. He is a recipient of 21 Best Paper Awards, the Fulbright Specialist Award in 2021, the IEEE Consumer Electronics Society Outstanding Service Award in 2020, the IEEE-CS-TCVLSI Distinguished Leadership Award in 2018, and the PROSE Award for Best Textbook in the Physical Sciences and Mathematics category in 2016. He has delivered 34 keynotes and served on 15 panels at various international conferences and served on the editorial board of several peer-reviewed international transactions/journals, including IEEE Transactions on Big Data (TBD), IEEE Transactions on Computer-Aided Design of Integrated Circuits and Systems (TCAD), IEEE Transactions on Consumer Electronics (TCE), and ACM Journal on Emerging Technologies in Computing Systems (JETC). He also served as the Editor-in-Chief (EiC) of the IEEE Consumer Electronics Magazine (MCE) from 2016 to 2021 and the Chair of the TPC on Very Large-Scale Integration (TCVLSI), IEEE Computer Society (IEEE-CS), during 2014-2018, and on the Board of Governors of the IEEE Consumer Electronics Society during 2019-2021. He serves on the steering, organizing, and program committees of several international conferences. He is the steering committee chair/vice-chair for the IEEE International Symposium on Smart Electronic Systems (IEEE-iSES), the IEEE-CS Symposium on VLSI (ISVLSI), and the OITS International Conference on Information Technology (OCIT). He has supervised 3 postdoctoral researchers, 22 Ph.D. dissertations, 29 M.S. theses, and 41 undergraduate projects.

\end{document}